\documentclass[apj]{emulateapj}
\slugcomment{Published in the Astrophysical Journal, 793, 19}

\usepackage{graphicx}
\usepackage{epstopdf}
\usepackage{stmaryrd}
\usepackage{amsmath}
\usepackage{ccaption}
\usepackage{multirow}
\usepackage{bm}
\usepackage{nicefrac}
\usepackage{comment}
\usepackage{textcomp}
\usepackage{color}
\definecolor{red}{rgb}{1,0,0}

\bibpunct[; ]{(}{)}{,}{a}{}{,}

\newcommand{\msun}{\ensuremath{M_{\odot}}}
\newcommand{\mstar}{\ensuremath{M_{\star}}}

\newcommand{\bdm}{\begin{displaymath}}
\newcommand{\edm}{\end{displaymath}}
\newcommand{\beq}{\begin{equation}}
\newcommand{\eeq}{\end{equation}}
\newcommand{\bit}{\begin{itemize}}
\newcommand{\eit}{\end{itemize}}
\newcommand{\ben}{\begin{enumerate}}
\newcommand{\een}{\end{enumerate}}
\newcommand{\bfi}{\begin{figure}[htb]}
\newcommand{\bpfi}{\begin{figure}[p]}

\newcommand{\lir}{\ensuremath{L_{\rm IR}}}
\newcommand{\lco}{\ensuremath{L'_{\rm CO}}}
\newcommand{\aCO}{\ensuremath{\alpha_{\rm CO}}}
\newcommand{\aCOMS}{\ensuremath{\langle\aCO\rangle_{\rm MS}}}

\newcommand{\mhtwo}{\ensuremath{M_{\rm H_2}}}
\newcommand{\mgas}{\ensuremath{M_{\rm gas}}}
\newcommand{\mmol}{\ensuremath{M_{\rm mol.}}}

\newcommand{\tdep}{\ensuremath{\tau_{\rm dep.}}}
\newcommand{\sfr}{\rm SFR}
\newcommand{\sfrMS}{\ensuremath{\langle\sfr\rangle_{\rm MS}}}
\newcommand{\ssfr}{\rm sSFR}
\newcommand{\ssfrMS}{\ensuremath{\langle\ssfr\rangle_{\rm MS}}}
\newcommand{\sfe}{\rm SFE}
\newcommand{\sfeMS}{\ensuremath{\langle\sfe\rangle_{\rm MS}}}

\shorttitle{Molecular gas in star-forming galaxies: simple scaling laws}
\shortauthors{Sargent et al.}

\begin{document}

%% LaTeX will automatically break titles if they run longer than
%% one line. However, you may use \\ to force a line break if
%% you desire.

\title{Regularity underlying complexity: a redshift-independent description of the continuous variation of galaxy-scale molecular gas properties in the mass-star formation rate plane}

\author{M.~T. Sargent\altaffilmark{1, 2, $\star$},
E. Daddi\altaffilmark{1},
M. B\'ethermin\altaffilmark{1},
H. Aussel\altaffilmark{1},
G. Magdis\altaffilmark{3},
H.~S. Hwang\altaffilmark{4},
S. Juneau\altaffilmark{1},
D. Elbaz\altaffilmark{1},
E. da Cunha\altaffilmark{5}
}

\altaffiltext{$\star$}{~E-mail: \texttt{Mark.Sargent@sussex.ac.uk}}

\altaffiltext{1}{~CEA Saclay, DSM/Irfu/S\'ervice d'Astrophysique, Orme des Merisiers, F-91191 Gif-sur-Yvette Cedex, France}
\altaffiltext{2}{~Astronomy Centre, Department of Physics and Astronomy, University of Sussex, Brighton, BN1 9QH, UK}
\altaffiltext{3}{~Department of Physics, University of Oxford, Keble Road, Oxford OX1 3RH, UK}
\altaffiltext{4}{~Smithsonian Astrophysical Observatory, 60 Garden Street, Cambridge, MA 02138, USA}
\altaffiltext{5}{~Max-Planck-Institut f\"ur Astronomie, K\"onigstuhl 17, D-69117 Heidelberg, Germany}

\begin{abstract}
Star-forming galaxies (SFGs) display a continuous specific star formation rate (sSFR) distribution which can be approximated by two log-normal functions: one encompassing the galaxy main sequence, the other a rarer, starbursting population. Starburst sSFRs can be regarded as outcome of a physical process (plausibly merging) taking the mathematical form of a log-normal boosting kernel that enhances star formation activity. We explore the utility of splitting the star-forming population into main-sequence and starburst galaxies -- an approach we term ``2-Star Formation Mode" (2-SFM) framework -- for understanding their molecular gas properties. Star formation efficiency (SFE) and gas fraction variations among SFGs take a simple redshift-independent form, once these quantities are normalized to the corresponding values for average main-sequence galaxies. SFE enhancements during starburst episodes scale supra-linearly with the SFR increase, as expected for mergers. Consequently, galaxies separate more clearly into loci for starbursts and normal galaxies in the Schmidt-Kennicutt plane than in (s)SFR versus \mstar\ space. Starbursts with large deviations ($>$10-fold) from the main sequence, e.g. local ULIRGs, are not average starbursts, but are much rarer events whose progenitors had larger gas fractions than typical main-sequence galaxies. Statistically, gas fractions in starbursts are reduced two- to threefold compared to their direct main-sequence progenitors, as expected for short-lived SFR boosts where internal gas reservoirs are depleted more quickly than gas is re-accreted from the cosmic web. We predict variations of the conversion factor \aCO\ in the SFR-\mstar\ plane and we show that the higher sSFR of distant galaxies is directly related to their larger gas fractions.
\end{abstract}


\section{Introduction}
\label{sect:intro}

Studies of star-forming galaxies (SFGs) over the last decade have revealed a positive and tight correlation between their current star formation rate (SFR) and stellar mass \mstar, which is intimately linked to the integral of the preceding star formation (SF) activity. Initially observed at low redshift \citep[e.g.,][]{brinchmann04, salim07, wyder07}, this ``star-forming main sequence (MS)" was soon shown to be present out to $z$\,$\sim$\,2 \citep{noeske07, elbaz07, daddi07b}. Subsequent work on the relation between SFR and \mstar\ in SFGs charted its evolution to $z$\,$\sim$\,2.5 using different SFR tracers and selection criteria \citep[e.g.,][]{damen09, dunne09, pannella09, santini09, kajisawa10,oliver10, elbaz11,  lee11, rodighiero11, wuyts11, whitaker12}, has explored factors affecting the exact shape and dispersion of the MS \citep[e.g.,][]{karim11, salmi12} and has traced it out to even higher redshifts $z$\,$\sim$\,3--4 \citep[e.g.,][]{daddi09, magdis10}. The existence of a scaling relation between SFR and \mstar\ throughout much of cosmic time implies that the (mass-dependent) assembly history of SFGs is characterized by a high degree of homogeneity and simplicity \citep[e.g.,][]{noeske07, bouche10, peng10, leitner12, behroozi13} and that the strong decline of the cosmic SFR density since $z$\,$\sim$\,2 \citep[e.g.,][and references therein]{reddy09, rodighiero10, karim11, magnelli11, cucciati12} reflects the uniform SFR evolution of the majority of the SFG population rather than a decreased frequency of episodic starburst (SB) events. Nevertheless, the study of SBs remains central to understanding the nature of interacting galaxies and the physics of merging events that may produce the most luminous sources at all redshifts \citep[henceforth abbreviated as `S12']{sargent12}.\\
SF activity at a rate which locally occurs only in strong SBs is common in massive MS galaxies in the distant universe. Hence alternatives to pure luminosity-selection are required for obtaining a census of bursty SF activity at high redshift, e.g., based on their position in the SFR versus \mstar\ plane \citep[e.g.,][]{rodighiero11, whitaker12} or based on their morphology \citep[e.g.,][]{kartaltepe12, kaviraj13}. The latter approach relies on the assumption that SBs are generally triggered by interactions between galaxies, as observed in the local universe \citep[e.g.,][]{sanders88, barton00}, while MS galaxies would represent a ``normal", secular channel of stellar mass growth in galaxies that is fueled by the steady accretion of cold, primordial gas \citep[e.g.,][]{bouche10}.\\
A series of studies on the star-forming population has improved our understanding of normal (MS) galaxies and SBs and is in qualitative agreement with this picture. Starbursting sources are more compact on average \citep[e.g.,][]{elbaz11, rujopakarn11} than MS galaxies, which have a stellar structure that is well described by exponential disks \citep[e.g.,][]{wuyts11, salmi12}. They display deficits in the intensity of infrared (IR) spectral features (e.g., polycyclic aromatic hydrocarbon (PAH) bands or in the far-IR [CII]-line; \citealp{elbaz11, graciacarpio11}), and have warmer IR spectral energy distributions \citep[SEDs; e.g.][and references therein]{heisler94, sandersmirabel96, chapman03, elbaz11, magdis11, bethermin12}. These are telltale features of intense and spatially concentrated SF as are expected to occur in interacting or merging galaxies where gravitational torques funnel gas to their centers \citep[e.g.,][]{mihos96, hopkins06}.\\
The efficiency with which gas is converted into stars in such settings may be up to an order of magnitude higher than in the extended gas reservoirs that fuel SF activity in normal galaxies out to $z$\,$\sim$\,2 \citep[e.g.,][]{daddi10b,genzel10, tacconi13}. This strong contrast in star formation efficiency (SFE\,$\equiv$\,SFR/$M_{\rm gas}$) is often taken as one of the most clear-cut manifestations of the existence of two distinct SF laws -- a secular mode in main-sequence galaxies and an SB mode characterized by short depletion timescales \citep[$\lesssim$100\,Myr, e.g.][and references therein]{solomonvdbout05}. Whether or not such a bimodality represents the physical reality has been questioned \citep[e.g.,][]{narayanan12} on the grounds of discrete ``concordance" values being assumed for the CO-to-H$_2$ conversion factor \aCO\ and, second, due to the expectation that SF laws at a basic level should be expressed in terms of volumetric quantities rather than observationally more easily accessible surface densities of SFR and gas \citep[e.g.,][]{krumholz12}. The lack of known sources with SFEs between those measured for normal disk galaxies and strong SBs could also be a selection effect: initial CO follow-up observations of high-redshift galaxies targeted only highly luminous sources experiencing ``bursty" SF \citep[submillimeter galaxies and QSOs; e.g.,][]{omont96, frayer98, walter03, greve05, maiolino07, tacconi08} and following improvements in the sensitivity of millimeter receivers \citep{chenu07, perley11}, dedicated studies of typical MS galaxies were undertaken \citep[e.g.,][]{daddi08, daddi10a, tacconi10, geach11, tacconi13, bauermeister13a}. If, as discussed in \citet{renaud12}, the gas density distribution function -- which reflects the turbulence-driven structure of the interstellar medium (ISM) --  is a crucial factor in determining the shape of SF laws, then intermediate SFEs should indeed occur in, e.g., minor mergers or in certain stages of galaxy interactions when the gas density distribution is not modified from the steady state as strongly as during final coalescence. However, a dichotomy in the distribution of SFEs could still occur if the timescales for such variations were short \citep[e.g.,][]{teyssier10,bournaud11a}.

\noindent In this paper we consider a large sample of local and high-redshift SFGs which we use to extend the ``2 Star-Formation Mode" (2-SFM) framework introduced in S12 to the molecular gas component of SFGs. The 2-SFM framework relies on basic observables (e.g., the evolution of specific star formation rate (sSFR) in MS galaxies or their stellar mass distribution) and correlations between observables (e.g. the star-forming MS or the Schmidt-Kennicutt (S-K) relation). Our goal is to describe how the SFE and the molecular gas content of galaxies are related to their location with respect to the MS, i.e., to their sSFR, which is the main diagnostic of ``starburstiness" within the 2-SFM framework. We will show that the population of massive SFGs that reside on the MS has similar molecular gas properties across a broad range of redshifts ($z$\,$\lesssim$\,3) and we will use our detailed description of the SB population and its SFR-``boosting" developed in Section \ref{sect:boostmath} to demonstrate how in the 2-SFM framework a bimodal behavior in terms of SFE arises naturally even in the absence of discrete SF laws for normal galaxies and SBs. The description of SFE in the SFG population developed in this paper forms the basis for the prediction of molecular gas mass functions and CO luminosity functions in a companion paper (M.T. Sargent et al., in prep.; henceforth Paper II).\\
The outline of this manuscript is as follows. Section \ref{sect:data} introduces the observational data set we use and how it was homogenized. We then employ this reference sample in Section \ref{sect:calib} to calibrate galaxy-scale SF laws -- both in terms of observables (\lir\ \& \lco) or intrinsic quantities (\sfr\ \& \mhtwo) -- for galaxies at low and high redshift. These calibrations depend on the adopted CO-to-H$_2$ conversion factor \aCO, and the corresponding systematics will also be assessed in Section \ref{sect:calib}. The mathematical description of the SB population in the 2-SFM framework, and its relation to MS galaxies is the focus of Section \ref{sect:2SFM} where we derive the distribution of the burst amplitudes -- the ``boost function" -- that transforms a theoretical population of pure main-sequence star forming galaxies into the observed distribution of sSFR. We discuss what physical mechanisms could produce this boost function and consider in particular the possible link between SBs and galaxy mergers. In Section \ref{sect:results} we will combine the redshift-independent, integrated S-K law derived in Section \ref{sect:calib} with the evolution of the sSFR distribution from S12 to construct prescriptions for the relative variation of molecular gas properties of normal and SB galaxies that are particularly simple (and self-similar) once they are referred to the properties of the average MS galaxy. Our results are presented in three main blocks: the SFE and gas fractions of MS galaxies and SBs are the subject of Section \ref{sect:SFE} and \ref{sect:fgas}, respectively; Section \ref{sect:XCO} focuses on the CO-to-H$_2$ conversion factor and its variation within the MS and among starbursting systems. We then discuss and summarize our findings in Sections \ref{sect:discussion} and \ref{sect:summary}.

\noindent Throughout this article we adopt the WMAP-7 cosmology \citep[$\Omega_m$\,=\,0.273, $\Omega_{\Lambda}$\,+\,$\Omega_m$\,=\,1 and $H_0$\,=\,70.4 km\,s$^{-1}$\,Mpc$^{-1}$;][]{larson11}. SFRs and stellar masses are given for a \cite{chabrier03} initial mass function\footnote{~Logarithmic masses and SFRs based on a \cite{salpeter55}, a \cite{kroupa01} and a \cite{baldryglazebrook03} IMF are converted to the Chabrier scale by adding -0.24\,dex, 0\,dex and 0.02\,dex, respectively.} (IMF). All literature values have been adapted accordingly. Metallicities are given on the \citet[][henceforth `KD02']{kewleydopita02} scale\footnote{~When necessary, metallicity information from the literature was converted to the KD02 calibration by means of the prescriptions in \citet{kewleyellison08}.} and (molecular) gas mass estimates include a 36\% correction for helium.

\section{Data}
\label{sect:data}

We discuss two different kinds of data sets in this section. To begin with (Section \ref{sect:moldata}), we describe individual, CO-detected SFGs which we will utilize to establish the basic scaling relations (e.g. for SFE or gas fractions) that link the molecular gas content of massive (\mstar\,$>$\,10$^{10}$\msun) SFGs to fundamental galaxy properties like (s)SFR or \mstar. In Section \ref{sect:GOODSgals} we introduce statistical samples of SFGs at $z$\,$\sim$\,1 and 2. These will subsequently be used (1) to visualize/simulate complete samples of galaxies that obey the aforementioned scaling relations, and (2) to extend the analysis to fainter galaxies where the validity of such scaling relations can be verified with image stacking.

\subsection{The Reference Sample of Individual Star-forming Galaxies}
\label{sect:moldata}

Our ``reference sample" of normal galaxies at redshifts $z$\,$\lesssim$\,3 comprises 131 sources from the recent literature (see Sections \ref{sect:MSdata_loz} \& \ref{sect:MSdata_hiz}). We complement these with local and high-redshift starbursts with measured CO-to-H$_2$ conversion factors (see Section \ref{sect:SBdata}). These are essential for a further investigation of the notion that the ``bimodality" of SF is particularly pronounced in terms of SFE \citep[e.g.,][]{daddi10a,genzel10}.

\subsubsection{Normal Galaxies: Low-redshift CO-detections}
\label{sect:MSdata_loz}

The HERACLES survey \citep{leroy08, leroy09, leroy13} targeted the CO($J$=2$\rightarrow$1) transition in nearby ($D$\,$\lesssim$\,15\,Mpc) THINGS galaxies \citep{walter08} with the IRAM 30\,m single-dish telescope. Here we select 20 galaxies with spiral galaxy morphology and stellar mass \mstar\,$\geq$\,$10^{10}$\,\msun\ from the HERACLES sample. Stellar masses (converted to the \citet{chabrier03} scale) and morphological information are taken from the compilations of \citet{skibba11} or \citet{leroy08, leroy09}, or from the NASA/IPAC Extragalactic Database (NED)\footnote{~\texttt{http://ned.ipac.caltech.edu}} if not listed in either of the former. Metallicity estimates for most of the selected HERACLES spirals are provided in \citet{moustakas10}. The IR (8-1000\,$\mu$m) luminosities attributed to the HERACLES galaxies are based on the photometry reported in \citet{dale07} and have been calculated following Equation 22 in \citet{draineli07}.\\
We augment the local MS galaxies from the HERACLES data set with a subset of galaxies from the first release of the COLD GASS survey \citep{saintonge11} for which an accurate IR luminosity could be calculated thanks to the presence of a counterpart in either the {\it IRAS} Faint Source Catalog \citep[v2;][]{moshir92} or the {\it AKARI}/Far-Infrared Surveyor \citep[FIS;][]{kawada07} all-sky survey Bright Source Catalog \citep[v1.0;][]{yamamura10}. An additional cut in stellar mass \citep[taken from][]{saintonge11} at \mstar\,=\,$10^{10}$\,\msun\ excluded all less massive sources. Among the 222 sources in the first COLD GASS data release 32 fulfill these criteria; they lie in the redshift range 0.025\,$<$\,$z$\,$<$\,0.05 and are all late-type galaxies with CO($J$=1$\rightarrow$0) fluxes measured by the IRAM 30\,m telescope at signal-to-noise ratio, $S/N$\,$>$\,4. IR luminosities for these galaxies were computed using the SED library of \citet{charyelbaz01} and allowing renormalization of the templates when fitting the reliable\footnote{~Flux quality flags are either ``high" or ``moderate" for {\it IRAS} sources and ``high" for {\it AKARI} sources.} {\it IRAS} or {\it AKARI} photometry at $\lambda_{\rm rest}$\,$\geq$\,30\,$\mu$m \citep[for details on the IR SED-fitting see][]{hwang10}.\\

\subsubsection{Normal Galaxies: Intermediate- \& High-redshift CO-detections}
\label{sect:MSdata_hiz}

CO-transitions in MS galaxies at $z$\,$>$\,0 have been targeted by \citet[CO($J$=1$\rightarrow$0) at $z$\,$\sim$\,0.4]{geach09, geach11}, \citet[CO($J$=2$\rightarrow$1) at $z$\,$\sim$0.5 \& 1.5]{daddi10a, daddi10b}, in the PHIBSS survey by \citep[][CO($J$=3$\rightarrow$2) at $z$\,$\sim$\,1.2 \& 2.3]{tacconi13}, and by \citet[CO($J$=3$\rightarrow$2) at $z$\,$\sim$\,3]{magdis12a} -- all using the IRAM Plateau de Bure interferometer -- and by the EGNoG survey \citep[CO($J$=1$\rightarrow$0) at 0.06\,$\lesssim$\,$z$\,$\lesssim$\,0.3]{bauermeister13a} using the Combined Array for Research in Millimeter-wave Astronomy. These observations have produced line flux measurements at $S/N$\,$\gtrsim$\,4 toward 70 of 79 observed galaxies (for the remaining galaxies 3\,$\sigma$ upper flux limits are available). The stellar masses of these sources are in the range 10$^{10}$\,$<$\,\mstar/\msun\,$\lesssim$\,$5\times10^{11}$, as determined by SED fitting of the near-UV to near-IR  broad-band photometry.\\
The derivation of SFRs varies among the different aforementioned studies. EGNoG galaxies have SFRs that are based on emission line fluxes and which were extracted from the MPA-JHU value-added catalog for SDSS DR7 \citep[see][for details]{bauermeister13a}. \citet{geach09, geach11} estimate the IR-luminosity from the flux of the 7.7\,$\mu$m emission, as constrained by the {\it Spitzer} IR spectrograph, while \citet{daddi10b} based the luminosity measurements for their $z$\,$\sim$\,0.5 sources on {\it Spitzer}/MIPS 24\,$\mu$m fluxes. SFR-estimates for $z$\,$\sim$\,1.5 sBzK galaxies presented in \citet{daddi10a} are an average of dust-corrected UV luminosities, mid-IR continuum luminosities from 24\,$\mu$m imaging, and Very Large Array (VLA) 1.4\,GHz radio continuum fluxes, all of which were found to give consistent SFR-estimates. \citet{magdis12a} adopted a similar averaging approach but were able to add {\it Herschel}/PACS and SPIRE photometry to constrain the dust-emission of their $z$\,$\sim$\,3 Lyman-break galaxies (LBGs). We adopted all of these SFR (\lir) measurements without further modifications, but in the interest of maximal sample homogeneity chose to re-compute \lir\ values for PHIBSS sources at $z$\,$\sim$\,1.2 as \citet{tacconi13} report SFRs based on a combination of extinction-corrected optical emission lines, UV flux and mid-IR photometry. To derive new \lir\ and {\sfr} estimates we searched the Far Infrared Deep Extragalactic Legacy \citep[FIDEL, PI: M. Dickinson, see also data description in][]{magnelli09} coverage of the Extended Groth Strip for associated {\it Spitzer}/MIPS 24\,$\mu$m detections and converted these to IR luminosities with the MS SFG IR SED of \citet{elbaz11}. SFRs for $z$\,$\sim$\,1.2 PHIBSS galaxies are the averages of the SFRs reported in \citet{tacconi13} and our IR-based estimates. Two of the 27 PHIBSS sources at $z$\,$\sim$\,1.2 have VLA 1.4\,GHz detections from the AEGIS20 catalog \citep{ivison07}, such that the original, the 24\,$\mu$m- and a radio-based SFR could be averaged. For PHIBSS galaxies at $z$\,$\sim$\,2.3 neither IR nor radio flux measurements were available; we hence used the original SFRs reported by \citet{tacconi13} which are based on extinction-corrected H$\alpha$ luminosities and which we converted to an IR-luminosity following \citet{kennicutt98b}. 

\begin{figure}
\epsscale{1.05}
\centering
\plotone{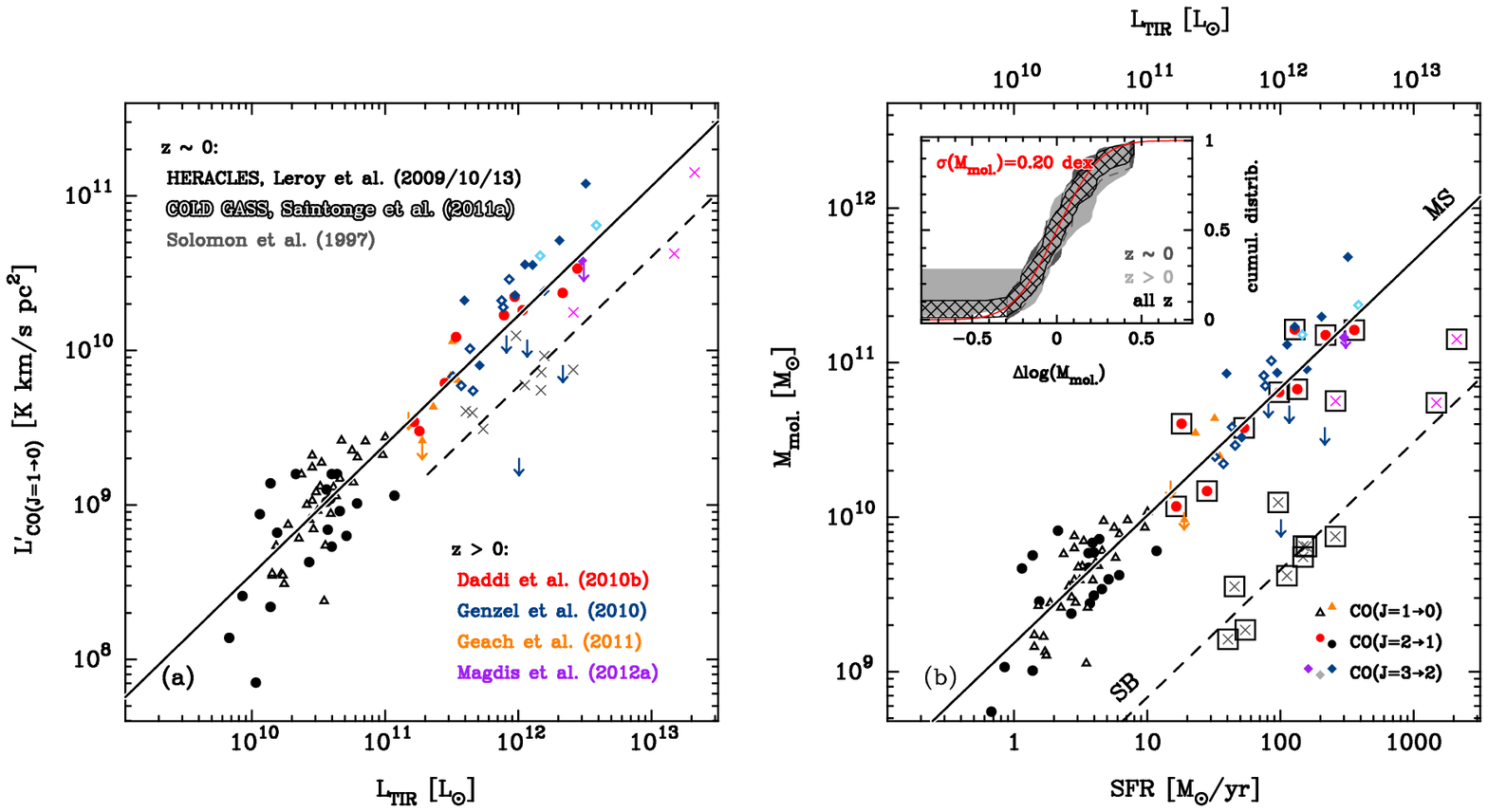}
\caption{\footnotesize Redshift dependence of physical properties for galaxies in our calibration sample (crosses -- starbursting galaxies; all other symbols -- massive MS galaxies). ({\it a}) stellar mass {\mstar}; ({\it b}) star formation rate {\sfr}; ({\it c}) offset from the mass- and redshift-dependent mean locus of the star-forming main sequence (MS) (s)SFR/(s)SFR$_{\rm MS}$ (see Appendix \ref{appsect:sSFR}); ({\it d}) star formation efficiency SFE; ({\it e}) normalized SFE, i.e. the efficiency normalized to the SFE that a galaxy of equal gas mass would have if it lay directly on the integrated S-K relation (see Figure \ref{fig:invSKcalib}(b)). Boxed symbols are used for galaxies with an observational constraint on the CO-to-H$_2$ conversion factor \aCO\ (see text for details); symbol coloring depends on the literature source (see legend above figure). The shape of the symbols reflects the CO-transition which was observed to infer molecular gas properties: triangles -- CO(J{=}1$\rightarrow$0); dots -- CO(J{=}2$\rightarrow$1); diamonds -- CO(J{=}3$\rightarrow$2).
\label{fig:sampintro}}
\end{figure}

\noindent To summarize, we have compiled a sample of 131 massive (\mstar\,$\geq$\,$10^{10}$\,\msun), MS galaxies\footnote{~Three of these -- one each from HERACLES, PHIBSS and the sample of \citet{geach11} -- have an sSFR excess larger than four, the frequently adopted threshold to separate normal from SB galaxies \citep[e.g.,][]{rodighiero11}, but a clearly disk-like morphology showing now indications of interactions. We have hence included them in our main-sequence reference sample.} with CO-detections of which 46\% are low-redshift systems ($z$\,$<$\,0.1) systems and the remaining 54\% redshifted to 0.1\,$<$\,$z$\,$<$\,3.2. Our sample is thus well-balanced between nearby and distant galaxies, as shown in Figure \ref{fig:sampintro} where we review the most important physical properties of our reference galaxies, e.g., {\mstar}, {\sfr} and (s){\sfr}-excess with respect to the star-forming MS.

\subsubsection{Starbursting Galaxies with Measured $\alpha_{\rm CO}$}
\label{sect:SBdata}

Starbursting galaxies with the same information as available for our reference sample of normal galaxies are listed in the recent study of \citet[SB-like galaxies at 2.3\,$<$\,$z$\,$<$\,4]{magdis12b} and in \citet[local starbursting {\it IRAS} ULIRGs]{solomon97}. An accurate assessment of the behavior of SFE during SB episodes -- one of the main aims of this paper -- requires an observational determination of the CO-to-H$_2$ conversion factor \aCO. In the case of the local ULIRGs, we thus restrict ourselves to nine objects -- VII Zw 31, Arp 193, Arp 220, Mrk 273, 00057+4021, 02483+4302, 10565+2448, 17208-0014, 23365+3604 -- with two independent measurements of \aCO: one based on dynamical constraints\footnote{~In the following we adopt a mass-to-light ratio \aCO\ that is given by the ratio between gas mass and dynamical mass within a $\sim$1-3\,kpc region encompassing both the inner, high-density nuclear disk/ring and an outer, lower-density disk (with volume filling factor $\sim$0.1 for the gas) of the ULIRGs modeled in \citet[see their Tables 3 \& 9]{downessolomon98}.} by \citet{downessolomon98} and the other based on large velocity gradient (LVG) radiative transfer modeling by \citet{papadopoulos12}.\\
The flux of the CO($J$=1$\rightarrow$0) transition toward our subsample of nine $z$\,$<$\,0.07 {\it IRAS}-detected ULIRGs was measured with the IRAM 30\,m single-dish telescope by \citet{solomon97}. We computed their IR luminosities (and thence SFRs) using spectroscopic redshifts from NED and all available {\it IRAS} photometry, following the standard recipes provided by \citet[their Table 1]{sandersmirabel96}. Stellar masses -- which are particularly important for the characterization of these systems in terms of sSFR, our prime indicator of ``starburstiness" -- have been published for some of the ULIRGs in our sample \citep[e.g.,][]{elbaz07, dacunha10, howell10, u12}, but to our knowledge no single study has done this consistently for all sources of interest. We hence re-estimated stellar masses for the \citet{downessolomon98} ULIRGs based on 2 Micron All Sky Survey $K$-band fluxes \citep{skrutskie06} and prescriptions for mass-to-light ratios, \mstar/$L_K$, as derived by \citet[their Equation 2]{arnouts07} and \citet[their Equation B2]{juneau11}. The stellar masses we adopt in the following for the \citet{downessolomon98} ULIRGs were obtained by averaging the estimates calculated according to these two prescriptions. They agree well with the available literature measurements (median offset 0.1\,dex).\\
In addition to the nine low-redshift SBs just described we also include the three high-$z$ sub-millimeter galaxies GN20 ($z$\,=\,4.05), SMMJ2135-0102 ($z$\,=\,2.325) and HERMES J105751.1+573027 ($z$\,=\,2.957) in our analysis. The recent determination of their conversion factor \aCO\ in \citet{magdis12b} by means of the \mgas/$M_{\rm dust}$-ratio technique \citep[see also][]{leroy11, magdis11} relied on: (i) a far-IR SED/dust emission that is accurately constrained by {\it Herschel} and millimetric continuum observations (see \citealp{magdis12b} for a detailed listing), and (ii) CO($J$=1$\rightarrow$0) line fluxes from the (J)VLA \citep[for GN20; see][]{carilli10, hodge12} and the Green Bank Telescope (for SMMJ2135-0102 and HERMES J105751.1+573027; see \citealp{swinbank10} and \citealp{riechers11}, resp.). All three sources have sSFR enhancements with respect to the MS of at least a factor three, as constrained by the optical to near-IR and IR SED-fitting of \citet{magdis12b}. Furthermore, their CO-to-H$_2$ conversion factors are systematically lower than that of the Milky Way, similar to the values typically measured in interacting local ULIRGs.

\subsection{Statistical Samples of Star-forming Galaxies in GOODS-South}
\label{sect:GOODSgals}

With the purpose of demonstrating the applicability of our recipes for computing molecular gas properties  for observed galaxy samples, based on individual measurements of stellar masses and SFRs, we use two samples of $K$-selected galaxies in the GOODS-S field, at $z$\,$\sim$\,1 and 2 taken from the work of \citet{daddi07a} and \citet[][see also \citealp{salmi12} for more details on the $z$\,$\sim$\,1 sample]{daddi07b}. The same samples were used in the recent papers by \citet{magdis12b} and \citet{mullaney12a}. We refer to the original papers for details of how stellar masses were derived, based on empirical recipes using colors and absolute luminosities. The SFRs of the galaxies at $z$\,$\sim$\,1 and 2 are based on 24\,$\mu$m and UV observations, respectively, and are known to compare well on average with other tracers including {\it Herschel}-based SFR measurements \citep{daddi07a, elbaz10, reddy12}.

\section{Galaxy-scale star formation laws: correlations between SFR and gas mass and the associated observables}
\label{sect:calib}

Recent reports \citep[e.g.,][]{daddi10b, genzel10} of a systematic offset between the power law relation linking the overall surface density of SFR and gas mass \citep[the Schmidt-Kennicutt (S-K) law $\Sigma_{\sfr}\,{\propto}\,\Sigma_{\rm gas}^n$;][]{schmidt59, kennicutt98a} of normal galaxies and SBs were highly influential in shaping the notion of ``bimodal" SF. These findings are subject to two systematic uncertainties. First, the measured offset between normal and SB galaxies depends on (potentially population-dependent) recipes for CO-to-H$_2$ conversion factors, which are hard to measure for a statistically significant number of SFGs, especially at $z$\,$\gg$\,0. Second, the sampling of the S-K plane obtained as a result of targeted CO follow-up observations toward selected SFGs is patchy. Constructing a reliable and statistically representative sampling of the distribution of SFGs in the S-K plane is thus not only important to explore different modes of SF. The observed S-K law is also often
referred to as a benchmark for the performance/validity of recipes for ISM processes in simulations \citep[e.g.,][]{robertsonkravtsov08, monaco12} and hence used to gauge our understanding of the underlying physics itself.\\
In this section we return to our literature compilation of low- and high-redshift MS galaxies with CO-detections that we presented in Sections \ref{sect:MSdata_loz} and \ref{sect:MSdata_hiz}. We re-measure the slope and dispersion of the galaxy-scale SF law and in doing so for the first time are able to incorporate \aCO\ measurements from the recent study of \citet{magdis12b} for a fraction of our reference sample of normal galaxies. Rather than using \sfr\ and gas mass surface densities, we consider the simpler relations between integrated quantities, namely the total \sfr\ and molecular gas mass or the corresponding observables, \lir\ and $L'_{{\rm CO}(J=1\rightarrow0)}$. For the rest of this article we will use ``\lco" as a shorthand for the line luminosity $L'_{{\rm CO}(J=1\rightarrow0)}$ of the first rotational transition of $^{12}$CO.

\begin{figure*}
\epsscale{1.13}
\centering
\plotone{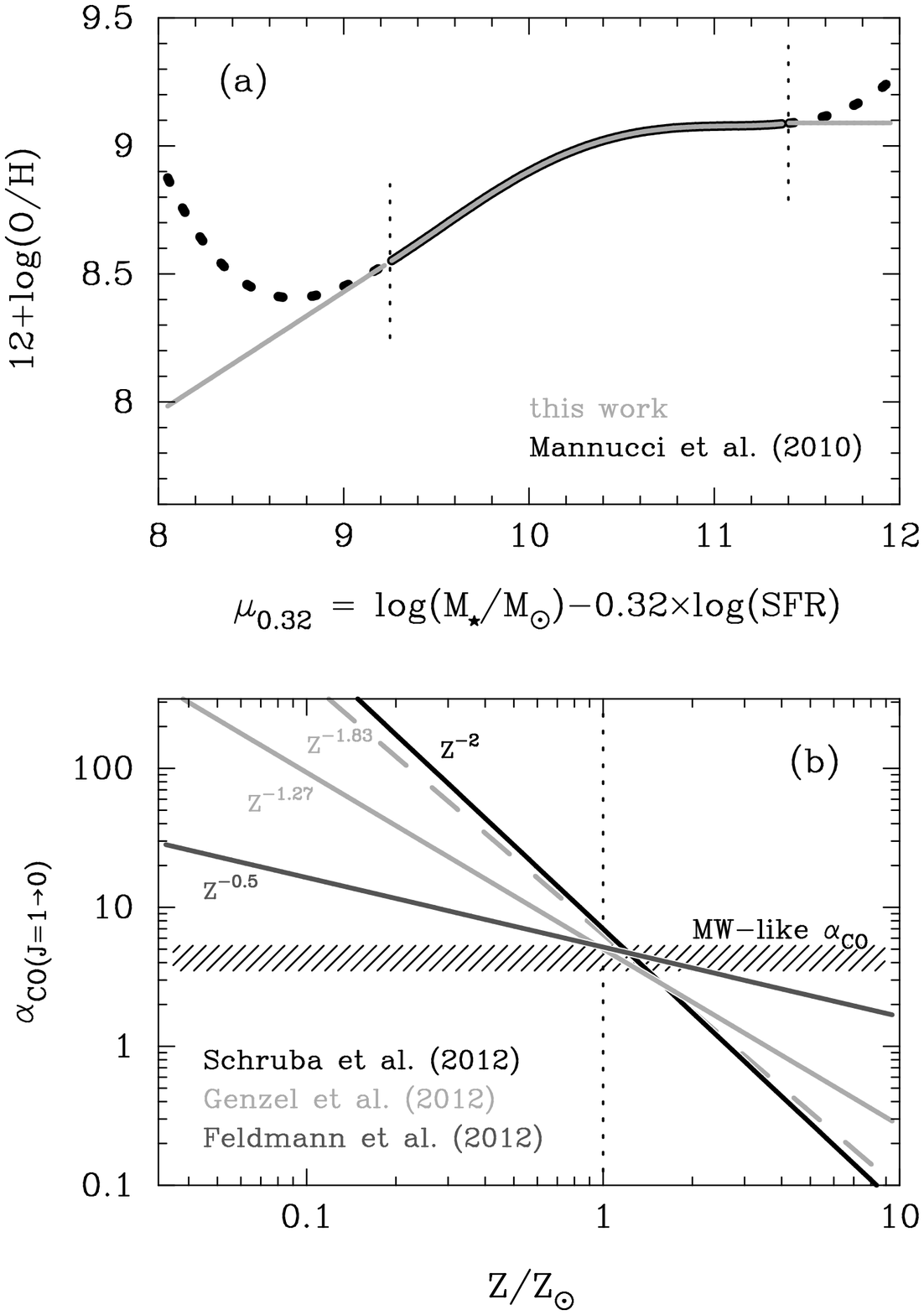}
\caption{\footnotesize Observed correlation between measures of star formation rate and molecular gas content for massive ($M_{\star}$\,$>$\,10$^{10}\,M_{\odot}$) main-sequence (MS) galaxies and starbursts (SBs) at low and high redshift. Low-redshift sources (filled/open black symbols -- normal galaxies; grey crosses -- starbursting (U)LIRGs) are from the HERACLES, GOLD GASS and {\it IRAS} surveys. Redshifted galaxies from the literature (see legend) are plotted in color ({\it pine green} -- 0.04\,$\lesssim$\,$z$\,$\lesssim$\,0.4; {\it orange} -- $z$\,$\sim$\,0.4; {\it red} -- $z$\,$\sim$\,0.55 \& $z$\,$\sim$\,1.5; {\it blue} -- $z$\,$\sim$\,1.2 \& $z$\,$\sim$\,2.3; {\it magenta} -- SBs at 2.3\,$<$\,$z$\,$<$\,4), with arrows indicating 3\,$\sigma$ upper limits for CO non-detections. Open blue and cyan symbols indicate modified SFR estimates (see section \ref{sect:MSdata_hiz} for details) based on radio and/or IR data for sources in \citet{tacconi13}. The shape of the symbol indicates which CO transition was detected toward the individual sources (cf. legend in lower right corner of panel {\it b}). ({\it a}) Correlation between infrared luminosity ($L_{\rm IR}$) and CO-luminosity ($L'_{\rm CO(J{=}1\rightarrow0)}$; standard excitation corrections -- e.g. \citet{dannerbauer09, leroy09} -- were applied to $J$\,$>$\,1 transitions) with the best-fitting relation derived for the MS galaxy sample plotted as a solid black line. The strong SBs considered here (cross symbols) are on average offset to higher \lir\-values by a factor of three (dashed line). ({\it b}) Inverse, integrated Schmidt-Kennicutt relation between SFR and molecular gas mass ($M_{\rm mol.}$), the latter having been derived based on either (i) observational determinations of $\alpha_{\rm CO}$ (available for sources with boxed symbols) or (ii) using a metallicity-dependent conversion factor (see text of Section \ref{sect:MgasvsSFR}). The dispersion about the best-fit linear trend (solid black line) for MS galaxies is approximately Gaussian with a dispersion $\sigma(M_{\rm mol.})$\,$\sim$\,0.2\,dex (red curve in inset). Dashed line -- offset locus with approx. 15 times higher SFE for strong SB galaxies.
\label{fig:invSKcalib}}
\end{figure*}

\subsection{$L'_{\rm CO(J{=}1\rightarrow0)}$ versus \lir}
\label{sect:LCOvsLIR}

We begin with the relation between the observables, \lir\ and \lco, that are the starting point for estimating {\sfr}s and the molecular gas content of SFGs. As described in Section \ref{sect:moldata} all galaxies considered in the following have stellar masses \mstar\,$\geq$\,10$^{10}$\,\msun. Current observations of BzK-selected MS galaxies \citep[e.g.,][]{dannerbauer09, aravena10} suggest that typical excitation corrections for the first two higher order transitions $J$=2$\rightarrow$1 and 3$\rightarrow$2 are $r_{21}$\,=\,0.8-0.9 and $r_{31}$\,$\simeq$\,0.5. Similarly, \citet{leroy09} find an average $J$=2$\rightarrow$1/$J$=1$\rightarrow$0 line ratio of 0.8 for nearby HERACLES galaxies and \citet{bauermeister13b} report $J$=3$\rightarrow$2/$J$=1$\rightarrow$0 line ratios of 0.46$\pm$0.07 for $z$\,$\sim$\,0.3 galaxies in the EGNoG sample. In Figure \ref{fig:invSKcalib}(a) we plot the accordingly corrected CO($J$=1$\rightarrow$0) luminosities of local (black and white symbols) and redshifted sources (color symbols) against their IR luminosity. We then fitted the CO line luminosity as a function of IR luminosity. While performing a regression of \lir\ on \lco\ would be more natural (as representing the relation between cause and effect, i.e. \mmol\ and \sfr, resp.) our choice is motivated by the aim to provide recipes for the molecular gas content and associated tracer emission beginning with the currently observationally more easily accessible \sfr\ measurements. A \citet[][hereafter ``BJ'"]{buckleyjames79} regression \citep[implemented as described in][]{isobe86}, which allows for a statistically correct treatment of the 3\,$\sigma$ upper detection limits for six galaxies from \citet{tacconi10} and \citet{geach11}, gives
\small
\begin{flalign}
&{\rm log}\left(\frac{L'_{{\rm CO}(J{=}1\rightarrow0)}}{\rm K\,km/s\,pc^2}\right) = \alpha_1 + \beta_1\,{\rm log}\left(\frac{L_{\rm IR}}{L_{\odot}}\right)~, \quad{\rm with} \label{eq:LCOvsLIR}\\[1ex]
&\left(\alpha_1; \beta_1\right) = (0.54\pm0.02; 0.81{\pm}0.03)\quad \text{for normal galaxies.} \nonumber
\end{flalign}
\normalsize
The dispersion about this best-fit trend line in the $y$-direction is 0.21\,dex. In performing the linear regression we have down-weighted sources detected in CO($J$=3$\rightarrow$2) by a factor of two due to the large excitation corrections $r_{31}$.\\
Under the assumption that SBs follow a correlation with identical slope, we use our reference sample of SB galaxies (see Section \ref{sect:SBdata}) to solve for the normalization of Equation \ref{eq:LCOvsLIR} that best reproduces their average offset. We find
\small
\begin{equation*}
\left(\alpha_1; \beta_1\right) = (0.08_{-0.08}^{+0.15}; 0.81)~,
\end{equation*}
\normalsize
i.e. an offset of 0.46\,dex or approx. a factor 2.9 with respect to the locus of MS galaxies. This similar systematic difference was already indicated by \citet[see their Figure 3]{solomon97} in their pioneering analysis of CO-emission in nearby ULIRGs. Local MS galaxies with the IR luminosities of starbursting ULIRGs are, however, exceedingly rare (e.g., S12) such that this difference could also have been explained by a double power-law nature of the SF law or a single, steeper relation (\lir\,$\propto$\,\lco$^{1.3}$) owing to different probability gas density distributions in mergers and normal galaxies \citep{narayanan08, juneau09}. The advent of CO line flux measurements for high-$z$ normal galaxies with ULIRG-luminosities has since added another piece of evidence in support of a systematic offset \citep[e.g.,][]{genzel10}.

\subsection{\mmol~versus SFR}
\label{sect:MgasvsSFR}

The integrated S-K law linking the molecular gas mass (\mmol) and SFR is expected to have a different slope or curvature than the correlation between logarithmic luminosities \lco\ and \lir\ unless the average CO-to-H$_2$ conversion factor 
\small
\begin{equation}
\alpha_{{\rm CO}(J=1\rightarrow0)} = \frac{\mmol}{L'_{{\rm CO}(J{=}1\rightarrow0)}} \nonumber
\end{equation}
\normalsize
is a constant. Evidence to the contrary has been presented in numerous observational studies, the most recent of which are \citet{leroy11}, \citet{schruba12}, \citet{genzel12} and \cite{remy-ruyer14} who show that there is a tendency for \aCO\ to decrease with metallicity both in local galaxies and in massive MS galaxies at $z$\,$<$\,2.5 in general. Note that \citet{sandstrom13} find a similar trend for decreasing $\alpha_{\rm CO}$ with increasing metallicity when considering spatially distinct regions within nearby galaxies. In the next paragraphs we discuss a scheme for assigning metallicity-dependent \aCO\ values to the normal SFGs in our reference sample (Section \ref{sect:statmetall}) and then proceed to fit the resulting relation between \sfr\ and \mmol\ in Section \ref{sect:invSKcalib}. We close this section with an assessment of the robustness of the galaxy-scale SF law obtained in this way (Section \ref{sect:invSKrobust}).

\subsubsection{Statistically Inferred CO-conversion Factors}
\label{sect:statmetall}

\begin{figure}
\epsscale{1.05}
\centering
\plotone{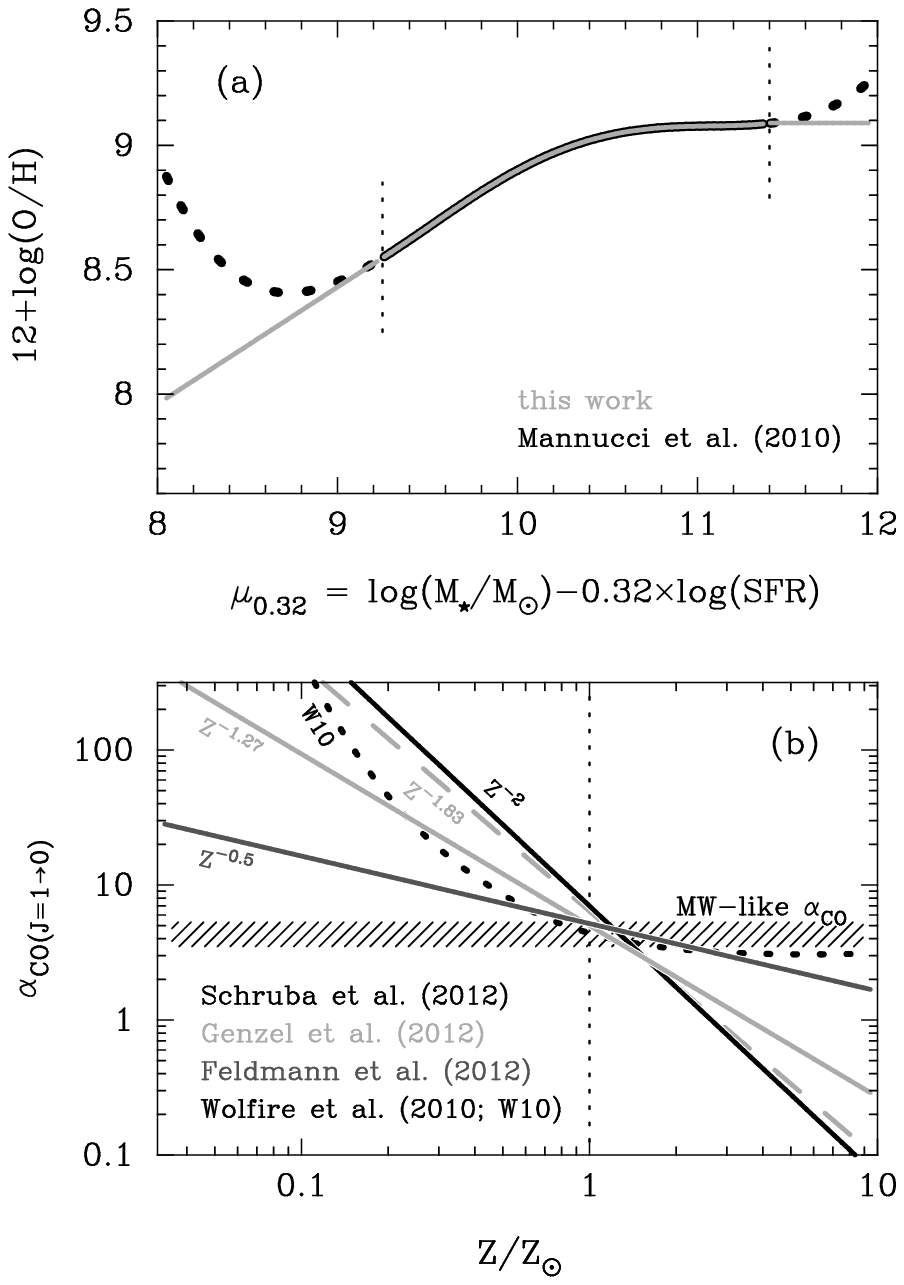}
\caption{\footnotesize Overview of recipes used to assign CO-to-H$_2$ conversion factors, \aCO, to observed and modeled galaxies based on their SFR and stellar mass. ({\it a}) Fundamental metallicity relation (FMR) as determined by \citet[][{\it thick line section}]{mannucci10} and continuation of the FMR assumed in our analytic-empirical modeling to regions of parameter space where the mathematical description of the plane proposed in \citet{mannucci10} diverges due to the absence of data. Note that the extension beyond $\mu_{0.32}$\,$\in$\,[9.25, 11.4] affects only a minority of the galaxies modeled in the present work. ({\it b}) Recently proposed relations between metallicity \citep[expressed in multiples of the solar metallicity $Z_{\odot}$ in the system of][]{kewleydopita02} and \aCO\ based on simulations \citep{feldmann12a} or observations \citep[for the latter study the fit derived for both low- and high-$z$ SFGs ({\it light grey, dashed}) and that for a sample restricted to $z$\,$\geq$\,1 SFGs ({\it light grey, solid}) is shown]{schruba12, genzel12}. The dotted line shows the shielding-based prescription of \citet{wolfire10} which we adopt here. All recipes predict a Milky-Way-like \aCO\ at approx. solar metallicity but diverge significantly at lower/higher enrichment due to the different measured slopes (see annotations beside trend lines).
\label{fig:FMPadapt}}
\end{figure}

For 9 of the 131 MS galaxies in our reference sample the CO-to-H$_2$ conversion factor has been measured by  \citet{magdis12b} and found to be broadly consistent with the trend between \aCO\ and metallicity $Z$ found for galaxy samples in the nearby universe. For the remaining 122 galaxies we assign a metallicity-dependent \aCO\ as explained below, and assume that metallicity can be deduced in a statistical sense from stellar mass \mstar\ and \sfr\ as proposed by \citet{mannucci10} and \citet{laralopez10}. In the following we use this ``fundamental metallicity relation" (FMR; see Figure \ref{fig:FMPadapt}(a)) as parametrized by \citet{mannucci10}. Since SBs constitute a small fraction of the star-forming population, the FMR primarily reflects the dependence of metallicity on \sfr\ and \mstar\ for MS galaxies. As the stellar mass and SFR of the CO-detections in our reference sample are known (within observational errors) we can use the FMR to statistically infer metallicities $Z({\rm SFR}, M_{\star})$, and thence CO-conversion factors $\alpha_{\rm CO}$ for each of the galaxies in our reference sample\footnote{~\citet{mannucci10} originally were only able to study the FMR at $z$\,$\gg$\,0 for massive (\mstar/\msun\,$\geq$\,10$^{10}$) field galaxies. New work has since extended the relation to lower stellar masses \citep{cresci12} and separately verified its validity in the cluster environment at $z$\,$\sim$\,1.4 \citep{magrini12}. The FMR is generally assumed to hold over the range 0\,$<$\,$z$\,$<$\,2.5, i.e. should apply to almost all galaxies in our reference sample. Beyond $z$\,$\sim$\,2.5 conflicting evidence for constancy \citep[e.g.,][]{dessaugesavadsky11, richard11, laralopez13, belli13} and evolution \citep[e.g.,][]{laskar11, sommariva12} of the FMR has been presented.}. This enables us to calibrate the \mmol\ versus SFR relation independently from the correlation between \lco\ and \lir\ derived in Section \ref{sect:LCOvsLIR}. Note that more general expectations for \aCO\ variations in the \mstar\ versus \sfr\ plane that account for (1) changes of the conversion factor within both the normal and starbursting galaxy population, and (2) the relative importance of these to classes of SFGs depending on the location in the plane, are the topic of Section \ref{sect:XCOmaps}.\\
A frequently used, first-order description of the metallicity dependence of \aCO\ is the single power-law
\small
\begin{equation}
{\rm log}(\aCO) = \nu + \xi\,{\rm log}\left(\nicefrac{Z}{Z_{\odot}}\right)~.
\label{eq:alphaofZ}
\end{equation}
\normalsize
Existing literature consistently reports the normalization $\nu$ of the \aCO\ versus metallicity relation in Equation \ref{eq:alphaofZ} to be such that Milky-Way-like conversion factors \aCO\,=\,4.4 $M_{\odot}$/(K\,km/s\,pc$^2$) are reached around solar metallicity. Measurements and expectations for the slope $\xi$ span a larger range which we illustrate in Figure \ref{fig:FMPadapt}(b). Analysis of the varying findings in the literature suggests that the slope measured depends on, e.g., the range of metallicities probed in different samples. This may be linked to a rapid steepening of the \aCO\ versus $Z$ relation that occurs around $Z$\,$\sim$\,1/3 to 1/2\,$Z_{\odot}$ \cite[e.g.,][]{leroy11, remy-ruyer14}. Models explicitly treating the shielding of CO by dust and (atomic and molecular) hydrogen can reproduce this behavior which arises because CO is more easily photodissociated at low metallicities \citep[see, e.g.,][and references therein]{bolatto13}. While a single power law as in Equation \ref{eq:alphaofZ} may adequately reproduce variations of \aCO\ among galaxies with approximately solar enrichment -- and hence among most of the massive SFGs in our reference sample -- we will also consider a broader range of metallicities in parts of Section \ref{sect:results}.\\
For this reason, we assign \aCO\ values to those 122 galaxies lacking an observational constraint with the following prescription from \citet{bolatto13}, which is based on modeling in \citet{wolfire10}:
\small
\begin{equation}
\frac{\aCO(Z')}{\aCO(Z'{=}1)} = {\rm exp}\left(\frac{+4.0\,\Delta A_V}{Z'\,\overline{A}_{V,{\rm MW}}}\right)\,{\rm exp}\left(\frac{-4.0\,\Delta A_V}{\overline{A}_{V,{\rm MW}}}\right)~.
\label{eq:alphaofZ_Wolfire}
\end{equation}
\normalsize
Here $Z'$\,=\,$\nicefrac{Z}{Z_{\odot}}$ is metallicity, normalized to solar abundance, and $\overline{A}_{V,{\rm MW}}$\,=\,5 is the mean visual extinction through a giant molecular cloud (GMC) at $Z_{\odot}$. $\Delta A_V$, the differential extinction between ISM regions where only the CO molecule or both CO and H$_2$ are found, is calculated as in Equation 26 in \citet{bolatto13}, except that we adopt an underlying double power-law relation between gas-to-dust ratio and metallicity in keeping with \citet{remy-ruyer14}. We assume that $\aCO(Z'{=}1)$\,=\,4.4\,$M_{\odot}$/(K\,km/s\,pc$^2$), i.e. a Millky-Way-like conversion factor at $Z_{\odot}$. With Equation \ref{eq:alphaofZ_Wolfire} the equivalent of the logarithmic slope $\xi$ as defined in Equation \ref{eq:alphaofZ} varies from -0.6 at $Z_{\odot}$ to a significantly steeper dependence at sub-solar metallicities as illustrated in Figure \ref{fig:FMPadapt}(b) (dotted line).

\subsubsection{Integrated Schmidt-Kennicutt Laws}
\label{sect:invSKcalib}

Multiplication of the excitation-corrected CO-luminosities of Figure \ref{fig:invSKcalib}(a) with (i) observed CO-to-H$_2$ conversion factors if available or (ii) the statistical CO-conversion factors discussed in the previous section provides a measure of the molecular gas mass for each of the MS galaxies in our reference sample. These measurements are plotted against their SFR in Figure \ref{fig:invSKcalib}(b). At the stellar masses considered here, the IR-excess \lir/$L_{\rm UV}$ of MS galaxies is in general $\sim$10 or larger \citep[e.g.,][]{whitaker12, pannella14, heinis14}, leading to a nearly 1:1 correspondence between \lir\ and SFR, as indicated by the lower and upper scale for the $x$-axis of Figure \ref{fig:invSKcalib}(b). We convert IR luminosities to SFRs following the prescription of \citet{kennicutt98b}.\\
BJ regression, applied to the data in Figure \ref{fig:invSKcalib}(b), returns a very similar logarithmic slope as for the correlation between \lco\ and \lir\ due to the relatively narrow range of metallicities spanned by our low- and high-$z$ data:
\small
\begin{flalign}
&{\rm log}\left(\frac{\mmol}{M_{\odot}}\right) = \alpha_{2,\,\sfr} + \beta_2\,{\rm log}\left(\frac{\sfr}{\nicefrac{\msun}{\rm yr}}\right)~, \quad{\rm with}\label{eq:MgasvsSFR}\\[1ex]
&\left(\alpha_{2,\,\sfr}; \beta_2\right) = \begin{cases} ~(9.22{\pm}0.02; 0.81{\pm}0.03) ~ \text{for normal galaxies}\\ ~(8.05_{-0.10}^{+0.29}; 0.81) ~\text{for strong starbursts.} \end{cases} \nonumber
\end{flalign}
\normalsize
Here the line parameters for the SBs in our reference sample were derived by solving for the normalization under the assumption of an identical, slightly sub-linear slope of the correlation between SFR and \mmol\ for both MS and SB galaxies. Alternatively, in units of \lir,
\small
\begin{flalign}
&{\rm log}\left(\frac{\mmol}{M_{\odot}}\right) = \alpha_{2,\,{\rm IR}} + \beta_2\,{\rm log}\left(\frac{L_{\rm IR}}{L_{\odot}}\right)~, \quad{\rm with}\label{eq:MgasvsLIR}\\[1ex]
&\left(\alpha_{2,\,{\rm IR}}; \beta_2\right) = \begin{cases} ~(1.14{\pm}0.02; 0.81{\pm}0.03) ~ \text{for normal galaxies}\\ ~(-0.03_{-0.10}^{+0.29}; 0.81) ~ \text{for strong starbursts.} \end{cases} \nonumber
\end{flalign}
\normalsize
The dispersion of the correlation is 0.20\,dex and almost equal for the low- and high-$z$ subsamples as shown in the inset panel of Figure \ref{fig:invSKcalib}(b) where we plot the \citet{kaplanmeier58} estimator for the cumulative distribution functions of the offsets $\Delta(\mmol)$ of the individual measurements from the best-fitting trend line in Equation \ref{eq:MgasvsSFR}. The average offset of the reference SBs with respect to the locus for normal galaxies is 1.17$_{-0.29}^{+0.10}$\,dex or roughly a factor 15 in terms of SFE\footnote{~In this section we have used the dynamical constraints from \citet{downessolomon98} to convert \lco\ to \mmol\ for the local starbursting ULIRGs. As discussed in Section \ref{sect:SB-XCO}, the average offset from the integrated S-K law does not change for this sample when we use the \aCO\ values given by \citet{papadopoulos12}.}. It should be emphasized, that this bimodality is arbitrary and merely reflects the properties of our small and incomplete selection of strong SBs. In Sect \ref{sect:2-SFM_SFE} we propose an empirical description of the SB population that allows for a continuous enhancement of SFE, depending on the importance of burst-induced SF activity.

\begin{figure*}
\epsscale{1.13}
\centering
\plotone{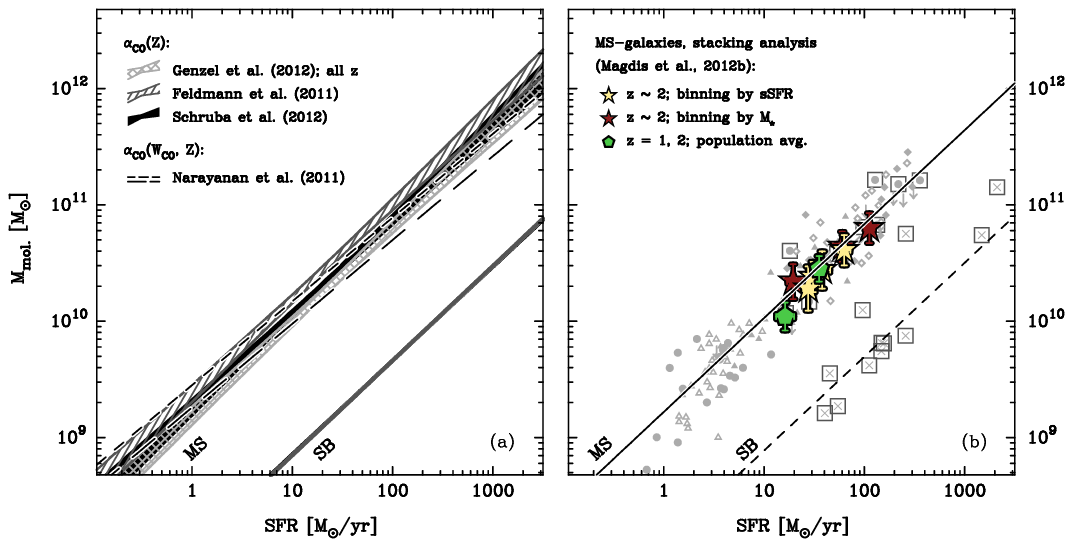}
\caption{\footnotesize Robustness assessment of the integrated inverse S-K relation calibrated in Figure \ref{fig:invSKcalib} (see also Equation \ref{eq:MgasvsSFR}). ({\it a}) Changes in the best-fit S-K relation, depending on the metallicity dependence, $Z^{\beta}$, of the conversion factor, \aCO, reported by recent observational \citep{genzel12, schruba12} and numerical \citep{narayanan11, feldmann12a} work. (Filled/hatched areas encompass the 95\% confidence region of the corresponding best-fit relations.) The additional dependence on the CO surface brightness, W$_{\rm CO}$, proposed by \citet{narayanan11} introduces an implicit dependence on galaxy size (long dashes -- CO-flux averaged over optical half-light radius; short dashes -- CO-flux averaged over two optical half-light radii). ({\it b}) Extension of high-$z$ S-K relations to the highest SFRs ($\sim$20\,$M_{\odot}$/yr) reached by the $z$\,$\sim$\,0 MS galaxies in our calibration sample. (Grey symbols in background reproduce data plotted in Figure \ref{fig:invSKcalib}.b). After inclusion of the stacked samples of $z$\,$\sim$\,2 MS galaxies from \citet[][colored symbols]{magdis12b}, the $z$\,$\sim$\,2 S-K relation is sampled over two orders of magnitude in SFR. The stacked samples bridge the gap between local and $z$\,$>$\,0 calibration sources and are aligned with the best-fit S-K relation determined using these, thereby providing evidence of a universal S-K relation for massive MS galaxies.
\label{fig:invSKrobust}}
\end{figure*}

\subsubsection{Robustness of Calibrated \mmol\ versus SFR Relation}
\label{sect:invSKrobust}

The integrated S-K law for MS galaxies which we calibrated in the previous section (Equation \ref{eq:MgasvsSFR}) constitutes a key ingredient for the description of the gaseous component of SFGs in the 2-SFM framework. It is thus essential to ascertain that the shape of our best-fit S-K law is not strongly dependent on assumptions made during the calculation of, e.g., gas masses.\\
A first potential cause of systematic uncertainty is the shape of the relation between \aCO\ and metallicity discussed in Section \ref{sect:statmetall}. The relation of our choice (see Equation \ref{eq:alphaofZ_Wolfire}), which is based on the models of \citet{wolfire10}, is equivalent to a continuously varying power law changing from an effective dependence \aCO\,${\propto}\,Z^{-1.4}$ to $Z^{-0.2}$ for metallicities 1/2\,$<$\,$Z/Z_{\odot}$\,$<$\,2. This is intermediate between the observed trends that range from $Z^0$ \citep[e.g.,][]{bolatto08} to $Z^{-2.7}$ \citep[e.g.,][]{israel97}. As representative examples of these different measurements\footnote{~Observational studies generally parametrize \aCO\ variations as a function of an absolute value of oxygen abundance rather than relative to the solar metallicity. The inferred oxygen abundances may vary significantly depending on the metallicity calibration and abundance diagnostic used \citep[e.g.,][]{kewleyellison08}. In addition to these systematic uncertainties concerning the normalization of the metallicity scale, transformations between metallicity systems following, e.g., the recipes in \citet{kewleyellison08} also change the curvature of the \aCO\ versus metallicity relation. As an example, a log-linear relation between \aCO\ and metallicity derived using the $R_{23}$ indicator \citep[e.g.,][]{pilyugin01, kewleydopita02} may become a convex or concave function of metallicity when converted directly to a system based on the N2 diagnostic \citep[e.g.,][]{denicolo02, pettinipagel04}. In view of these complications we chose to renormalize all literature determinations of the \aCO\ versus metallicity relation considered here to solar metallicity (see Figure \ref{fig:FMPadapt}(b)). In terms of oxygen abundance, solar enrichment generally corresponds to a value of log(O/H)+12\,$\approx$\,8.7, but this may change somewhat depending on the metallicity calibration \citep[e.g. log($Z_{\odot}$)+12\,=\,8.9 in the scale of][]{kewleydopita02}} we show  in Figure \ref{fig:invSKrobust}(a) the SF laws that we obtain when applying the simulation-based recipe of \citet[\aCO\,$\propto\,Z^{-0.5}$]{feldmann12a}, and those observationally determined by \citet[\aCO\,$\propto\,Z^{-1.27}$]{genzel12} and \citet[\aCO\,$\propto\,Z^{-2}$]{schruba12}, to the normal galaxies in our reference sample (where we again have assigned metallicities using the FMR). We find that the slope of the different S-K laws are very similar and that their normalization only varies by 0.2\,dex, such that they are all consistent within uncertainties among each other and also with our best-fit SF law as given by Equation \ref{eq:MgasvsSFR}. We also tested the prescription \citet{narayanan11} developed based on their simulations of disks and mergers. These authors parametrize \aCO\ as a function of metallicity and CO surface brightness, W$_{\rm CO}$, which introduces an implicit dependence on galaxy size. Using optical size measurements from the literature\footnote{~Optical half-light radii for MS galaxies in our reference sample were derived using the following literature sources: \citet{leroy08, leroy09} for HERACLES galaxies; \citet{foersterschreiber09, foersterschreiber11} and \citet{genzel10} for SINS galaxies; \citet{daddi10a} for CO-detected sBzK galaxies. No size information was available for the $z$\,$\sim$\,0.4-0.6 galaxies from \citet{geach11} and \citet{daddi10b}, nor for galaxies in the COLD GASS sample \citep{saintonge11}.} and the CO-fluxes of our reference galaxies we calculated CO-to-H$_2$ conversion factors and thence molecular gas masses following \citet{narayanan11}. Our subsequent fit to the data showed that -- for reasonable assumptions about the relative spatial distribution of optical and CO emission, see Figure \ref{fig:invSKrobust}(a) -- the \citet{narayanan11} prescription leads to a SF law that is a bit shallower than our preferred integrated S-K relation (\mmol\,$\propto$\,\sfr$^{0.72\pm0.04}$ versus \sfr$^{0.81\pm0.03}$, cf. Equation \ref{eq:MgasvsSFR}) but still agrees within 1.5\,$\sigma$. The good general agreement between all these different recipes is due to the flatness of the mass-metallicity relation at \mstar\,$>$\,10$^{10}$\,\msun.\\
A second source of systematic uncertainty is our assumption that massive galaxies at all redshifts align along a single integrated S-K law relating their SFR and molecular gas mass. By combining high-redshift galaxies from the PHIBSS survey with COLD GASS data \citet{tacconi13} recently presented an alternative scenario of parallel and linear S-K laws that are characterized by an SFE that increases with redshift. \citet{tacconi13} based their gas surface mass densities on CO($J$=3$\rightarrow$2) fluxes for $z$\,$>$\,1 SFGs and uniformly applied a Milky-Way-like conversion factor to all sources in their sample, regardless of stellar mass, SFR, and redshift. Both the strong excitation corrections applied to the high-$z$ SFGs that dominate the high-$\Sigma_{\rm gas}$ regime of the S-K relation and the universal \aCO\ are in principle uncertain enough to bring about seemingly systematic shifts between the high- and low-redshift galaxy population in the S-K plane. Given the fact that there is little overlap in $L_{\rm IR}$ between our own subsamples of low- and high-redshift galaxies (see, e.g., Figure \ref{fig:sampintro}(b)), we cannot rule out a series of offset and conceivably also curved SF laws. By including average \sfe\ constraints from the stacking analysis of \citet{magdis12b}, however, it is possible to bridge the luminosity gap between $z$\,$>$\,2 and local SFGs, as shown in Figure \ref{fig:invSKrobust}(b).  In combination with the BM/BX-selected galaxies of \citet{tacconi10}, the $z$\,$\sim$\,2 S-K relation thus spans one and a half orders of magnitude and is seen to extend continuously into the parameter space of intermediate-redshift (0.4\,$<$\,$z$\,$<$\,0.6) and local disks without evidence for a discontinuity. When considering individual detections, the 0.1\,$<$\,$z$\,$<$\,0.4 galaxies of \citet{geach09, geach11} and the EGNoG sample \citep{bauermeister13a} are aligned with the $z$\,$\sim$\,1.2 and 1.5 sample of \citet{tacconi10} and \citet{daddi10a}. Moreover, the skew and scatter around our universal S-K law are very similar for the subsets of $z$\,$\sim$\,0 and $z$\,$>$\,0 galaxies in our reference sample (see inset of Figure \ref{fig:invSKcalib}(b)). Based on these observations, we conclude that the assumption of a single, slightly sub-linear relation between \sfr\ and \mmol\ is presently a valid working hypothesis. This is also in line with recent work by \citet{feldmann13} who shows that in semi-analytic models a redshift-invariant and approximately linear SF--gas relation is able to reproduce several observed galaxy properties, including gas fractions, metallicities, UV luminosity functions, and the cosmic SF history.

\section{The inner workings of the 2-SFM framework: Main-sequence galaxies and boosted, Starbursting Sources}
\label{sect:2SFM}

While the S-K relation we found in Section \ref{sect:calib} for MS galaxies is very well defined, it is much less obvious which concrete form of the SF law should be used to describe the sparse and scattered SB data. Our approach to interpreting the incomplete information on these sources will be to statistically link them to a synthetic and complete SB population where we are able to relate the SB properties to the pre-starburst (MS) state. In the following we describe the ``2 Star-Formation Mode" (2-SFM) framework which is the basis for establishing this link.

\subsection{Basic Ingredients and Successes of the 2-SFM Framework}
\label{sect:2SFMintro}

2-SFM is a simple and self-consistent scheme for the prediction of basic properties of the SFG population that relies on basic observables (e.g. the evolution of sSFR in MS galaxies or their stellar mass distribution) and uses their mathematical description (e.g., the Schechter function parameterization of the stellar mass distribution or slope and normalization of the MS) to produce an analytico-empirical description of the statistical properties of SFGs. It can be both predictive (see, e.g., the indirect measurement of the evolution of molecular gas mass functions in Paper II) or help to (re)interpret existing measurements \citep[e.g., IR luminosity functions or source counts; see S12,][]{bethermin12, gruppioni13}.\\
We introduced the 2-SFM framework in S12 where we demonstrated that the observational constraints on the $z$\,$\lesssim$\,2.5 IR luminosity functions can be reproduced based on only three observables: (i) the redshift evolution of the stellar mass function for SFGs, (ii) the evolution of the sSFR of MS galaxies, and (iii) a double log-normal decomposition of the sSFR distribution at fixed stellar mass into contributions (assumed redshift- and mass-invariant) from MS and SB activity. The split into (overlapping but offset) (s)SFR distributions associated with MS and SB activity is based on the distributions of sSFR published for massive (\mstar/\msun\,$>$\,10$^{10}$) SFGs at $z$\,$\sim$\,2 published by \citet{rodighiero11}. The assumption that this double log-normal decomposition of the (s)SFR distribution is invariant with stellar mass and redshift leads to a good agreement with IR-observables (a mild decrease of the importance of the SB component by $<$50\% between $z$\,$<$\,1 and 0 leads to additional small improvements; see discussion in S12 and \citealp{bethermin12}).\\
The distinction between ``normal" SFGs (implicitly assumed to be growing their stellar mass in a secular mode on the star-forming MS) and starbursting galaxies is central to the 2-SFM framework and of particular interest since it yields observationally verifiable predictions of the notion that SF is a bimodal process at low and high redshift. In this vein \citet{bethermin12} assigned a characteristic \citep[albeit redshift-dependent; see][]{magdis12b} IR SED to MS and SB galaxies and showed that this simple approach is capable of reproducing the IR/radio source counts (incl. new {\it Herschel} counts) at 24--1100\,$\mu$m and 1.4\,GHz. Given the sensitivity of the source counts, this observation evidences that the 2-SFM framework provides a valid description of the dust emission from SFGs out to at least $z$\,$\sim$\,4, (i.e. over 84\% of the age of the universe). The 2-SFM description of the IR-properties of SFGs has also provided testable predictions which were verified in recent work, e.g. the redshift distribution of SCUBA-2 450\,$\mu$m sources \citep{geach13} and the redshift distribution of lensed 1.4\,mm sources detected with the South Pole Telescope \citep[see][Figure 9]{weiss13}. The good agreement of the predictions with the latter measurement suggests that the basic ingredients of the 2-SFM framework (e.g. the minor role of SBs) remain applicable out to $z$\,$\sim$\,6.

\subsection{Boosting of Main-sequence Galaxies: Mathematical Description}
\label{sect:boostmath}

Encouraged by the successful reproduction of the IR properties of the SFG population we now further develop the 2-SFM framework with the primary aim of using it for a predictive analysis of the molecular gas properties of SFGs at high redshift. In preparation for this we revisit the key ingredient of the 2-SFM framework -- the double log-normal decomposition of the (s)SFR distribution at fixed stellar mass. Analogously to S12 we write this (s)SFR distribution as the sum of two log-normal distribution functions $\mathcal{G}$ describing MS (MS) and starburst (SB) galaxies, respectively:
\small
\begin{equation}
p{\rm (sSFR)}|_{M_{\star}} = {\rm \mathcal{G}_{MS}(sSFR)} + {\rm \mathcal{G}_{SB}(sSFR)}~.
\label{eq:DGintro}
\end{equation}
\normalsize
The log-normal shape of the sSFR distribution of MS galaxies, $\mathcal{G}_{\rm MS}$, is observationally established by numerous studies on independent data sets and covering different redshift ranges \citep[e.g.,][C. Schreiber et al., in prep.]{rodighiero11, guo13}. Our assumption that the excess population of high-sSFR objects is drawn from a second log-normal distribution $\mathcal{G}_{\rm SB}$ is more uncertain, and its parameters are less well constrained (see Section 2.2 in S12). In addition to being fully consistent with observations, the main appeal of our choice of a log-normal $\mathcal{G}_{\rm SB}$ is that this implies the simplest possible relation (see also discussion at the end of Section \ref{sect:boostfct} and in Section 6.1.1) between normal galaxies and SBs: SB activity is the consequence of a stochastic process that at any given time acts on only a small subset of the MS population. The natural outcome of this is a random resampling of the parent (MS) distribution, i.e. a second log-normal. Furthermore, note that with this parameterization ``starburstiness" is not an all-or-nothing property, but that the {\it Ansatz} in Equation \ref{eq:DGintro} naturally leads to a continuous spectrum of burst-bearing sources ranging from those with strongly boosted SF activity to others with only a mild enhancement (see Section \ref{sect:boostspec}).\\

\subsubsection{The Boost Function: basic properties}
\label{sect:boostfct}

Both the burst-bearing and the normal galaxy population are described by an amplitude $A_{\rm X}$ with units of [Mpc$^{-3}$ dex(sSFR)$^{-1}$], a dispersion $\sigma_{\rm X}$ (units: [dex(sSFR)]) and a mode $\langle$sSFR$\rangle_{\rm X}$ (X\,$\in$\,\{MS, SB\}). In particular, the MS-distribution has the functional form
\small
\begin{flalign}
\mathcal{G}&_{\rm MS}(\ssfr) = \nonumber \\
& A_{\rm MS}\,{\rm exp}\left(-\frac{[{\rm log(sSFR)}-{\rm log(\langle sSFR\rangle_{MS})}]^2}{2\sigma_{\rm MS}^2}\right)~,\quad{\rm or} \nonumber \\
& A_{\rm MS}\,{\rm exp}\left(-\frac{x^2}{2\sigma_{\rm MS}^2}\right) \label{eq:lognormMS}
\end{flalign}
\normalsize
if, for the sake of brevity, we introduce an sSFR $x$\,$\equiv$\,log(sSFR/${\rm \langle sSFR\rangle_{MS}}$) that is normalized to the stellar mass- and redshift-dependent average sSFR of MS galaxies, ${\rm \langle sSFR\rangle_{MS}}$. Similarly, for starbursting sources, we write
\small
\begin{flalign}
\mathcal{G}&_{\rm SB}(\ssfr) =\nonumber \\
& A_{\rm SB}\,{\rm exp}\left(-\frac{[{\rm log(sSFR)}-{\rm log(\langle sSFR\rangle_{SB})}]^2}{2\sigma_{\rm SB}^2}\right) \nonumber \\ 
& A_{\rm SB}\,{\rm exp}\left(-\frac{[{\rm log(sSFR)}-\{{\rm log(\langle sSFR\rangle_{MS})}+B_{SB}\}]^2}{2\sigma_{\rm SB}^2}\right) \nonumber \\ 
& A_{\rm SB}\,{\rm exp}\left(-\frac{[x-B_{\rm SB}]^2}{2\sigma_{\rm SB}^2}\right)~. \label{eq:lognormSB}
\end{flalign}
\normalsize
Here $B_{\rm SB}$ is the offset between the peak position of the MS and SB component of the sSFR distribution. In our interpretation it represents the average sSFR enhancement -- or boost -- brought about by the burst-inducing process. In the following we will assume that there is a process (to be discussed in Section \ref{sect:boostcontext}) that boosts the SF activity of an (s)SFR-dependent fraction of MS galaxies with initial sSFR distribution $\mathcal{G}_{\rm MS}^0$. $\mathcal{G}_{\rm MS}^0$ is identical with the log-normal distribution in Equation \ref{eq:lognormMS}, except for a higher normalization $A_{\rm MS} \rightarrow A_{\rm MS}^0 = A_{\rm MS}\times \nicefrac{I(\mathcal{G}_{\rm MS}+\mathcal{G}_{\rm SB}:\,x)}{I(\mathcal{G}_{\rm MS}:\,x)}$. (Here $I(f: x)$ stands for the integral of the function $f$ over the range $x$\,$\in$\,]$-\infty,\infty$[.)

\begin{figure}
\epsscale{1.05}
\centering
\plotone{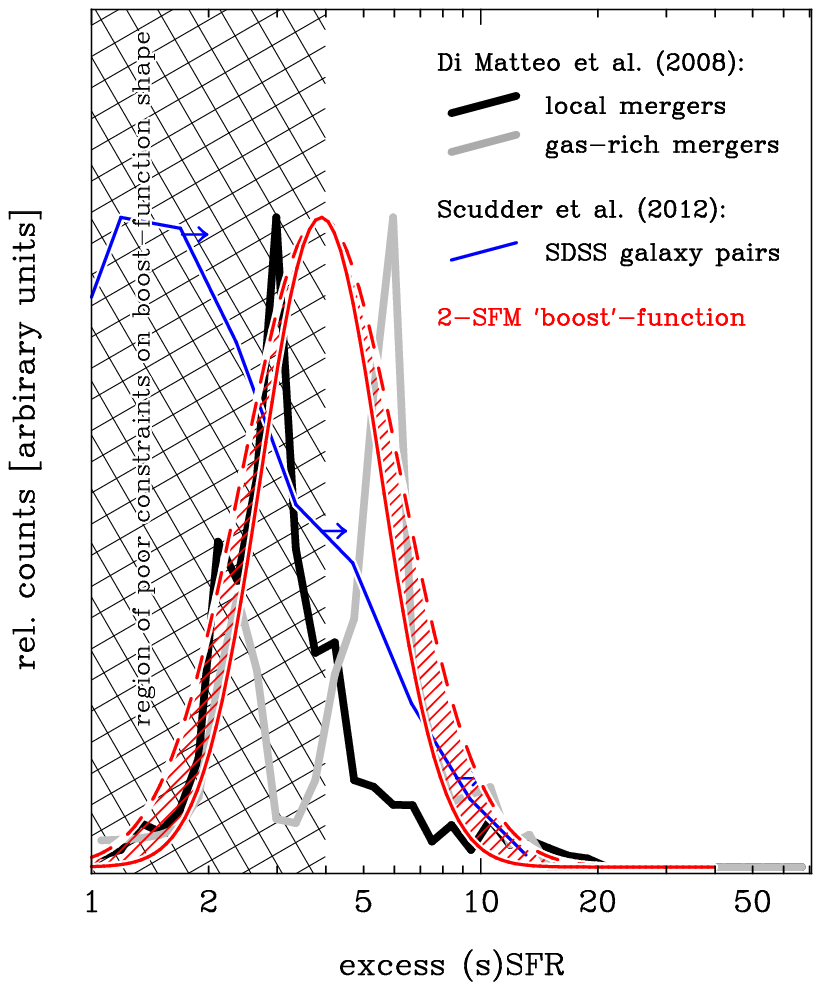}
\caption{\footnotesize Comparison of the 2-SFM boost function (i.e., the spectrum of (s)SFR-perturbations that move normal galaxies off the MS and into a SB state) with measured and simulated distributions of SFR enhancements induced by galaxy interactions. Red (solid) -- ``direct" boost function; red (dashes) -- boost function including explicit correction for merger statistics (see Section \ref{sect:corrboost}). In the cross-hatched area the shape of the 2-SFM boost function is not known accurately (see discussion in Section \ref{sect:boostcontext}). The distribution of SFR enhancements in massive (\mstar\,$>$\,10$^{10}$\,\msun) members of $z$\,$\in$\,[0.02,0.15] SDSS galaxy pairs measured by \citet[blue line]{scudder12} does not include mergers that have already undergone final coalescence and hence represents a lower observational limit to the total local SFR excess distribution caused by interactions. Black and light grey lines -- average of peak boosts (relative to isolated galaxies) measured in SPH and grid-based N-body simulations \citep[cf.][]{dimatteo08} of local mergers and gas-rich mergers with gas fraction similar to those of $z$\,$\sim$\,2 MS galaxies.
\label{fig:boostcomp}}
\end{figure}

\noindent The ``boost function" describes the spectrum of perturbations that MS galaxies suffer. It is effectively a convolution kernel that transfers galaxies from the MS- to the SB-distribution:
\small
\begin{eqnarray}
\mathcal{G}_{\rm SB}(x) &=& (\mathcal{G}_{\rm MS}^0 * \mathcal{BK})(x) \nonumber \\
&=& \int^{\infty}_{-\infty} \mathcal{G}_{\rm MS}^0(y)\,\mathcal{BK}(x-y)\,dy~. \label{eq:bostkerndef}
\end{eqnarray}
\normalsize
The boost function kernel ($\mathcal{BK}$) is obtained using the convolution theorem which links the Fourier transforms\footnote{~For the one-dimensional Fourier transform and its inverse we use the following convention
\begin{eqnarray*}
\widehat{f}(k) &=& \int^{\infty}_{-\infty} f(x)e^{-ikx}\,dx \\
f(x) &=& \frac{1}{2\pi}\int^{\infty}_{-\infty} \widehat{f}(k)e^{ikx}\,dk ~.
\end{eqnarray*}
With this definition the Fourier transforms of the log-normal distributions for starbursting sources and the unperturbed MS become
\begin{eqnarray*}
\widehat{\mathcal{G}}_{\rm SB}(k) &=& e^{ikB_{\rm SB}}A_{\rm SB}\sqrt{2\pi}\sigma_{\rm SB}e^{-\left(\frac{\sigma_{\rm SB}k}{\sqrt{2}}\right)^2}~,\quad{\rm and}\\
\widehat{\mathcal{G}}_{\rm MS}^0(k) &=& A_{\rm MS}^0\sqrt{2\pi}\sigma_{\rm MS}e^{-\left(\frac{\sigma_{\rm MS}k}{\sqrt{2}}\right)^2}~.\\
\end{eqnarray*}
}
\small
\begin{equation}
\widehat{\mathcal{G}}_{\rm SB}(k) = \widehat{\mathcal{G}}_{\rm MS}^0(k) \times \widehat{\mathcal{BK}}(k)~,
\label{eq:convtheor}
\end{equation}
\normalsize
and by then applying the inverse Fourier transform:
\small
\begin{equation}
\mathcal{BK}(x) = \frac{1}{2\pi}\int_{-\infty}^{\infty}\frac{\widehat{\mathcal{G}}_{\rm SB}(k)}{\widehat{\mathcal{G}}_{\rm MS}^0(k)}e^{ikx}\,dk~.
\label{eq:invfourier}
\end{equation}
\normalsize
Equations \ref{eq:convtheor} and \ref{eq:invfourier} will not always have an analytical solution. In the present case however, since both parent and resultant distribution -- $\mathcal{G}_{\rm MS}^0$ and $\mathcal{G}_{\rm SB}$, resp. -- are log-normal, the boost function kernel also has this functional form:
\small
\begin{equation}
\mathcal{BK}(x) \equiv \mathcal{G}_{\rm BK}(x) = C_{\rm BK}\,{\rm exp}\left(-\frac{[x-\langle x\rangle_{\rm BK}]^2}{2\sigma_{\rm BK}^2}\right)~. \label{eq:Gaussboost}
\end{equation}
\normalsize
By explicitly solving Equation \ref{eq:bostkerndef}, the three free parameters of the boost function in Equation \ref{eq:Gaussboost} are found to be
\small
\begin{eqnarray}
C_{\rm BK} &=& \frac{A_{\rm SB}\sigma_{\rm SB}}{A_{\rm MS}^0\sigma_{\rm MS}}\frac{1}{\sqrt{2\pi}\sigma_{\rm BK}}~,\quad {\rm with}\nonumber \\ \label{eq:dirboostpars}
\sigma_{\rm BK} &=& \sqrt{\sigma_{\rm SB}^2-\sigma_{\rm MS}^2}~,\quad {\rm and}\\
\langle x\rangle_{\rm BK} &=& B_{\rm SB}~.\nonumber
\end{eqnarray}
\normalsize
The shape of the boost function with free parameters given by expressions \ref{eq:dirboostpars} is shown as a solid red line in Figure \ref{fig:boostcomp} and compared to distributions of SFR enhancements reported for simulated and observed interacting galaxies.

\noindent When writing down Equation \ref{eq:bostkerndef} we implicitly assume that a statistical link exists between the MS population and high-sSFR outliers. Such a link is expected on the grounds that SBs are transient events, meaning that these galaxies must be drawn from the larger population of active (normal) SFGs. Defining a boost kernel which is independent of an initial distribution of SF activity would hence necessarily involve some arbitrary assumptions. The model we propose here is thus as simple a scenario as one can imagine. It is important to realize that we did not choose the shape of the boost kernel $\mathcal{BK}$ {\it a priori}. It is the consequence of the fact that the cross-section of the MS has a log-normal shape and that the high-SFR tail can be modelled well by the addition of a second, shifted log-normal distribution
as shown in S12.\\
It is natural to expect that the process which statistically/physically links starbursting and normal galaxies is galaxy interactions and merging. In Section \ref{sect:boostcontext} we discuss in detail whether theory or observations can provide supporting evidence for such a straightforward connection. To summarize, some properties of the 2-SFM boost function are suggestively reminiscent of SFR enhancements in observations and simulations. Other aspects do not conform to the expectations of what a realistic boost distribution should look like if it accounts for, e.g., minor interactions, interactions including passive galaxies and the fact that an observational snapshot of the SB population will catch different objects in different phases of the burst. This difference, however, could be entirely due to the impossibility of statistically distinguishing between normal and only weakly starbursting galaxies in a direct fit to the \ssfr\ distribution; the conventional view that SBs are often tied to galaxy interactions hence remains a viable scenario and we now consider a modification to the boost function that  we expect to apply for the idealized case that all SB events are triggered by merging.

\subsubsection{Statistical Correction for Paired, Ante-merger Galaxies}
\label{sect:corrboost}

In our presentation of the 2-SFM boost function we have so far skipped issues that would complicate an immediate interpretation that is based purely on a mathematical description of the problem. A first and strong simplifying assumption is that the boost function is mass- and redshift-independent. However, as discussed in S12 and \citet{bethermin12}, current observations so far are consistent with this hypothesis. The sSFR distributions of MS galaxies and starbursting sources are snapshots that provide no direct information on the time scale over which SF in MS galaxies is boosted (and over which they, possibly having undergone a merger, later fall back onto the relation or drop below it). The relative redshift-independence of the boost function that is suggested by observations implies that the flux of galaxies into and out of the SB component of the double log-normal distribution also should not evolve strongly with redshift. If the SFR enhancements were always the result of galaxy merging, then this would motivate a modification to the boost function. We regard each pair of merging galaxies as a single system that is composed of two MS galaxies. Because the parent \ssfr\ distribution  $\mathcal{G}_{\rm MS}(\ssfr)$ is by assumption independent of mass and symmetric, both galaxies involved are drawn from the same distribution and their average SFR will follow a log-normal function that is centered on the same \ssfr\ as the original distribution $\mathcal{G}_{\rm MS}(\ssfr)$ but narrowed by a factor $\sqrt{2}$. For the boost function this implies a broader distribution of SFR enhancements while the peak location of the boost spectrum remains identical (see also distribution plotted with a dashed red line in Figure \ref{fig:boostcomp}):
\small
\begin{eqnarray}
\sigma_{\rm BK} &=& \sqrt{\sigma_{\rm SB}^2-\left(\nicefrac{\sigma_{\rm MS}}{\sqrt{2}}\right)^2}~,\quad {\rm and}\nonumber \\ 
\langle x\rangle_{\rm BK} &=& B_{\rm SB}~\qquad\qquad\qquad, \quad \text{such that} \\ \label{eq:corrboostpars}
C_{\rm BK} &=& \frac{A_{\rm SB}\sigma_{\rm SB}}{A_{\rm MS}^0\sigma_{\rm MS}}\frac{1}{\sqrt{2\pi}\sigma_{\rm BK}}~.\nonumber
\end{eqnarray}
\normalsize
In the following we will refer to this version of the boost function as ``boost function including an explicit correction for mergers" (as opposed to the boost function described by Equations \ref{eq:dirboostpars} and henceforth called: ``direct boost function"). The principle findings of this paper are valid irrespective of the choice of boost function, but for the sake of legibility we will only show results obtained with the direct boost function where plotting both alternatives would reduce rather than improve clarity.

\begin{figure*}
\epsscale{1.13}
\centering
\plotone{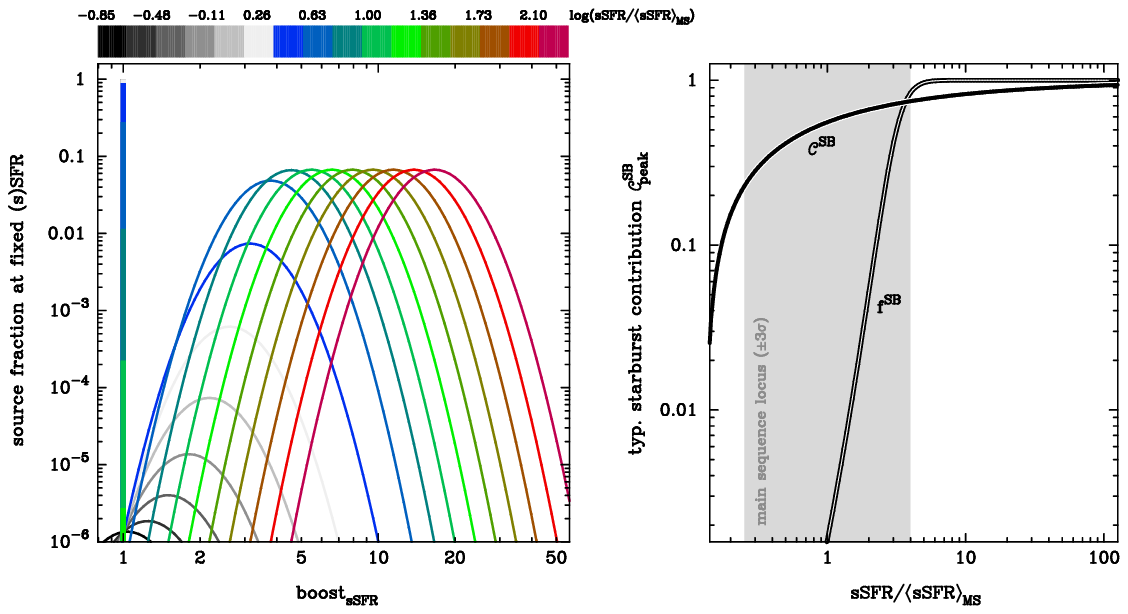}
\caption{\footnotesize Illustration of the properties of the continuously varying, sSFR-dependent 2-SFM boost distribution (i.e. the distribution of (s)\sfr\ enhancements of starbursting galaxies; see Section \ref{sect:corrboost}). Left -- log-normal distribution of boosts (see eqs. \ref{eq:boostpeak}--\ref{eq:boostshape2}) of SFGs with a specific sSFR. All distributions are normalized to the total number of sources at a given sSFR and colors vary according to the value of sSFR, where sSFR is referenced to the MS average. For example, at \ssfr/\ssfrMS\,=\,10 the median boost is approx. seven-fold and 95\% of the SB population have boosts in the range 2.5-17. Starbursts with \ssfr/\ssfrMS\,=\,\nicefrac{1}{2}, on the other hand, have on average experienced weak boosting by only approx. 30\% and 95\% of such systems have boosts that are smaller than a factor three. The relative number of secularly evolving galaxies (plotted at boost\,=\,1 with a bar of appropriate height) and of boosted sources is given by the burst-bearing fraction f$^{\rm SB}$. Right -- variation of the starbursting fraction f$^{\rm SB}$ with normalized sSFR, compared to the evolution of the typical fractional contribution, $\mathcal{C}^{\rm SB}$, of the burst-induced activity to the total SFR of boosted sources at a given sSFR. (Here `typical fractional contribution' is defined as the contribution of a source located at the sSFR-dependent peak of the boost distribution.) Directly on the MS locus (\ssfr/\ssfrMS\,=\,1) starbursting sources are rare (f$^{\rm SB}$\,$<$\,2\textperthousand) but in those systems that have experienced boosting the contribution of the burst-activity to the total \sfr\ is significant ($\mathcal{C}^{\rm SB}$\,$\sim$\,47\%). At an \ssfr\ excess of \ssfr/\ssfrMS\,$\simeq$\,3, above which SBs are more numerous than normal galaxies (f$^{\rm SB}$\,$\geq$\,0.5), the SB contribution to the total \sfr\ of boosted sources is already clearly dominant ($\mathcal{C}^{\rm SB}$\,$\sim$\,71\%). These numbers are based on the ``direct" boost function that uses the best-fit parameters of the double log-normal \ssfr\ decomposition in S12 (see also Equation \ref{eq:DGintro} in this paper).
\label{fig:boostdistrib}}
\end{figure*}

\subsubsection{The Continuously Varying Boost Distribution}
\label{sect:boostspec}

As previously mentioned, the ``typical" sSFR increase $B_{\rm SB}$ of starbursting sources is approx. a factor of four, but the average boost varies as a function of (s)SFR, as does the (relative) number of sources undergoing burst-like activity. For example, a source with measured sSFR twice as large as the (redshift- and stellar-mass-dependent) MS average could either display this excess simply due to a larger than average gas fraction and without having suffered any triggering, it could have experienced a modest boost, or -- with a lesser probability -- it could initially have been a gas-poor, low-(s)SFR outlier to the MS which has been strongly boosted. In the following we quantify these variations that are a consequence of the convolution in Equation \ref{eq:bostkerndef}.\\
To find the ``typical" boost of burst-bearing sources at a given sSFR$_0$ (or log(sSFR$_0$/$\langle$sSFR$\rangle_{\rm MS}$)\,=\,$x_0$) we consider the integrand in Equation \ref{eq:bostkerndef},
\small
\begin{equation}
\mathcal{G}_{\rm MS}^0(x)\,\mathcal{G}_{\rm BK}(x_0-x) = \mathcal{G}_{\rm MS}^0(x_0-b_{\ssfr})\,\mathcal{G}_{\rm BK}(b_{\ssfr})~. \label{eq:logboostintro}
\end{equation}
\normalsize
Here we introduced a variable for the logarithmic boost, $b_{\ssfr}$\,=\,$x_0-x$\,$\equiv$\,log(sSFR$_0$/sSFR), in order to be able to directly locate the peak of the boost distribution, $b_{\ssfr}^{\rm max}$, by solving the minimization problem:
\small
\begin{equation}
\frac{\partial}{\partial b_{\ssfr}} \bigg\{ \mathcal{G}_{\rm MS}^0(x_0-b_{\ssfr})\,\mathcal{G}_{\rm BK}(b_{\ssfr})\bigg\} \doteq 0~. \label{eq:boostpeakproblem1}
\end{equation}
\normalsize
Given the properties of the exponential function this is equivalent to requiring
\small
\begin{equation}
\frac{\partial}{\partial b_{\ssfr}} \bigg\{ \frac{-\left(x_0-b_{\ssfr}\right)^2}{2\sigma_{\rm MS}^2}\bigg\} + \frac{\partial}{\partial b_{\ssfr}} \bigg\{ \frac{-\left(b_{\ssfr}-B_{\rm SB}\right)^2}{2\sigma_{\rm BK}^2} \bigg\} \doteq 0~, \label{eq:boostpeakproblem2}
\end{equation}
\normalsize
an equation which has the solution
\small
\begin{equation}
b_{\ssfr}^{\rm max} = \frac{B_{\rm SB}+x_0\left(\nicefrac{\sigma_{\rm BK}}{\sigma_{\rm MS}}\right)^2}{1+\left(\nicefrac{\sigma_{\rm BK}}{\sigma_{\rm MS}}\right)^2}~. \label{eq:boostpeak}
\end{equation}
\normalsize
To determine the shape of the boost spectrum which peaks at $b_{\ssfr}^{\rm max}$ we consider the product of the two Gaussians in Equation \ref{eq:logboostintro}
\small
\begin{flalign}
\mathcal{G}_{\rm MS}^0&(x_0-b_{\ssfr})\,\mathcal{G}_{\rm BK}(b_{\ssfr}) = \\ \label{eq:boostshape1}
&A_{\rm MS}^0\,C_{\rm BK}\,{\rm exp}\left(-\Bigg\{\frac{[x_0-b_{\ssfr}]^2}{2\sigma_{\rm MS}^2} + \frac{[b_{\ssfr}-B_{\rm SB}]^2}{2\sigma_{\rm BK}^2}\Bigg\}\right) \nonumber
\end{flalign}
\normalsize
and examine the exponent
\small
\begin{equation}
\frac{[x_0-b_{\ssfr}]^2}{2\sigma_{\rm MS}^2} + \frac{[b_{\ssfr}-B_{\rm SB}]^2}{2\sigma_{\rm BK}^2} \nonumber
\end{equation}
\normalsize
which can be re-written as
\small
\begin{eqnarray}
\frac{1}{2\frac{(\sigma_{\rm MS}\sigma_{\rm BK})^2}{\sigma_{\rm MS}^2 + \sigma_{\rm BK}^2}}\Bigg\{b_{\ssfr}^2 &-& 2\,\frac{x_0\sigma_{\rm BK}^2+B_{\rm SB}\sigma_{\rm MS}^2}{\sigma_{\rm MS}^2 + \sigma_{\rm BK}^2}\,b_{\ssfr} + ...\nonumber \\
&...& + \frac{x_0^2\sigma_{\rm BK}^2+B_{\rm SB}^2\sigma_{\rm MS}^2}{\sigma_{\rm MS}^2 + \sigma_{\rm BK}^2}\Bigg\}~. \label{eq:boostshape2}
\end{eqnarray}
\normalsize
This is again a quadratic form, implying that the boosts at fixed normalized sSFR $x_0$ are distributed log-normally with a width $\sigma_b$\,=\,$\sqrt{(\sigma_{\rm MS}\sigma_{\rm BK})^2/(\sigma_{\rm MS}^2 + \sigma_{\rm BK}^2)}$ that is independent of \ssfr.

\noindent On the left-hand side of Figure \ref{fig:boostdistrib} we visualize with different colors the changes in the boost distribution for sSFRs ranging from $\sim$0.1 to two hundred times the characteristic MS value, \ssfrMS. Note that the \ssfr\ variations take place within a given bin of stellar mass and that the integral over the boost distributions at all sSFRs would give a total boost-distribution that is equal to the boost function plotted with the red solid line in Figure \ref{fig:boostcomp}. This is not immediately obvious based on Figure \ref{fig:boostdistrib} where we have scaled all \ssfr-dependent boost distributions such that they give the fraction of sources with boost $b_{\ssfr}$ in a specific \ssfr\ bin. This representation highlights the evolution of the fraction f$^{\rm SB}$ of starbursting sources (given by the ratio of the two log-normal curves in Equation \ref{eq:DGintro}; see also Figure \ref{fig:boostdistrib}, right) with \ssfr\ while simultaneously compensating for the variation of the total number of sources across the width of the MS. The amplitude of the log-normal boost distributions thus grows until \ssfr/\ssfrMS\,$\simeq$\,8 where the number of MS galaxies becomes insignificant with respect to the number of starbursting sources (see also flattening of the evolution of f$^{\rm SB}$ in the right-hand panel of Figure \ref{fig:boostdistrib}).\\
An alternative quantity which traces the increasing importance of SB activity at successively higher sSFRs is the typical fractional contribution, $\mathcal{C}^{\rm SB}$\,=\,$\nicefrac{(\sfr-\sfr_{\rm MS,\,init.})}{\sfr}$\,=\,$1-10^{-b_{\ssfr}}$, of the burst-induced activity to the total SFR of boosted sources. (Here $\sfr_{\rm MS,\,init.}$ is the \sfr\ of the galaxy in the MS state prior to boosting.) $\mathcal{C}^{\rm SB}$ is complementary to the starbursting fraction of the population, f$^{\rm SB}$, in that it describes the impact of the boost on an individual galaxy, while f$^{\rm SB}$ provides the number of galaxies of a given \ssfr\ within the 2-SFM framework that are subject to such boosting. The relative variation of these two quantities is compared in the right-hand panel of Figure \ref{fig:boostdistrib}.\\
While the 2-SFM boost kernel has the effect of increasing the star-formation activity for the vast majority of the objects affected, its lower tail formally allows for ``negative" boosts (boost\,$<$\,1; i.e. suppression of star formation activity). This only occurs for a tiny fraction of $<$0.1\textperthousand\ and 2\textperthousand\ of the SFG population for the direct and merger-corrected boost function, respectively. Our illustration of the sSFR dependence (see color scale) of boost distributions in the left-hand panel of Figure \ref{fig:boostdistrib} shows that the mode of these distributions is located at positive boosts at all normalized {\ssfr/\ssfrMS}\,$\gtrsim$\,0.1. Below this sSFR (which lies about 4-5\,$\sigma_{\rm MS}$ below the average MS locus) a majority of boost-bearing sources experience a suppression of activity. The exact location of this transition is a somewhat arbitrary mathematical consequence of our choice to describe the boosting-process with a log-normal kernel (but see final paragraph of Section \ref{sect:boostfct} for why this is a reasonable {\it Ansatz}). From the observational perspective, however, this arbitrariness and the details of the boost demographics in general on and below the MS locus are inconsequential because (a) the number of boosted galaxies is negligibly small compared to the dominant MS population in this regime (f$^{\rm SB}$\,$\ll$\,10\% over much of the MS, cf. right-hand panel of Figure \ref{fig:boostdistrib}) and (b) it is exceedingly hard to distinguish a galaxy with a small boost from a normal MS galaxy. For example, the rare, burst-bearing sources at the lower envelope of the MS typically are predicted to have small SFR enhancements (e.g. by 30\% at -2\,$\sigma_{\rm MS}$). At an identical positive offset from the MS locus (+2\,$\sigma_{\rm MS}$), typical SFR enhancements by about a factor three\footnote{~These numbers are for the boost function including the explicit correction for merger statistics. For the direct boost function the typical boost $b_{\ssfr}^{\rm max}$ varies more slowly across the MS (see Equation \ref{eq:boostpeak} and Figure \ref{fig:boostconvert}).} are expected. This ambiguity between normal MS galaxies and modest SBs reflects that fact that the lower half of our boost kernel is not well constrained by observations, as highlighted by the hatched area in Figure \ref{fig:boostcomp}. The strengths of the 2-SFM approach thus lie in describing the properties of high-activity outliers rather than characterizing objects that are in practice indistinguishable from normal SFGs.

\section{Results}
\label{sect:results}

In the following we will show how a simpler understanding of the molecular gas properties of SFGs at all redshifts $z$\,$<$\,3 emerges when variations of, e.g. SFE or gas fraction, about the typical value of MS galaxies are considered. To be able to establish such normalized trends requires a reliable prescription for the evolution of slope and normalization of the star-forming MS with redshift. In Appendix \ref{appsect:sSFR} we parametrize the evolution of sSFR for MS galaxies as a smoothly varying function of redshift and stellar mass (see Equation \ref{eq:sSFRevoeq}) which we fit to a compilation of sSFR-data from the literature. We find that on average a sSFR versus \mstar\ relation \ssfr\,$\propto$\,$M_{\star}^{\nu}$ with exponent $\nu$\,$\simeq$\,-0.2 reproduces the systematic shift between the sSFR evolution of galaxies in different mass bins out to $z$\,$\sim$\,3. The extrapolation of the sSFR evolution to higher redshift (as briefly proposed for the discussion of gas fraction evolution in Section \ref{sect:fgas_evo}) is speculative, since constraints on the shape of the MS at $z$\,$>$\,3 are much sparser.\\
We begin this section with our new description of SFE-variations between normal and starbursting galaxies (Section \ref{sect:SFE}) and then discuss gas fractions variations and evolution plus empirical recipes for the CO-to-H$_2$ conversion factor (Sections \ref{sect:fgas} and \ref{sect:XCO}, resp.).

\subsection{Simple Recipes for Star Formation Efficiency in Massive Star-forming Galaxies}
\label{sect:SFE}

\subsubsection{Star Formation Efficiency in Normal and Starbursting Galaxies: Observations}
\label{sect:obs_SFE}

\begin{figure*}
\epsscale{1.13}
\centering
\plotone{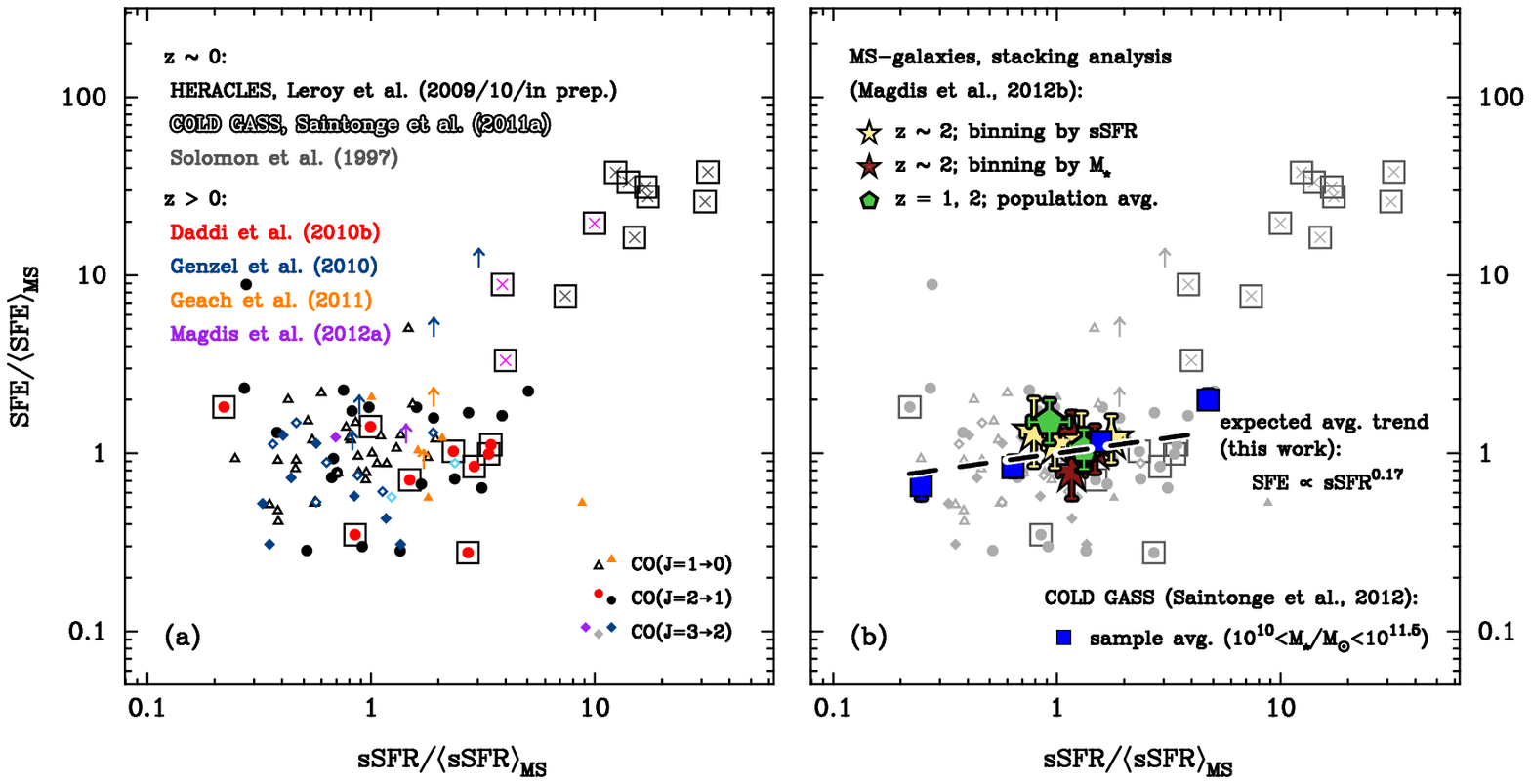}
\caption{\footnotesize Dependence of SFE (see scale on right) and gas depletion time $\tau_{\rm dep.}$ (scale on left) on SFR for individual galaxies in our calibration sample (panel {\it a}) and for stacked galaxies from \citet[][panel {\it b}]{magdis12b}. (All data and symbols as in Figure \ref{fig:invSKcalib}.b and \ref{fig:invSKrobust}.b).
\label{fig:SFEvsSFR}}
\end{figure*}

The non-linearity of the integrated S-K law found in Section \ref{sect:invSKcalib} implies a residual dependence of the gas depletion time, \tdep\,=\,\mmol/\sfr, and its inverse, the SFE, on SFR. Using our fit for MS galaxies from Equation \ref{eq:MgasvsSFR} we obtain
\small
\begin{eqnarray}
{\rm log}\left(\frac{\tdep}{\rm Gyr}\right) &=& (\alpha_{2,\,\sfr}{-}9) + (\beta_2{-}1)\,{\rm log\left(\frac{\sfr}{\nicefrac{\msun}{\rm yr}}\right)}\\
&=& 0.22(\pm0.02) - 0.19(\pm0.03){\times}{\rm log\left(\frac{\sfr}{\nicefrac{\msun}{\rm yr}}\right)} \nonumber
\end{eqnarray}
\normalsize
and
\small
\begin{eqnarray}
{\rm log\left(\frac{\sfe}{\rm Gyr^{-1}}\right)} &=& (1{-}\beta_2)\,{\rm log\left(\frac{\sfr}{\nicefrac{\msun}{\rm yr}}\right)} - (\alpha_{2,\,\sfr}{-}9) \\
 &=& 0.19(\pm0.03){\times}{\rm log\left(\frac{\sfr}{\nicefrac{\msun}{\rm yr}}\right)} - 0.22(\pm0.02)~. \nonumber
\label{eq:absSFE_MS}
\end{eqnarray}
\normalsize
The dispersion about this characteristic value is $\sim$0.2\,dex (see Section \ref{sect:MgasvsSFR}). Figure \ref{fig:SFEvsSFR} illustrates how the galaxies from our reference sample (see Section \ref{sect:moldata}) and the stacked samples of \citet{magdis12b} scatter around this average trend which -- due to the general redshift evolution of \ssfr\ in SFGs -- implies a roughly two-fold decrease of \tdep\ between $z$\,$\sim$\,0 and $z$\,$\sim$\,2 for galaxies of \mstar/\msun\,=\,$4{\times}10^{10}$, which contribute most to the cosmic SFRD over this period \citep{cowiebarger08, gilbank11, karim11}. The variation between the depletion times of 1-2\,Gyr in local spiral galaxies \citep[e.g.,][]{leroy08, bigiel11} and the approx. 0.5-1\,Gyr determined for BM/BX- and BzK-selected galaxies at 1.5\,$<$\,$z$\,$<$\,2.5 \citep[e.g.,][]{daddi10b, tacconi13} is much smaller than the difference between normal galaxies and strong SBs (see Figure \ref{fig:sampintro}(d) and offset, dashed locus in Figure \ref{fig:SFEvsSFR}).

\begin{figure*}
\epsscale{1.13}
\centering
\plotone{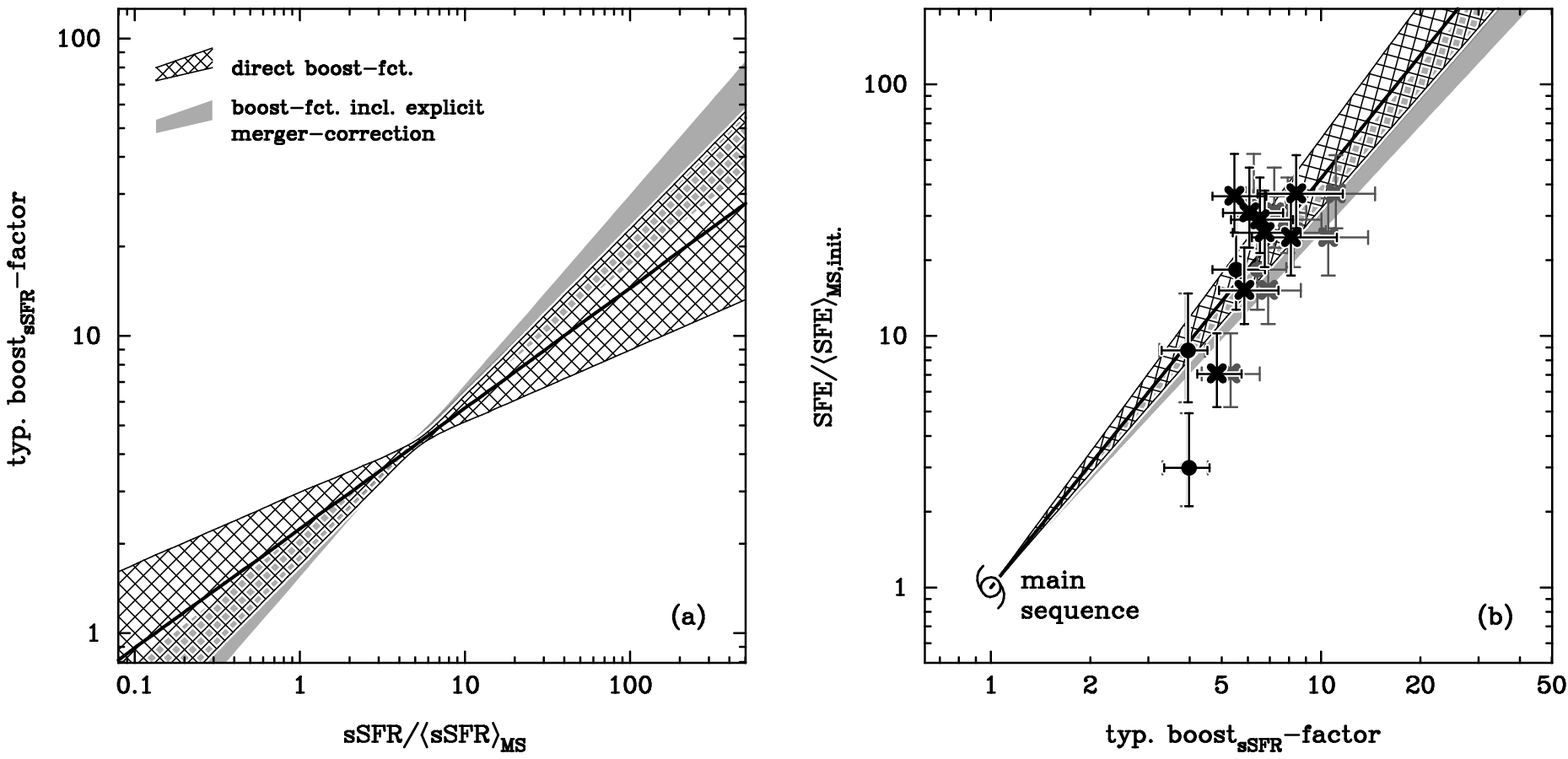}
\caption{\footnotesize Star formation efficiency (SFE) versus sSFR for selected main-sequence (MS) galaxies and starbursts (SBs) at $z$\,$\lesssim$\,3 ({\it left}) and for stacked galaxies from \citet[][{\it right}]{magdis12b}. When normalized to the characteristic MS value, $\langle\,.\,\rangle_{\rm MS}$, a homogeneous behavior of MS galaxies at all redshifts becomes visible: SFEs vary very little within the MS while starbursting sources display enhanced SFEs that lead to their excess (s)SFR. (All data and symbols as in Figure \ref{fig:SFEvsSFR}. In panel (b) we also show sample SFR-averages for local COLD GASS galaxies \citep{saintonge12}, binned by \ssfr\ excess.)
\label{fig:normSFE}}
\end{figure*}

\noindent We attempt to correct for the implicit redshift dependence of SFE by considering a renormalized efficiency (see Figure \ref{fig:sampintro}(e)). For each galaxy in our reference sample the normalization constant, \sfeMS, is the SFE that a galaxy of equal gas mass would have if it lay directly on the inverse S-K relation given by Equation \ref{eq:MgasvsSFR}. In Figure \ref{fig:normSFE} we then plot the normalized \sfe/\sfeMS\ as a function of the normalized sSFR, \ssfr/\ssfrMS, which in the 2-SFM framework is a good measure of starburstiness. The stellar mass- and redshift-dependent MS average \ssfrMS\ is calculated according to Equation \ref{eq:sSFRevoeq}.\\
With this choice for the representation of the data, MS galaxies occupy the same region of parameter space regardless of their redshift. Our small reference sample of SB galaxies, on the other hand, is clearly offset from the MS population in the plane of normalized \sfe\ and \ssfr. Note that we assign statistically estimated CO-to-H$_2$ conversion factors based on a \mstar- and SFR-dependent metallicity $Z(\sfr,\mstar)$ to the majority of the normal galaxies in our reference sample, but that \aCO\ has been directly measured for our subsample of starbursting galaxies. While the scatter of the normal galaxies about the MS average \sfeMS\ is thus model-dependent, the strong \sfe\ excess found for SBs is not an artefact of, e.g., assuming {\it a priori} Milky-Way- and ULIRG-like conversion factors for MS galaxies and SBs, respectively. Figure \ref{fig:normSFE} shows that the \ssfr\ and \sfe\ excess of the strongest SBs in the sample are of a similar order of magnitude. This suggests that there is some kind of link between the \sfr\ enhancement (or ``boost" in the terminology of Section \ref{sect:2SFM}) and the increased \sfe\ that SBs display. The empirical calibration of this relation is the topic of the next section.

\subsubsection{Star Formation Efficiency in Normal and Starbursting Galaxies: the 2-SFM Description}
\label{sect:2-SFM_SFE}

Stellar mass and SFR are fundamental parameters in the 2-SFM description of SFGs. The tight and apparently redshift-independent integrated Schmidt-Kennicutt law found in Section \ref{sect:MgasvsSFR} links the SFR and gas mass of MS galaxies and hence provides a straightforward recipe to extend the 2-SFM framework to their molecular gas properties. At fixed stellar mass, in which case \sfr/\sfrMS\,$\equiv$\,\ssfr/\ssfrMS\ we can write
\small
\begin{equation}
{\rm log}\left(\frac{\sfe}{\sfeMS}\right) = (1-\beta_2)\times{\rm log}\left(\frac{\ssfr}{\ssfrMS}\right)
\label{eq:normSFE_MS}
\end{equation}
\normalsize
for the relation between normalized {\sfe}s and {\ssfr}s. This slow variation across the spread of the MS with \sfe\,$\propto$\,\ssfr$^{0.19\pm0.03}$ is superimposed on the data in Figure \ref{fig:normSFE}(b).

\begin{figure*}
\epsscale{1.13}
\centering
\plotone{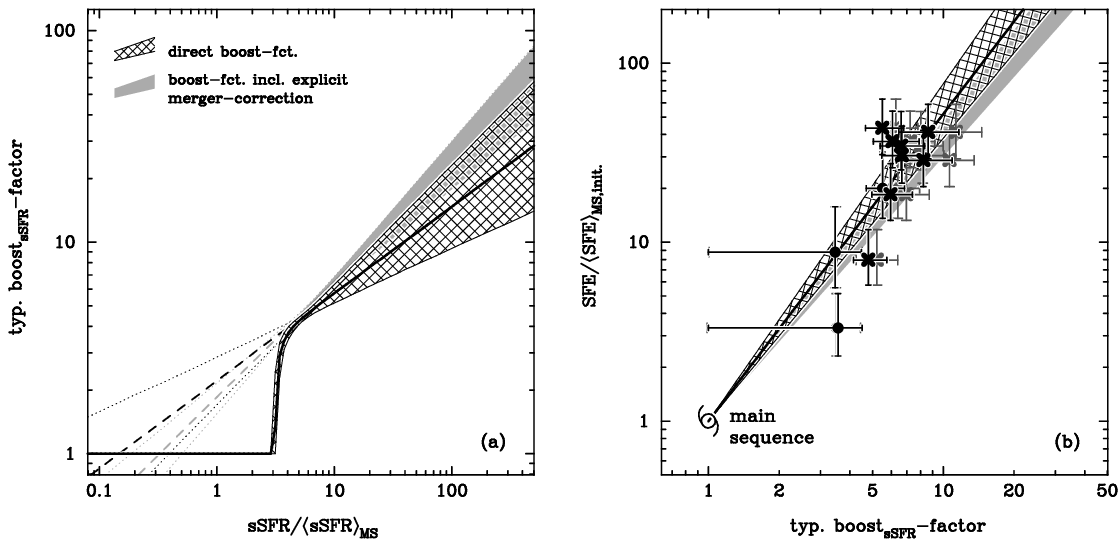}
\caption{\footnotesize Illustration of the two steps needed to establish a link between the star formation efficiency SFE of starbursting galaxies (normalized to the average efficiency of MS galaxies with a given H$_2$-mass, $\langle$SFE$\rangle_{\rm MS}$) and their (s)SFR boost within the 2-SFM framework. ({\it a}) Variation of the median SFR boost of SFGs as a function of their sSFR (normalized to the \mstar- and redshift-dependent value of MS galaxies, {\ssfr}/{\ssfrMS}). At {\ssfr}/{\ssfrMS}\,$\sim$\,3, where the starbursting sub-population begins to outnumber MS galaxies, the average boost rapidly rises from unity (i.e. no SFR enhancement) to values significantly larger than one. The ``typical" boost of burst-bearing SFGs rises continuously after this step, in keeping with the evolution of the peak of the sSFR-dependent spectra of boosts shown in Figure \ref{fig:boostdistrib}. Two cases are considered: (1) boost function including explicit correction for merger statistics ({\it grey shading}, cf. Section \ref{sect:corrboost}), and (2) ``direct" boost function ({\it hatched area}). The shaded/hatched areas mark the 1\,$\sigma$-error on the typical boost, and were derived based on the uncertainties associated with the decomposition of the sSFR distributions of massive $z$\,$\sim$\,2 SFGs into two log-normal components (for MS and starbursting galaxies, resp.; see S12 or the schematic representation in Figure \ref{fig:sSFRnSFEdistrib}). Bold lines at the core of the highlighted confidence regions trace the best-fit variation of the typical boost values. The formal continuation to {\ssfr}/{\ssfrMS}\,$\ll$\,3 of the trend line following the evolution of the typical boost of the burst-bearing sub-population only is plotted with a fine dashed line (associated uncertainties are indicated with dots). Starburst galaxies in this sSFR range are not observable in practice (see discussion in text of Section \ref{sect:2-SFM_SFE}). ({\it b}) Empirical calibration of the relation between SFE enhancement and boost amplitude using a reference sample of selected SB galaxies with measured \aCO\ \citep{solomon97, magdis12b} and the relation between normalized sSFR and the typical boost of panel ({\it a}). The reference sample of SBs is located with respect to the $x$-axis using the trend lines of panel ({\it a}) for the 2-SFM boost function including explicit correction for merger statistics ({\it dark grey symbols}) and for the ``direct" 2-SFM boost function ({\it black symbols}). (Crosses/dots are used for local/high-$z$ SBs from \citet{solomon97} and \citet{magdis12b}, respectively; 1\,sigma-error bars plotted account for observational uncertainty on \sfr, \mstar\ and \mmol\ but not systematic uncertainties, see discussion in Section \ref{sect:2-SFM_SFE}.) The shaded/hatched areas span the 68\% confidence region for a power-law relation between excess SFE and boost amplitude as parameterized in Equation \ref{eq:SFEcalib_SB} and passing through the MS locus highlighted schematically.
\label{fig:boostconvert}}
\end{figure*}

\noindent The 2-SFM framework distinguishes between MS galaxies and starbursting systems that support an elevated level of SF activity compared to what is assumed to be an initial, pre-burst state where such galaxies were indistinguishable from the large population of secularly evolving, normal SFGs. Having derived a prescription that links the SFE of MS galaxies to their offset from the MS in Equation \ref{eq:normSFE_MS} we now seek a similar relation for the starbursting fraction of the population. We adopt the following parameerization to describe the SFE of SBs:
\small
\begin{equation}
{\rm log}\left(\frac{\sfe}{\sfe_{\rm MS,\,init.}}\right) = \gamma_{\sfe} \times b_{\ssfr}
\label{eq:SFEcalib_SB}
\end{equation}
\normalsize
where $b_{\ssfr}$ is the logarithmic boost introduced in Equation \ref{eq:logboostintro} and $\sfe_{\rm MS,\,init.}$ is the \sfe\ in the MS state, prior to the onset of the burst-activity. Since we refer the \sfe\ to this initial state by definition no additional normalization constant is required in Equation \ref{eq:SFEcalib_SB}. Observationally, the amount of boosting that the SB galaxies in our reference sample have experienced is obviously unknown. We shall thus assume that their SFR enhancements correspond to the median boost (i.e. the peak location of the boost distribution, $b_{\ssfr}^{\rm max}$; see Equation \ref{eq:boostpeak}) which is expected for sources with an \ssfr\ excess as determined for these SBs. In Section \ref{sect:boostspec} we derived the boost spectrum at fixed \ssfr\ excess (or deficit), \ssfr/\ssfrMS, and calculated the shifting of its peak $b_{\ssfr}^{\rm max}$ with the normalized \ssfr. We reproduce the average trends in Figure \ref{fig:boostconvert}(a) for both the direct boost function and the boost function including an explicit correction for mergers. Since the latter scenario assumes an \ssfr\ distribution of paired, ante-merger galaxies that is narrower (see explanations in Section \ref{sect:corrboost}), a larger boost is required on average to reach a given \ssfr\ excess. This fact is reflected in a steeper slope of the corresponding boost versus \ssfr/\ssfrMS\ relation in Figure \ref{fig:boostconvert}(a). A conspicuous feature of this plot is the jump in average boost values at {\ssfr}/{\ssfrMS}\,$\sim$\,3, which is a direct consequence of the rapidly changing fraction of starbursting sources f$^{\rm SB}$ (see Figure \ref{fig:boostdistrib}, right). On the locus of the MS, most galaxies have not undergone any boosting, but at {\ssfr}/{\ssfrMS}\,$>$\,3 the starbursting sub-population begins to outnumber MS galaxies. This jump only occurs when the entire SFG population is taken into account. Average boost values for the starbursting subpopulation lie along the low-{\ssfr} extension of the power-law trend at high sSFR excesses (see fine dashed lines in Figure \ref{fig:boostconvert}(a)). In practice, burst-bearing sources with MS-like sSFRs are strongly outnumbered by normal SFGs (see Figure \ref{fig:boostdistrib}) and would in any case blend in with them because moderate SBs and ``regular", secular star-formation activity are hard to tell apart. The shaded/hatched areas in Figure \ref{fig:boostconvert}(a) indicate the uncertainty on the average relation between boost and normalized \ssfr, estimated with a full accounting of the errors on (and covariance between) the parameters of the 2-SFM double log-normal decomposition of the \ssfr\ distribution (see also Figure 1 in S12).\\
In Figure \ref{fig:boostconvert}(b) we show the result of assigning representative boost values to the SBs in our sample and then fitting Equation \ref{eq:SFEcalib_SB}. In accordance with our assumption that they experienced the median burst expected for an object with their \ssfr\ excess, we equate \sfe/\sfeMS\ -- the \sfe\ normalized to the average MS value -- and \sfe/$\sfe_{\rm MS,\,init.}$ -- the \sfe\ excess with respect to the pre-boost, MS state of each individual galaxy -- for these sources. The $y$-axis values of the SBs in Figure \ref{fig:boostconvert}(b) are thus identical to those in Figure \ref{fig:normSFE}. The $x$-axis values correspond to the {\ssfr}-dependent average boost\footnote{~As boosts between 4 and 10 have been inferred for our reference SBs, the pre-burst sSFRs of these galaxies are statistically expected to have been in the range 0.8--3 \ssfr/\ssfrMS.}, which varies as shown in Figure \ref{fig:boostconvert}(a). The associated errors span the 1\,$\sigma$ range of possible average boosts resulting from the uncertainty on the {\ssfr} measurements of the SBs. (For example, the {\ssfr} errors of high-$z$ GN20 and SMMJ2135-0102, which both formally lie offset from the MS by $\sim$0.6 dex, are such that we cannot exclude that they in truth have zero SFR enhancement; this leads to their large and strongly asymmetric error bars in Figure \ref{fig:boostconvert}(b).) We emphasize that our small reference sample of SBs does not provide sufficient statistics to justify the functional form of Equation \ref{eq:SFEcalib_SB}. Here we simply use this data to derive the best fit given the preceding choice of a plausible parameterization; in this context, the linear relation proposed in Equation \ref{eq:SFEcalib_SB} is the simplest possible form that can be envisaged. The best-fitting values of the slope are $\gamma_{\sfe}$\,=\,1.72$_{-0.14}^{+0.17}$ and 1.58$\pm$0.10 for the direct and merger-corrected boost function, respectively. The quoted 1\,$\sigma$ errors reflect the observational uncertainty on SFR, \mstar\ and \mmol\ (i.e. \lco\ \& \aCO) and on the boost inferred -- as plotted in Figure \ref{fig:boostconvert}(b) for our reference SBs -- but not the systematic uncertainties related to the calibration of the average \ssfr\ and \sfe\ of MS galaxies, nor those related to the functional form of Equation \ref{eq:SFEcalib_SB}. It is interesting to note that \citet{dimatteo07} find \sfe\ enhancements in merger simulations that exceed the boost in \sfr. The behavior of the simulations thus qualitatively matches the supra-linear relation inferred here, according to which \sfe/$\sfe_{\rm MS,\,init.}$\,$\propto$\,(boost)$^{\gamma_{\sfe}}$ with $\gamma_{\sfe}$\,$>$\,1.\\
A non-linearity of this kind is virtually inevitable in order to self-consistently match the {\sfe}s of strong SBs at the \ssfr\ they display. If, instead, we were to describe the \sfe\ of SBs as a linear mixing of two distinct S-K laws (one for normal galaxies and one for SBs) based on the SB contribution to the total \sfr\ of boosted sources ($\mathcal{C}^{\rm SB}$, see Figure \ref{fig:boostdistrib}(b)), the high {\sfe}s observed for, e.g., local ULIRGs could only be reproduced when assuming a ``template" S-K law for SBs, which is offset to higher \sfe\ than observationally seen for even the most extreme sources (cf. ``strong starburst" case of Equation \ref{eq:MgasvsSFR}).

\begin{figure*}
\epsscale{1.13}
\centering
\plotone{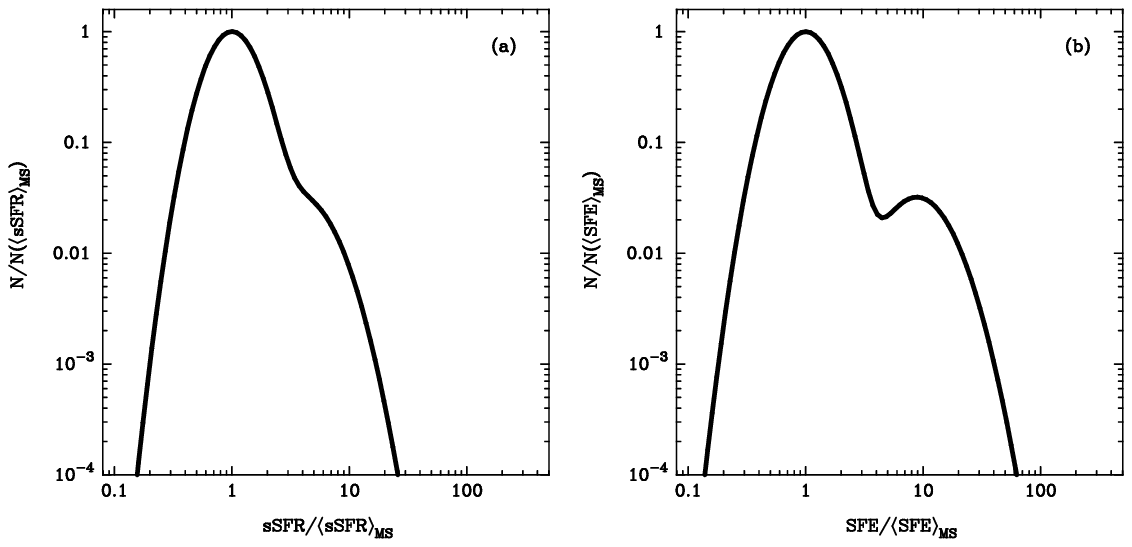}
\caption{\footnotesize Schematic comparison of the \ssfr\ distribution at fixed stellar mass ({\it a}) with the \sfe\ distribution at fixed molecular gas mass ({\it b}). The curves are representative examples, computed using the best-fit parameters of the double log-normal decomposition of the \ssfr\ distribution (Equation \ref{eq:DGintro}) and of the power-law relation fitted to the dependence of \sfe\ on boost amplitude for starbursting galaxies (Equation \ref{eq:SFEcalib_SB}). Associated uncertainties are not shown. 
\label{fig:sSFRnSFEdistrib}}
\end{figure*}

\noindent An immediate consequence of the inferred supra-linear relation between \sfe\ and \sfr\ enhancements in SB events is that the distributions of the SFG population with respect to (s)\sfr\ and \sfe\ are qualitatively different. (Here we consider the \sfe\ distribution at fixed gas mass.) We illustrate this in Figure \ref{fig:sSFRnSFEdistrib} for the case of the best-fitting double log-normal decomposition (eqs. \ref{eq:lognormMS} and \ref{eq:lognormSB}) and \sfe\ versus boost relation for SBs (Equation \ref{eq:SFEcalib_SB}). While the \ssfr\ distributions of MS galaxies and SBs blend, the two subpopulations are more strongly separated in terms of \sfe. In comparison with the distribution of galaxies in the \mmol\ versus \sfr\ plane in Figure \ref{fig:invSKcalib}(b), we see that the bimodality predicted by the 2-SFM framework is less pronounced. So far, CO follow-up observations of SFGs have generally explicitly targeted either strong SBs \citep[e.g.,][]{solomon97, greve05, riechers06, ivison11} or MS galaxies \citep[e.g.,][]{leroy08, daddi10a, tacconi10, geach11, magdis12a, tacconi13}. This selective observing strategy has thus very likely artificially deepened the expected trough between the MS and SB component in Figure \ref{fig:sSFRnSFEdistrib}(b) into the genuine gap that is seen in Figure \ref{fig:invSKcalib}(b). It should be clear that the detailed shape of the transition from the MS component of the SFE distribution to the offset SB component is uncertain to the same extent as we have only poor constraints on the lower part of the boost function (cf. Figure \ref{fig:boostcomp}). However, our conclusion that the SFE distribution should be broader than the sSFR-distribution of SFGs is inevitable if the scaling between the SFE enhancement and SFR boost of SBs is supra-linear as we find here.

\noindent We now have all ingredients to provide an empirical prescription for the \sfe\ of starbursting systems. In analogy to the expression for normal galaxies in Equation \ref{eq:normSFE_MS}, we write the recipe for SBs in terms of a normalized SFE,
\small
\begin{eqnarray}
{\rm log}\left(\frac{\sfe}{\sfeMS}\right) &=& {\rm log}\left(\frac{\sfe}{\sfe_{\rm MS,\,init.}}\right) + {\rm log}\left(\frac{\sfe_{\rm MS,\,init.}}{\sfeMS}\right) \nonumber\\
 &=& \gamma_{\sfe}\times b_{\ssfr} + (1{-}\beta_2)\nonumber\\
 &&\hspace{1.7truecm}\times{\rm log}\left(\frac{\ssfr_{\rm MS,\,init.}}{\ssfrMS}\right)~,
 \label{eq:normSFE_SB}
\end{eqnarray}
\normalsize
which is a combination of the \sfe\ of the initial, MS state prior to boosting (second summand; see also Equation \ref{eq:normSFE_MS}) and the burst-induced enhancement of this initial \sfe\ (first summand; see also Equation \ref{eq:SFEcalib_SB}). It is important to realize that Equation \ref{eq:normSFE_SB} stands for SFE changes in individual SBs rather than describing the sSFR dependence of the average SFE excess of the whole starbursting population. The latter trend will be discussed in the next paragraph.\\
By incorporating the prescriptions in Equation \ref{eq:normSFE_MS} and \ref{eq:normSFE_SB} we can predict the variation of \sfe\ throughout the entire \mstar\ versus \sfr\ plane and for the entire SFG population (as opposed to individually for the subpopulation of normal galaxies and SBs). In doing so, we will again make the simplifying assumption that all recipes are independent of stellar mass, which reduces the problem to a calculation of the evolution of \sfe\ with (s)\sfr. The prediction involves several steps which we detail here in bulletized format for maximal clarity and in preparation of analogous procedures for the variation of gas fractions and \aCO\ in Sections \ref{sect:fgas} and \ref{sect:XCO}, respectively:
\begin{itemize}
\item The hypothetical population of SFGs is distributed in \mstar\ and \sfr\ according to the double log-normal distribution in Equation \ref{eq:DGintro}.
\item At each point in the \mstar\ versus \sfr\ plane an \ssfr-dependent fraction f$^{\rm SB}$ of galaxies will fall into the SB category (see Figure \ref{fig:boostdistrib}, right) and hence require a different recipe for the computation of \sfe\ than is applied to MS sources.
\item For MS galaxies of a given \ssfr, the \sfe\ is computed with Equation \ref{eq:normSFE_MS}.
\item At each \ssfr, relations \ref{eq:boostpeak} to \ref{eq:boostshape2} from Section \ref{sect:boostspec} allow us to construct the spectrum of boosts $b_{\ssfr}$ and to infer the former pre-burst efficiencies, $\sfe_{\rm MS,\,init.}$, of the burst-bearing systems. Their {\sfe}s then follow from Equation \ref{eq:normSFE_SB}.
\item A ``typical" \sfe\ -- here we use the median -- is calculated for the joint population of normal and starbursting galaxies. It reflects the relative importance of the two subpopulations at a given location in the \mstar\ versus \sfr\ plane.
\end{itemize}
\begin{figure*}
\epsscale{1.13}
\centering
\plotone{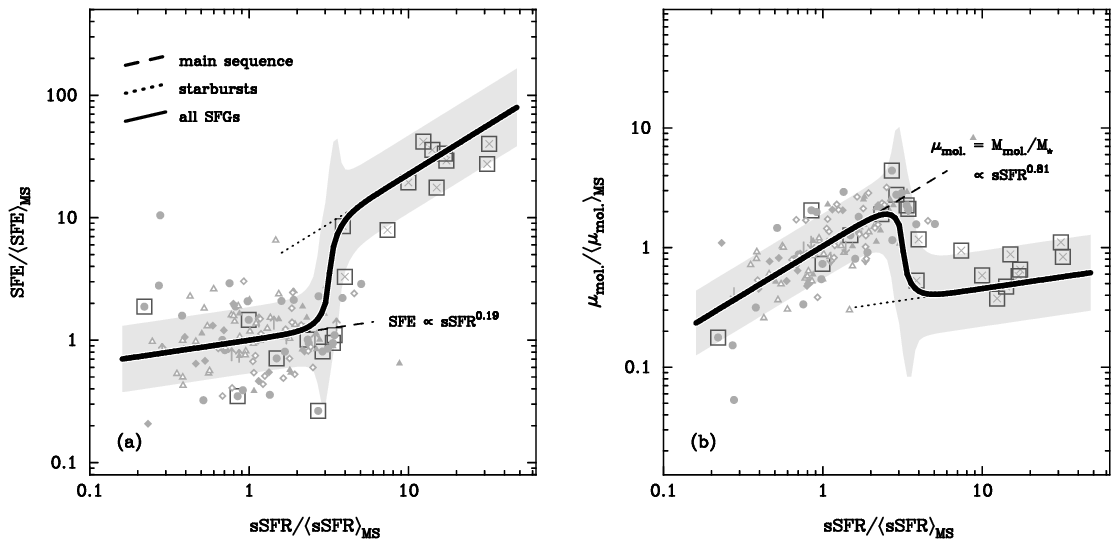}
\caption{\footnotesize Average variation of normalized SFE ({\it left}) and molecular gas mass fraction $\mu_{\rm mol.}$\,$\equiv$\,$M_{\rm mol.}/M_{\star}$ ({\it right}) with normalized sSFR, as predicted by the 2-SFM framework. Data points plotted in grey in the background are as in Figs. \ref{fig:normSFE} and \ref{fig:normfgas}. Dashed line -- expected average trend for MS galaxies (sSFR dependence as annotated adjacent to line); dotted line -- expected average trend for starbursting galaxies, plotted only in the sSFR-range where SBs represent more than 10\% of all SFGs (`direct' boost function assumed); solid line -- average evolution for the total population of SFGs; light grey shading -- expected 1\,$\sigma$ scatter around average {\sfe} variation. The scatter rises strongly over the fairly small range of sSFR where MS and starbursting galaxies occur in roughly equal numbers (see also Figure \ref{fig:boostdistrib}). The average evolution of the total population traces the variation of the median SFE ($\mu_{\rm mol.}$) of the combined SFE distribution ($\mu_{\rm mol.}$-distribution) of normal and starbursting galaxies. Its variation thus reflects the relative importance of MS and SB galaxies with offset from the MS locus \ssfrMS. When normalized to the typical MS value, the predicted trends do not depend on redshift due to the simple power-law relations between SFR and molecular gas mass that are assumed in the 2-SFM framework (see Sections \ref{sect:2-SFM_SFE} and \ref{sect:obs_fgas} for details).
\label{fig:loon}}
\end{figure*}
Figure \ref{fig:loon}(a) shows that, beginning at the lower edge of the MS locus, the median SFE of the total SFG population initially rises slowly. Boost-bearing sources are exceedingly rare throughout most of the MS so the median, normalized SFE-value in this regime is entirely determined by the S-K relation for normal galaxies. SBs become the dominant component of the SFG population at around 3\,$<$\,{\ssfr/\ssfrMS}\,$<$\,4 where the global median abruptly jumps to join the trend of steadily rising {\sfe} for the starbursting subpopulation (dotted line). The evolution of the median after this point\footnote{~The average \sfe\ of starbursting systems is predicted to have a slightly sub-linear dependence on \ssfr. This is the consequence of convolving the supra-linear evolution of \sfe\ with boost amplitude (see Equation \ref{eq:SFEcalib_SB}) with the shallow dependence of average boost amplitude on \ssfr\ (see Figure \ref{fig:boostconvert}).} reflects the increasing boost amplitudes that are required to reach the highest {\ssfr}s. Even though the 2-SFM framework assumes a full continuum of \sfr\ and \sfe\ enhancements, the changing population mix between normal and starbursting galaxies thus has lead to a ``bimodal" behavior of the \sfe. This general trend and the magnitude of the jump are insensitive to the details of the boost kernel mathematics because we have explicitly calibrated the relation between boost and SFE enhancement on the data (see Figure \ref{fig:boostconvert}(b)) and because the sSFR of our reference SBs is set by observations. The exact shape of the jump hence depends on the parameters of the double log-normal decomposition and also on the SFE dispersion of MS galaxies. In plotting the average trend in Figure \ref{fig:loon}(a) we have assumed a scatter of 0.2\,dex, in accordance with the measurement made on Figure \ref{fig:invSKcalib}.
\begin{figure*}
\epsscale{1.13}
\centering
\plotone{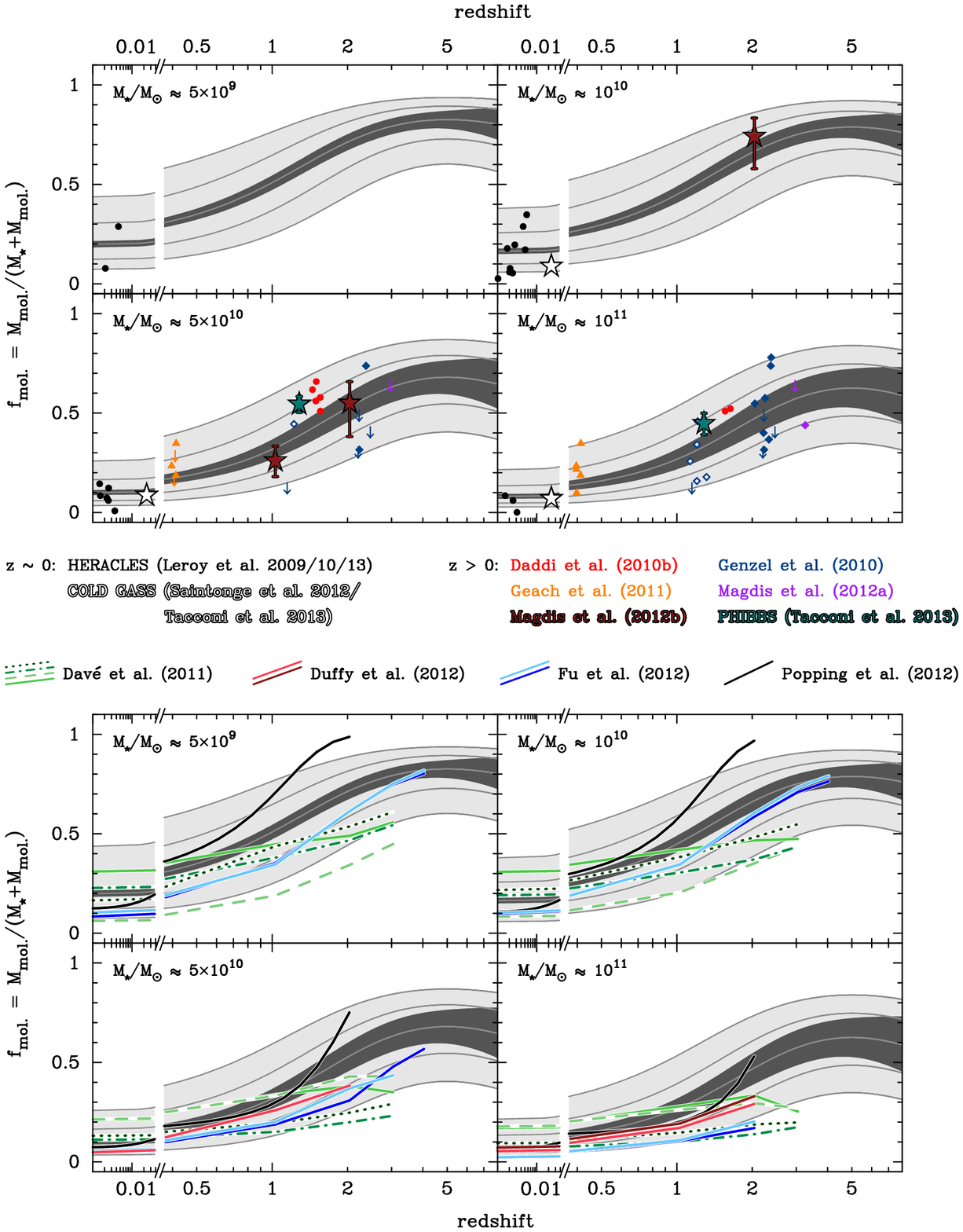}
\caption{\footnotesize Molecular gas mass fraction ($\mu_{\rm mol.}$\,$\equiv$\,$M_{\rm mol.}/M_{\star}$) versus sSFR for selected main-sequence (MS) galaxies and starbursts (SBs) at $z$\,$\lesssim$\,3 ({\it left}), as well as for high- and low-redshift sample averages from \citet{magdis12b} and COLD GASS \citep{saintonge12}, resp. ({\it right}). All measurements are normalized to the $M_{\star}$- and redshift-dependent average of the MS and all symbols and data are identical to those used in Figure \ref{fig:SFEvsSFR}. At all redshifts, the gas fractions of normal galaxies rise uniformly across the MS while starbursting sources have a gas content that is somewhat lower than the MS average.
\label{fig:normfgas}}
\end{figure*}
The scatter (indicated with light grey shading in Figure \ref{fig:loon}) abruptly increases in the transition region with its mixture of MS and SB galaxies and then at \ssfr/\ssfrMS\,$\gtrsim$\,4 is predicted to settle to a constant value that is somewhat larger than the 0.2\,dex of the MS locus as it simultaneously reflects (a) the spectrum of initial MS states that end up at a given \ssfr\ excess by virtue of their different \sfr\ boosts, and (b) of the dispersion in {\sfe}s in the pre-burst MS state. The depicted average trend also depends on the relation between \sfe\ and boost. In Figure \ref{fig:loon}(a) we show the prediction for the case of the ``direct" boost function (the \sfe\ evolution for starbursting galaxies would steepen when using the boost function including an explicit correction for mergers) and do so only for the best-fit value of $\gamma_{\sfe}$. We refrain from estimating formal errors for this average trend since the systematic uncertainties (for example those pertaining to the choice of a function relating \sfe\ excess and boost as in Equation \ref{eq:SFEcalib_SB}) in any case strongly outweigh these.\\
Note that, by working with (s){\sfr}s that are normalized to the average MS value rather than absolute quantities, we have removed all dependence on redshift. Our prediction for the variation of \sfe\ can thus be summarized by a single track, as shown in Figure \ref{fig:loon}(a). It is generalizable to the full range of redshifts and stellar masses by (1) multiplying (s){\ssfr}s by the normalization constant, \ssfrMS, which is given by Equation \ref{eq:sSFRevoeq}, and by (2) using Equation \ref{eq:absSFE_MS} with \sfr\,=\,\mstar\,$\times$\,\ssfrMS\ to obtain the normalization, \sfeMS, and thence the absolute value of \sfe. While this provides a conveniently simple and flexible basis for describing {\sfe} variations in SFGs over a large fraction of Hubble time, these trends do not represent fundamental scaling laws. Our observationally motivated purpose here is to derive the plausible variations of molecular gas-related quantities in the {\sfr}-\mstar\ plane, since this parameter space is an important focus of current literature, e.g., when discussing the properties of SFGs and AGN in general \citep[e.g.,][]{elbaz11, whitaker12, mullaney12b, rosario13}, and in particular also gas and dust properties \citep[e.g.,][]{magdis12b, saintonge12, santini14, magnelli14}. We do not propose that the {\sfe} versus {\ssfr} trends presented in this work are new laws superseding, e.g., the S-K relation. Instead, the dependence of {\sfe} on {\ssfr} highlighted in Figure \ref{fig:loon} is a consequence of the different loci (see Figure \ref{fig:sSFRnSFEdistrib}) of ``normal" and starbursting galaxies in the S-K plane (\mmol\ versus {\sfr}). The {\sfe} trends in {\sfr}-\mstar\ space thus jointly reflect the balance between the relative number of starbursting (``boosted") and normal galaxies at a given \mstar\ and {\sfr} {\it and} the {\sfe} of these two classes of galaxies.

\subsection{Molecular Gas Fractions in Star-forming Galaxies}
\label{sect:fgas}

\subsubsection{Gas Fractions in Normal and Starbursting Galaxies}
\label{sect:obs_fgas}

The predictions of the previous section for the average variation of \sfe\ in SFGs with offset from the MS locus are equivalent to a variation of the molecular gas mass $M_{\rm mol.}$\,=\,$\Big\{\mstar\,{\times}\,\ssfr(\mstar$,$z)\Big\}$/\sfe. By construction, the stellar mass \mstar\ in this expression is known in the 2-SFM approach, implying that we can directly compute the molecular gas mass to stellar mass ratio, $\mu_{\rm mol.}$\,$=$\,$M_{\rm mol.}$/\mstar, as a function of \ssfr. To obtain predictions that are independent of redshift and stellar mass we again consider normalized quantities, \ssfr/\ssfrMS\ and $\mu_{\rm mol.}/\langle\mu_{\rm mol.}\rangle$, in Figure \ref{fig:loon}(b). The bimodal behavior of the average \sfe\ evolution leads to two distinct regimes: (1) a nearly linear increase of the gas fraction across the MS, and (2) for strong SBs, gas fractions that vary more slowly with \ssfr\ and are somewhat lower than the gas fraction of the average MS galaxy. (Note that as in Figure \ref{fig:loon}(a), gas fraction variations within the starbursting population are again shown for the direct boost function; using the boost function that includes the explicit correction for mergers would result in a shallower trend.) As in the case of \sfe\ (see Figure \ref{fig:loon}(a)), the transition between these two regimes is almost step-like and characterized by a large dispersion in $\mu_{\rm mol.}$. We explore the link between gas fractions during an SB episode and prior to the onset of burst-activity in more detail in section \ref{sect:fgas_in_SBs}.\\
We compare these expectations of the 2-SFM framework with real data in Figure \ref{fig:normfgas}. While our reference sample of SBs is too small to quantitatively constrain any residual variation of gas fractions at high {\ssfr}s, the predicted rise of gas fractions across the MS is well sampled by the reference sample of normal galaxies. As discussed in Section \ref{sect:fgas_evo}, the gas fractions inferred for 90\% of the MS galaxies involve an assumption about the metallicity dependence of \aCO. Nevertheless, they show no systematically different behavior than the eight sources (boxed symbols in Figure \ref{fig:normfgas}) for which measurements of \aCO\ exist. A very similar slope (\ssfr$^{0.9}$) was measured by \citet{magdis12b} for stacked samples of MS galaxies divided into bins of \ssfr\ excess in which the gas mass was constrained via the far-IR dust emission. Stacking-based measurements of $\mu_{\rm mol.}$ at $z$\,=\,1, and 2 by \citet{magdis12b} are shown in Figure \ref{fig:normfgas}(b) and found to coincide with the 2-SFM predictions and gas fractions determined on an individual basis for galaxies in our reference sample. \citet{saintonge12} were able to sample molecular gas mass fraction variations over a larger range in \ssfr\ which extends to significantly below the star forming MS and also slightly into the SB regime. In their local COLD GASS data set the average $\mu_{\rm mol.}$ scales as approx. \ssfr$^{0.7}$. A significant deviation from the 2-SFM predictions for normal galaxies is only seen in their highest sSFR-bin which lies in the transition region between MS and starbursting outliers and may hence be expected to reflect the transition to the lower gas fractions in starbursting galaxies (see also the indication of an SFE increase in the COLD GASS data set at \ssfr/\ssfrMS\,=\,4.5 shown in Figure \ref{fig:normSFE}(b)).

\subsubsection{Gas Fraction Evolution Across Cosmic Time}
\label{sect:fgas_evo}

\begin{figure*}
\epsscale{0.93}
\centering
\plotone{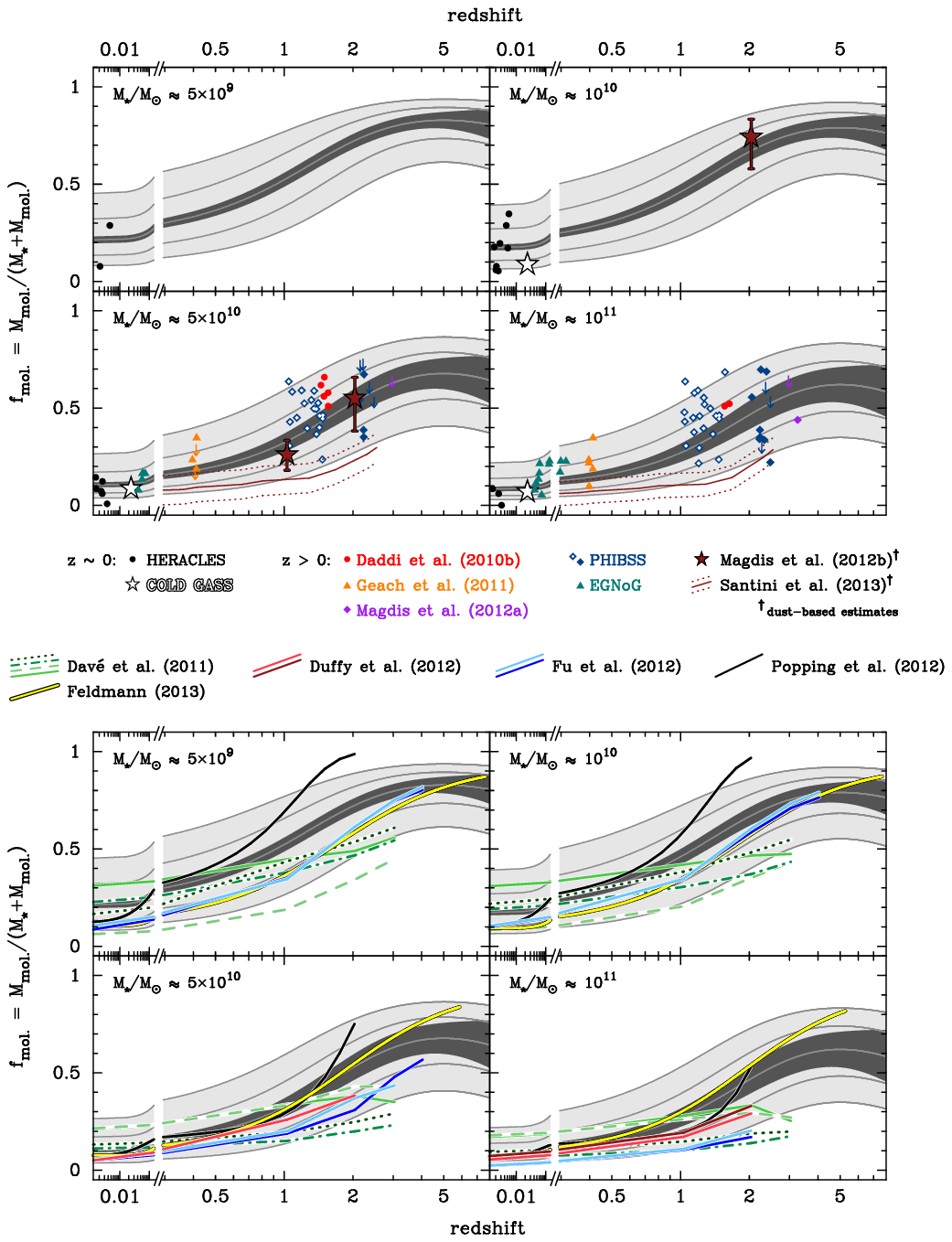}
\vspace{1truecm}
\caption{\footnotesize Redshift evolution of the molecular gas fraction, $f_{\rm mol.}$, of main-sequence (MS) galaxies in four different stellar mass scales, as predicted by the 2-SFM framework based on the evolution of the sSFR and of the integrated S-K relation (see Equation \ref{eq:fgasrecipe}). Upper half of figure -- comparison with literature data (taken to have stellar masses within at most a factor two of the mass scale used for the analytical predictions); lower half -- comparison with predictions from numerical simulations and semi-analytical modeling. The dark grey shading illustrates the uncertainty ($\pm$1\,$\sigma$; reflects the uncertainty of the \ssfr\ evolution according to Equation \ref{eq:sSFRevoeq} and of the integrated S-K law in Equation \ref{eq:MgasvsSFR}) on the evolution of $f_{\rm mol.}$ for a typical MS galaxy. Light grey areas illustrate the predicted dispersion of gas fractions. The evolution for galaxies offset by +2/+1/0/-1/-2\,$\sigma$ from the average MS locus are additionally highlighted (uppermost to lowermost medium grey line). The good agreement between predictions and data in the upper half of the figure -- i.e. the observation of a synchronous evolution of sSFR and $f_{\rm mol.}$ -- is consistent with the evolution of the gas reservoirs in normal galaxies being the primary driver of the cosmic sSFR evolution. Star-shaped symbols indicate gas fractions determined with stacking \citep{magdis12b} or sample averaging \citep{saintonge12}. In the upper half of the figure two examples of indirect estimates of the gas fraction evolution based on dust mass measurements \citep{magdis12b, santini14} are plotted in dark red. Colors and line styles used to represent different simulation predictions in the lower half of the figure are: dotted/dot-dashed/dashed/solid green lines for the vzw/cw/nw/sw (momentum-conserving/constant/no/slow wind) scenario in \citet{dave11}; red/scarlet lines for the ``L050N512"/``L100N512" realizations in \citet{duffy12}; light/dark blue lines for ``prescription 1"/``prescription 2" in \citet[H$_2$ fraction depending on local cold gas surface density and metallicity/H$_2$ fraction depending on ISM pressure]{fu12}, and for the assumption of two different SF laws for regions where atomic and molecular gas dominates, resp. \citep[see][]{bigiel08}; yellow line for predictions from \citet{feldmann13}.
\label{fig:avfgasevo}}
\end{figure*}

The well-defined relations between \mstar\ and \sfr, and between \sfr\ and H$_2$-mass allow for a straightforward prediction of the redshift evolution of the molecular gas fraction, $f_{\rm mol.}$, in normal galaxies. Using Equation \ref{eq:MgasvsSFR} we can write
\small
\begin{eqnarray}
f_{\rm mol.} &\equiv& \frac{M_{\rm mol.}}{M_{\rm mol.} + \mstar } = \frac{1}{1+\nicefrac{\mstar}{({\rm const.}\times\sfr^{\beta_2})}}\nonumber \\
&=& \frac{1}{1+\frac{\mstar^{1-\beta_2}}{\rm const.}\times\ssfr^{-\beta_2}}~,
\label{eq:fgasrecipe}
\end{eqnarray}
\normalsize
where const.\,=\,10$^{\alpha_{2,\,\sfr}}$. If we insert the average {\ssfr}(\mstar,\,$z$) of MS galaxies (parameterized as in Equation \ref{eq:sSFRevoeq}) into this equation we obtain an evolutionary trend, which we plot in the upper half of Figure \ref{fig:avfgasevo} for four different stellar masses in the range $5{\times}10^9$\,$<$\,\mstar/\msun\,$<$\,10$^{11}$. The gas fractions of normal galaxies at $z$\,$\lesssim$\,3 predicted in this manner are in excellent agreement with literature data based on CO line flux data, which suggests that the gas content of secularly evolving SFGs is an important driver of the cosmic \ssfr\ evolution. Supporting evidence for this tight link between the evolution of \ssfr\ and the gas fraction of MS galaxies was recently provided by an analysis of more than 50 SFGs at 1\,$<$\,$z$\,$<$\,3 with CO-flux measurements from the PHIBSS survey in \citet[blue filled and open diamonds in our Figure \ref{fig:avfgasevo}]{tacconi13}. \citet{combes13}, on the other hand, hold a joint redshift evolution of both \sfe\ and gas fractions by roughly similar amounts to be responsible for the cosmic evolution of star formation activity. This apparent discrepancy can be explained by the distinctly different behaviour of normal galaxies and SBs with respect to SFE and $f_{\rm mol.}$, which we illustrate in Figure \ref{fig:loon}, and by the fact that the 0.2\,$<$\,$z$\,$<$\,1 ULIRGs analyzed by \citet{combes11, combes13} have generally large \ssfr\ excesses \citep[see Figure 9 in][]{combes13}, while their local reference sample overlaps with the MS population. These authors hence tend to compare fairly low-efficiency $z$\,=\,0 systems with $z$\,$>$\,0.2 starbursting galaxies (thereby overestimating the importance of \sfe\ for sSFR evolution) which on average have lower gas fractions than equally massive normal galaxies (leading to an underestimate of the gas fraction evolution).\\
The gas-to-dust ratio technique, which is more efficient than CO follow-up in terms of observing time requirements, has become increasingly popular for indirect estimation of the ISM content of high-redshift galaxies \citep[e.g.,][]{magdis11, magdis12b, magnelli12, santini14, scoville14}. In the upper half of Figure \ref{fig:avfgasevo} we have plotted the redshift evolution of gas fractions measured by \citet{magdis12b} and \citet{santini14} using this approach (see dark red stars and lines, resp.). The results from these two studies differ by a factor two, suggesting that this approach is currently still subject to systematic uncertainties. Such systematics could be caused by field-to-field variance as proposed by \citet{santini14} or different corrections for environmental crowding in the stacks of {\it Herschel} photometry that are used to determine average dust-continuum fluxes for high-$z$ galaxy samples. The fact that metallicity estimates are central to the application of the gas-to-dust ratio technique implies that this method is also affected by the systematic offsets between different metallicity calibrations \citep[e.g.,][]{kewleyellison08}.

\noindent The link between \ssfr\ and gas fraction also manifests itself at fixed redshift as a variation of \ssfr\ across the MS (see Figure \ref{fig:normfgas}). This is equivalent to stating that the dispersion of the MS can be at least partially ascribed to different gas fractions. In a recent morphological study of $z$\,$\sim$\,1 disk galaxies \citet{salmi12} reported that systems with clumpy substructure are found to be systematically offset to higher values of \ssfr\ than their smoother counterparts. Since clumps are a telltale signature of violent disk instabilities in gas-rich high-redshift galaxies \citep[e.g.,][]{agertz09, ceverino10, foersterschreiber11, swinbank11, wuyts12} this observation thus provides independent evidence for increasing gas fractions within the MS.

\noindent Following the publication of simulation-based recipes for the calculation of \aCO\ by \citet{narayanan11}, the same authors have recently questioned \citep{narayanan12} the reliability of the high gas fractions reported in the literature. (Alternative predictions for \aCO\ that make use of the observed relations between \lco\ and \lir\ and \sfr\ and H$_2$-mass are presented in Section \ref{sect:XCO}.) Various additional predictions for gas fraction evolution based on numerical simulations \citep{dave11, duffy12} or semi-analytical modeling \citep{fu12,feldmann13} are shown in the lower half of Figure \ref{fig:avfgasevo}. These predictions generally lie within the range of gas fractions expected in the 2-SFM framework for galaxies located on the MS, but -- with the exception of the predictions in \citet{feldmann13} -- they show a tendency for a shallower redshift evolution of $f_{\rm mol.}$ than observed through CO-line flux measurements. Rather than being the consequence of incompatible assumptions for the calculation of \aCO\, these differences compared to the 2-SFM predictions might reflect the well-known problem that both semi-analytical models \citep[e.g.,][]{fontanot09} and cosmological hydrodynamical simulations \citep[e.g.,][]{weinmann12} tend to produce too many stars too early in the history of the universe, especially in lower mass galaxies. The resulting accelerated exhaustion of gas reservoirs would then likely lead to lower gas fractions than we predict using the 2-SFM approach. \citet{popping12} used a similar, empirically motivated approach as the one proposed here to indirectly infer the gas content of both late- and early-type galaxies at $z$\,$<$\,2. Here we show the redshift evolution of $f_{\rm mol.}$ these authors derive for MS galaxies with the stellar masses plotted in Figure \ref{fig:avfgasevo}. Their expectations are in good agreement with ours at $z$\,$\lesssim$\,1 but begin to differ from the 2-SFM predictions at higher redshift, probably due to incompleteness in their lower-mass (\mstar/\msun\,$<$\,10$^{11}$) galaxy samples.

\begin{figure}
\epsscale{1.2}
\centering
\plotone{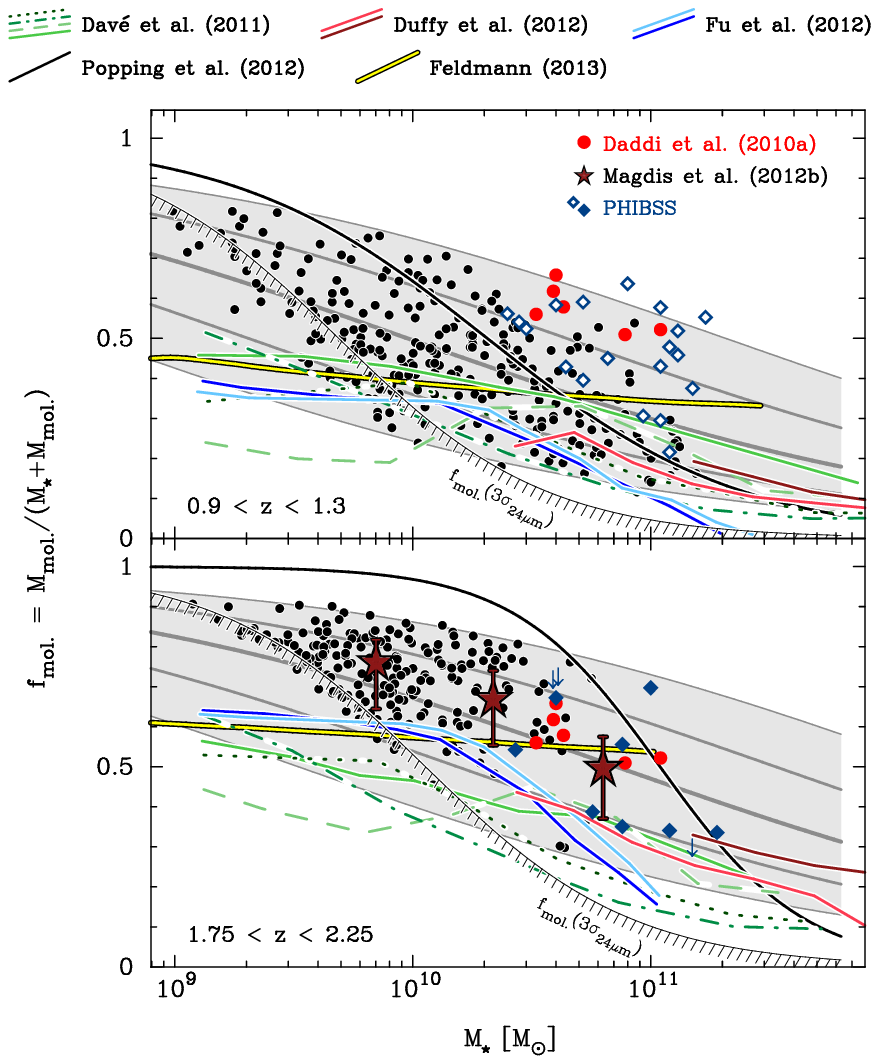}
\caption{\footnotesize Variation of the molecular gas fraction, $f_{\rm mol.}$, of main-sequence (MS) galaxies with stellar mass in two different redshift bins ($z$\,$\sim$\,1 -- {\it top}; $z$\,$\sim$\,2 -- {\it bottom}). Selected MS galaxies from \citet{daddi10a} and the PHIBSS survey \citep{tacconi13}, as well as average gas fractions determined in the stacking analysis of \citet{magdis12b} are plotted in color. Predictions from the 2-SFM framework and recent theoretical/numerical work in the literature are plotted with the identical symbols and color scheme as used in Figure \ref{fig:avfgasevo}. Black dots -- indirect measurements of $f_{\rm mol.}$ for 24\,$\mu$m-selected GOODS-S galaxies based on the inverse S-K relation calibrated in Section \ref{sect:MgasvsSFR}. The hatched line indicates the completeness limit of the statistical GOODS-S sample, as determined by the depth of the 24\,$\mu$m imaging and the average redshift of the two bins displayed.
\label{fig:fgasscatter}}
\end{figure}

\noindent We end this section by plotting explicitly in Figure \ref{fig:fgasscatter} the \mstar\ dependence of the molecular gas fraction which was already visible in Figure \ref{fig:avfgasevo} as a vertical offset between the evolutionary trends. Two panels with the 2-SFM predictions for $z$\,$\sim$\,1 and 2 are shown. In both cases the range of expected gas fractions $f_{\rm mol.}$\,=\,$M_{\rm mol.}$/(\mstar+$M_{\rm mol.}$) at fixed stellar mass can be significant, e.g., for a stellar mass of $5{\times}10^{10}$\,\msun\ a $\Delta f_{\rm mol.}$ of approx. 0.5 is expected, depending on whether the source is located at a positive or negative offset of 2\,$\sigma_{\rm MS}$ with respect to the MS. We illustrate this considerable scatter with our two statistical samples of GOODS-S galaxies (see Section \ref{sect:GOODSgals}), to which we apply Equation \ref{eq:MgasvsSFR} to indirectly infer gas masses. In spite of the large dispersion, a clear trend of decreasing gas fractions with increasing stellar mass is seen. Semi-analytical models and simulations predict either qualitatively similar, albeit somewhat shallower trends \citep[e.g.,][]{dave11, duffy12, fu12} or virtually no dependence at all of gas fractions on stellar mass \citep[e.g.,][]{feldmann13}.

\subsection{The CO-to-H$_2$ Conversion Factor $\alpha_{\rm CO}$}
\label{sect:XCO}

In Section \ref{sect:2-SFM_SFE} we showed how the \ssfr-dependent SB demographics of the 2-SFM approach led to a nearly step-like variation of the \sfe\ even if starbursting galaxies are treated as a continuous extension of normal galaxies, with depletion times that decrease in proportion to the burst strength (referred to as ``boost" throughout this paper). The actual value of \sfe\ is tightly linked to the CO-to-H$_2$ conversion factor, \aCO. In the past it has been common practice to adopt one of two discrete, ``consensus" values when estimating molecular gas masses for high-redshift galaxies: \aCO\,=\,4.4\,\msun\,(K\,km/s\,pc$^2)^{-1}$ (the conversion factor that is found to apply to GMCs in the Milky Way; e.g. \citealp{bolatto08, abdo10}) for normal galaxies and 0.8\,\msun\,(K\,km/s\,pc$^2)^{-1}$, a representative average for local starbursting ULIRGs \citep[e.g.,][]{downessolomon98}. Although the collectively high luminosities of distant, CO-detected galaxies suggested that their \aCO\ values should be ULIRG-like, the first actual estimates of \aCO\ for $z$\,$>$\,1 disk galaxies (based on both dynamical arguments as in \citealp{daddi10a} or on the gas-to-dust ratio approach implemented in \citealp{magdis11}) were all broadly consistent with a Galactic \aCO. Going a step further, \citet{genzel12} subsequently were able to show that the conversion factor of high-$z$ MS galaxies scales with gas-phase metallicity in a similar manner as the negative power laws observed for local galaxies \citep[e.g.,][]{wilson95, israel97,  boselli02, leroy11, schruba12}.\\
The physics of the multiphase ISM that ultimately determines the exact value of \aCO\ is complicated, regardless of whether the emission from individual star-forming regions \citep[e.g.,][]{glover11} or from larger scales even up to integrated emission are considered \citep[e.g.,][]{narayanan11, feldmann12a, papadopoulos12}. A common feature of all these theoretical or numerical calculations is a dependence of \aCO\ on metallicity. As motivated in Section \ref{sect:MgasvsSFR}, here we adopt a shielding-based prescription for \aCO\ from \citet{wolfire10}, in which conversion factors vary weakly with metallicity around solar abundance but then increase quickly at $Z/Z_{\odot}$\,$<$\,$\nicefrac{1}{2}$. The fact that SBs in the 2-SFM framework preserve a memory of their former MS state means that our predicted SB \aCO\ values also depend on metallicity but in a more complicated way which is detailed in the following.

\begin{figure*}
\epsscale{1.13}
\centering
\plotone{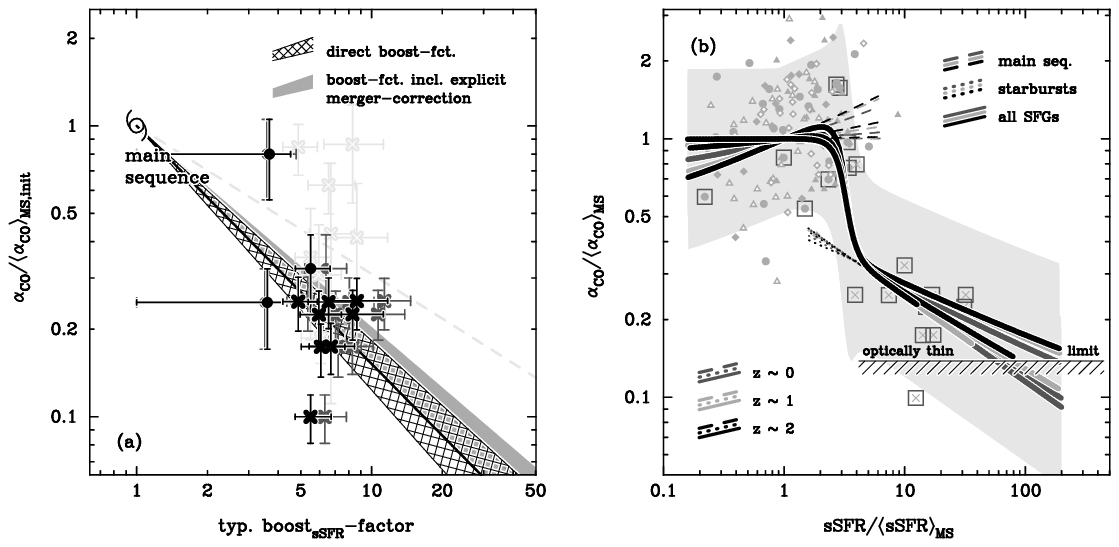}
\caption{\footnotesize ({\it a}) Empirical calibration of the decrement in \aCO\ with the amount of (s)\sfr\ boosting experienced by starbursts (SBs; all symbols and data as in Figure \ref{fig:boostconvert}(b)). The decrement is referred to the CO-to-H$_2$ conversion factor, $\langle$\aCO$\rangle_{\rm MS,\,init.}$, that would be expected for each SB galaxy if it were a secularly evolving, average, normal galaxy with the same molecular gas fraction (see Section \ref{sect:SB-XCO} for details). Starburst \aCO\ measurements plotted in black are from dynamical modeling in \citet{downessolomon98}, light grey symbols are for \aCO\ values derived by \citet{papadopoulos12} using a two-phase LVG model. (Crosses/dots are used for local/high-$z$ SBs from \citet{solomon97} and \citet{magdis12b}, respectively; 1\,sigma-error bars plotted.) The shaded/hatched areas span the 68\% confidence region for a power-law relation between SB \aCO\ and boost amplitude as parametrized in Equation \ref{eq:XCOcalib_SB} and when fitting to \aCO\ measurements from \citet{downessolomon98}. The light grey dashed line shows the best-fitting power-law relation inferred based on the \aCO\ values from two-phase LVG modeling. ({\it b}) Predicted variation of average, normalized \aCO\ with sSFR excess for normal MS galaxies ({\it dashed lines}), SBs ({\it dotted lines}; ``direct" boost function assumed) and the total population of SFGs ({\it solid lines}). Data points plotted in grey in the background are as in Figure \ref{fig:loon} (for visualization purposes in this figure an artificial dispersion has been added to galaxies with statistical estimates of \aCO, i.e., for all sources which are not plotted with boxed symbols; see text for details). Due to the non-linear dependence of the gas-phase metallicity on SFR and stellar mass \citep[as parameterized by the FMR of][]{mannucci10} the average \aCO\ trends predicted by the 2-SFM framework are both stellar mass- and redshift dependent. In a given redshift bin (color-coded as shown in the upper right corner) the shallowest variation across the MS (i.e. in the range $\nicefrac{1}{6}$\,$<$\,\ssfr/\ssfrMS\,$<$\,6) occurs for the most massive of the three stellar mass bins considered here (\mstar/\msun\,=\,5$\times$10$^{9}$, 5$\times$10$^{10}$ and 5$\times$10$^{11}$), while the steepest variation occurs for the least massive bin. At high excesses of sSFR the \aCO\ values are predicted to decrease the most steeply in the highest and most slowly in the lowest stellar mass bin plotted. Over the range 0\,$<$\,$z$\,$<$\,2 shown here, the variation with stellar mass is expected to be more significant than that with redshift. At the highest \ssfr\ excesses (boosts) the \aCO\ values in SBs plausibly asymptotically approach the lower limit set by optically thin CO line-emission. 
\label{fig:XCOvsboost}}
\end{figure*}

\subsubsection{Conversion Factors for Starbursts: Empirical Calibration of Boost Dependence}
\label{sect:SB-XCO}

As for the \sfe\ (see Equation \ref{eq:SFEcalib_SB}) we assume that \aCO\ varies smoothly\footnote{~The functional form of Equation \ref{eq:XCOcalib_SB} is not merely motivated by its symmetry with Equation \ref{eq:SFEcalib_SB}, it also reflects the expectation that the state of the ISM evolves continuously, e.g. depending on the strength of the tidal forces which may enhance the amplitude of the turbulent motions during galaxy-galaxy interactions \citep[e.g.,][]{bournaud11b}. Larger velocity gradients and higher temperatures, which are generally characteristic of the turbulent and dense starbursting ISM, both lower \aCO\ while higher column densities increase the CO-to-H$_2$ conversion factor \citep[see, e.g., Equation A4 in][]{papadopoulos12}.} with the boost of an SB,
\small
\begin{equation}
{\rm log}\left(\frac{\aCO}{\alpha_{\rm MS,\,init.}}\right) = \gamma_{\aCO} \times b_{\ssfr}~,
\label{eq:XCOcalib_SB}
\end{equation}
\normalsize
and use our sample of reference SBs (cf. Section \ref{sect:SBdata}) to determine the most suitable value of $\gamma_{\aCO}$, given this choice of parameterization. Boosts $b_{\ssfr}$ are assigned as in Section \ref{sect:2-SFM_SFE} and \aCO\ values are given in \citet{downessolomon98} or \citet{magdis12b} for each SB in the reference sample. $\alpha_{\rm MS,\,init.}$ corresponds to the conversion factor of a MS galaxy with the same SFR as a reference SB, but with \lco\ and \mhtwo\ given by the inverse integrated S-K relations in Equation \ref{eq:LCOvsLIR} and \ref{eq:MgasvsSFR}, respectively. Since we refer the \aCO\ to this initial state by definition no constant term is required in Equation \ref{eq:XCOcalib_SB}. Solving for $\gamma_{\aCO}$ we obtain $\gamma_{\aCO}$\,=\,-0.82$_{-0.09}^{+0.08}$ (1\,$\sigma$ errors quoted) for the case of the direct boost function and $\gamma_{\aCO}$\,=\,-0.75$\pm$0.06 for the merger-corrected boost function. The boost dependence for the two scenarios is shown in Figure \ref{fig:XCOvsboost}(a), together with the SB data used for the fit. The relatively slow, sub-linear decline of \aCO\ with boost amplitude (i.e. SFR enhancement) according to \aCO\,$\propto$\,(boost)$^{\gamma_{\aCO}}$ implies that to reach ULIRG-like values of the conversion factor, which are about \nicefrac{1}{5} of the typically assumed Milky-Way-like \aCO\,=\,4.4\,\msun\,(K\,km/s\,pc$^2)^{-1}$, a boost by a factor of 8--10 is expected according to the 2-SFM description.\\
Several of the \aCO\ measurements for our reference SBs deviate more strongly from the average trend between boost and conversion factor than was the case for the relation  between \sfe\ and boost we calibrated in Figure \ref{fig:boostconvert}(b). As a consistency check we hence used the results of the LVG radiative transfer modeling by \citet{papadopoulos12} of all nine starbursting local ULIRGs in our reference sample to re-derive the logarithmic slope $\gamma_{\aCO}$ in Equation \ref{eq:XCOcalib_SB}. We find that \aCO\ values determined with one-phase radiative transfer models are on average consistent with the dynamical estimates of \citet{downessolomon98}, such that the resulting slope is almost identical to the previously measured one: $\gamma_{\aCO}$\,=\,-0.85$_{-0.09}^{+0.08}$ (direct boost function) and $\gamma_{\aCO}$\,=\,-0.78$\pm$$_{-0.06}^{+0.05}$ (boost function corrected for merger-statistics). CO-to-H$_2$ conversion factors inferred with two-phase (for high- and low-excitation gas) ISM models are generally higher (see light grey crosses in Figure \ref{fig:XCOvsboost}(a)), leading to a shallower slope $\gamma_{\aCO}$\,=\,-0.51$_{-0.24}^{+0.10}$ (-0.46$_{-0.23}^{+0.09}$) for the direct (merger-corrected) boost function. Given the good agreement between the former two estimates of $\gamma_{\aCO}$ we have adopted the dynamically constrained CO-to-H$_2$ conversion factors reported in \citet{downessolomon98} throughout this paper.

\begin{figure*}
\epsscale{1.03}
\centering
\plotone{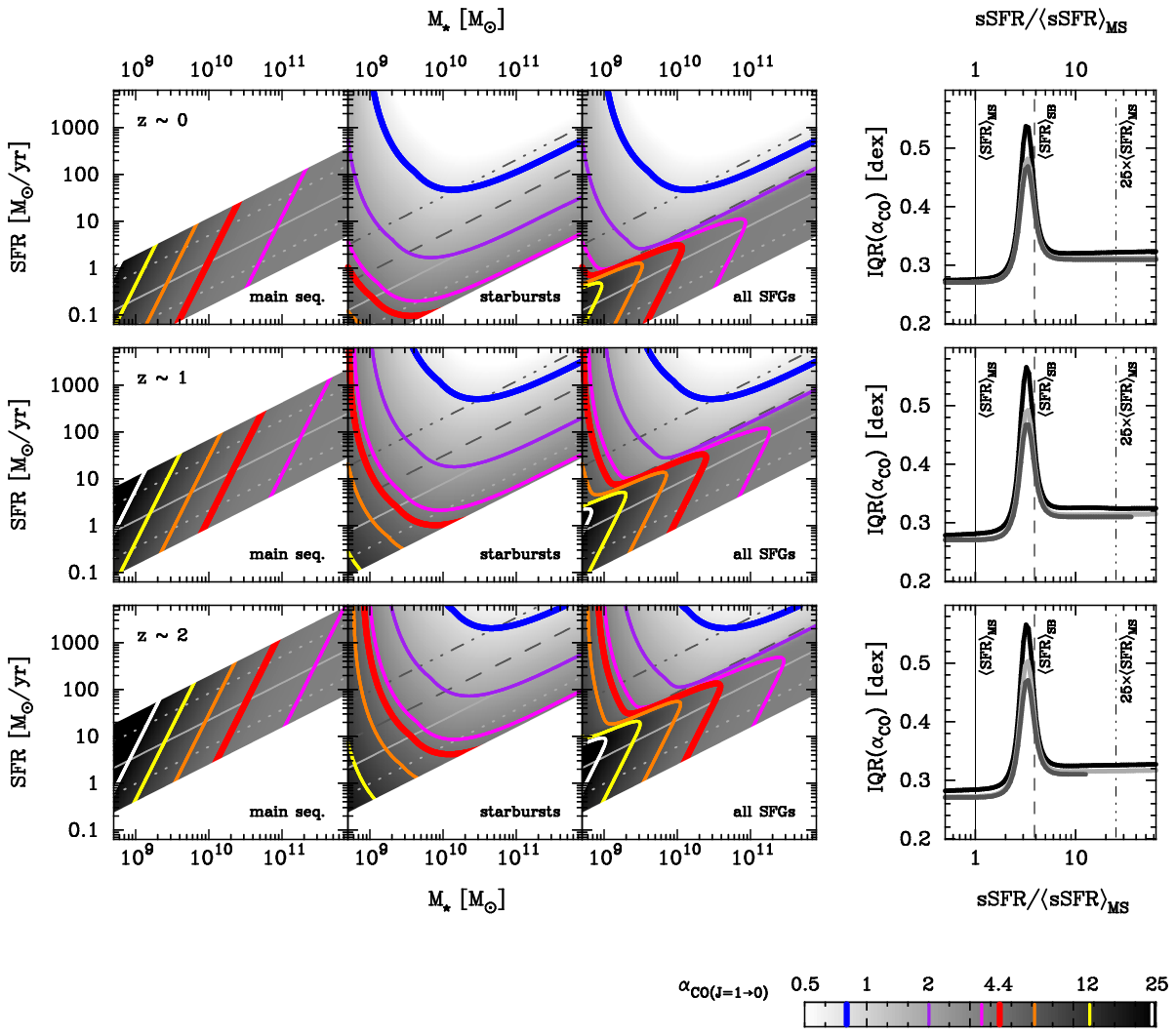}
\caption{\footnotesize Predicted variations of the CO-to-H$_2$ conversion factor, \aCO\ (for the $J$\,=\,1\,$\rightarrow$\,0 transition and assuming the `direct' boost function), in the SFR-\mstar\ plane for main-sequence (MS) galaxies, starbursting galaxies and for the combined population of normal SFGs and starbursts (SBs; {\it first, second \& third column}, resp.) at $z$\,$\sim$\,0, 1 and 2 ({\it from top to bottom}). \aCO\ variations are mapped within $\pm$5 times the dispersion ($\sigma_{\rm MS}$) of the MS for the normal galaxies and between -5\,$\sigma_{\rm MS}$ and arbitrarily high sSFR excesses for SB galaxies and the total SFG population (columns 2 \& 3). Lines of constant \aCO\,=\,0.8, 2, 3.5, 4.4, 6.5, 12 \& 24 are superimposed in blue, purple, magenta, red, orange, yellow and white (standard values of a ULIRG and Milky Way conversion factor -- \aCO\,=\,0.8 \& 4.4 -- are highlighted with bold lines). Values plotted in columns 1--3 represent the median for the respective (sub-)population. Grey diagonal lines trace lines of constant sSFR at: $\pm$3\,$\sigma_{\rm MS}$ -- dotted; \ssfrMS\ -- solid; $\langle{\ssfr}\rangle_{\rm SB}$ -- long dashes; 25$\times$\ssfrMS\ -- dash-dot-dot. Column 4 shows the scatter of \aCO\ (visualized here by the interquartile range IQR of logarithmic \aCO\ values) around the average trends for the total SFG population (cf. col. 3) at stellar mass \mstar/\msun\,=\,5$\times$10$^{9}$ ({\it black}), 5$\times$10$^{10}$ ({\it light grey}) and 5$\times$10$^{11}$ ({\it dark grey}). Vertical lines correspond to the lines of constant sSFR plotted in columns 1--3. The dispersion rises strongly over a fairly small range of sSFR where f$^{\rm SB}$\,$\sim$\,50\% (see also Figures \ref{fig:boostdistrib} and \ref{fig:XCOvsboost}(b)).
\label{fig:XCOmap}}
\end{figure*}

\noindent We can now write an expression relating the conversion factor of an SB in general to the MS average:
\small
\begin{eqnarray}
{\rm log}\left(\frac{\aCO}{\aCOMS}\right) &=& {\rm log}\left(\frac{\aCO}{\alpha_{\rm MS,\,init.}}\right) + {\rm log}\left(\frac{\alpha_{\rm MS,\,init.}}{\aCOMS}\right) \nonumber\\
 &=& \gamma_{\aCO}\times b_{\ssfr} + \lbrack f\left(\sfr_{\rm MS,\,init.}\right) \nonumber\\
 &&\hspace{2.5truecm}{-}f\left(\sfrMS\right) \rbrack~.
\label{eq:normXCO_SB}
\end{eqnarray}
\normalsize
Here the first term -- which describes the \aCO\ deficit of SBs with respect to the pre-boost, MS state of each individual galaxy -- 
corresponds to Equation \ref{eq:XCOcalib_SB}. The second term relates the pre-boost, MS state to the MS average via the relation between \aCO\ and metallicity (Equation \ref{eq:alphaofZ_Wolfire}). Eq. \ref{eq:normXCO_SB} strongly resembles Equation \ref{eq:normSFE_SB} for the normalized SFE of SBs but is nevertheless different in that the expression describing variations within the MS,
\small
\begin{eqnarray}
{\rm log}\left(\frac{\aCO}{\aCOMS}\right) &=& {\rm log}\left(\frac{{\rm exp}\left[\frac{\Delta\mathcal{A}_V(Z)}{\nicefrac{Z}{Z_{\odot}}}\right]\,{\rm exp}^{-\Delta\mathcal{A}_V(Z)}}{{\rm exp}\left[\frac{\Delta\mathcal{A}_V(\langle Z\rangle_{\rm MS})}{\nicefrac{\langle Z\rangle_{\rm MS}}{Z_{\odot}}}\right]\,{\rm exp}^{-\Delta\mathcal{A}_V(\langle Z\rangle_{\rm MS})}}\right) \nonumber \\
 &=& \frac{1}{{\rm ln}(10)} {\bigg [} \Delta\mathcal{A}_V(Z)\left(\frac{Z_{\odot}}{Z}-1\right) \nonumber \\
 &&\hspace{1truecm}- \Delta\mathcal{A}_V(\langle Z\rangle_{\rm MS})\left(\frac{Z_{\odot}}{\langle Z\rangle_{\rm MS}}-1\right){\bigg ]} \nonumber\\
 &=& \lbrack f\left(\mu_{0.32}\right){-}f\left(\langle\mu_{0.32}\rangle_{\rm MS}\right)\rbrack \nonumber\\
 &\equiv& \lbrack f\left(\sfr\right){-}f\left(\sfrMS\right) \rbrack~,
 \label{eq:normXCO_MS}
\end{eqnarray}
\normalsize
has higher order terms in log(SFR). This is caused both by the term $\Delta\mathcal{A}_V(Z)$\,=\,$\nicefrac{4\,\Delta A_V(Z)}{\overline{A}_{V,{\rm MW}}}$ which is a function of the metallicity-dependent gas fraction \citep[here we adopt the double power law dependence on metallicity of][]{remy-ruyer14}, and by the fact that our statistical estimates of metallicity involve the FMR parameter $\mu_{0.32}$\,=\,log(\mstar)$-0.32\times$log(SFR) defined in \citet{mannucci10}. The normalized logarithmic conversion factor for MS galaxies, ${\rm log}\left(\nicefrac{\aCO}{\aCOMS}\right)$, can no longer be written as a function $f\left(\nicefrac{\sfr}{\sfrMS}\right)$ of normalized SFR (or, equivalently, \ssfr\ when considering a fixed bin of stellar mass). This was possible, however, in the case of SFE and $f_{\rm mol.}$ (see Sections \ref{sect:2-SFM_SFE} and \ref{sect:obs_fgas}) and led to a redshift- and mass-independent recipe for the evolution of the population average of these quantities with \ssfr/\ssfrMS. In Figure \ref{fig:XCOvsboost}(b) we plot\footnote{~All predictions shown in Figs. \ref{fig:XCOvsboost}(b) and  \ref{fig:XCOmap} assume the ``direct" boost function and the corresponding best-fit value of $\gamma_{\aCO}$.} (fine dashes) the predicted variation of the median \aCO\ of MS galaxies for three stellar mass bins (\mstar/\msun\,=\,5$\times$10$^{9}$, 5$\times$10$^{10}$ and 5$\times$10$^{11}$) and three different redshifts ($z$\,=\,0, 1, 2). Due to the higher order (s)\sfr\ terms in Equation \ref{eq:normXCO_MS} these trends are no longer redshift- and mass-independent; while \aCO\ values for SFGs vary little across the MS in our highest mass bin (\mstar/\msun\,=\,5$\times$10$^{11}$), evolution by approx. a factor two is predicted between $\pm$4\,$\sigma_{\rm MS}$ for stellar masses \mstar\,$\sim$\,5$\times$10$^9$\,\msun. For the mass and redshifts considered here, the mass dependence of the average trends at fixed redshift is more pronounced than the redshift dependence at fixed mass. Note that although we assume the relation between boost and \aCO\ decrement for SBs (see Equation \ref{eq:XCOcalib_SB}) to be independent of redshift and stellar mass, the predicted average \aCO\ trends for starbursting sources nevertheless vary with stellar mass and redshift. This is a consequence of the mass- and redshift-dependency found for the ``parent" MS population. For example, the fact that at low masses conversion factors are predicted to rise across the MS implies that at fixed boost-dependent \aCO\ decrement the \aCO\ of high-\ssfr\ SBs will be higher than for the highest mass bins where \aCO\ values of normal galaxies on the MS are expected to be virtually constant. As a final comment on the description of \aCO\ variations for starbursting sources we should point out that in practice the conversion factor cannot decrease indefinitely (as formally implied by Equation \ref{eq:normXCO_SB}) but that optically thin CO line-emission sets a lower limit (see Figure \ref{fig:XCOvsboost}(b)). Assuming local thermal equilibrium and a gas temperature of 40-60\,K for SB sources, we estimated \aCO\ in the optically thin approximation using standard formulae \citep[see, e.g., Appendix A1 in][]{bryantscoville96} and obtained values ranging between 0.45 and 0.75\,\msun\,(K\,km/s\,pc$^2)^{-1}$. Values at the lower (higher) end of this range are generally predicted for lower (higher) redshift sources due to the evolution of the temperature of the cosmic microwave background, and with an additional contribution from the likely quite mild evolution of the dust temperature in SBs \citep[e.g.,][]{bethermin12}. In relative terms, at all redshifts 0\,$<$\,$z$\,$<$\,2.5 this is about 10\%--20\% of the \aCO\ values expected for massive MS galaxies if their conversion factors also increase with redshift because of the general evolution of the population toward lower metallicity.\\
The median \aCO\ of the total population -- computed analogously to the bulletized procedure sketched in Section \ref{sect:2-SFM_SFE} -- is plotted with thick, solid lines in Figure \ref{fig:XCOvsboost}(b). The exact shape of the transition between MS and SB regime depends both on the assumed scatter of the FMR and the dispersion of \aCO\ at fixed metallicity \citep[see, e.g.,][]{mannucci10, genzel12, schruba12}. For the present case we assume these to be 0.05\,dex and 0.2\,dex, respectively, which leads to a step-like decrease by about a factor 2-3 at an \ssfr\
excess \ssfr/\ssfrMS\,$\sim$\,3-4 with respect to the MS average. Just as for the predicted slope of the normalized \aCO\ versus \ssfr\ relations for normal and SB galaxies, this jump changes with redshift and stellar mass. In the next section we thus provide a more complete mapping of expected \aCO\ variations for SFGs.

\subsubsection{$\alpha_{\rm CO}$: Predicted Variations in the SFR-\mstar\ Plane}
\label{sect:XCOmaps}

To conclude this section on empirical recipes for the CO-to-H$_2$ conversion factor \aCO\ we map its predicted variation within the \mstar\ versus \sfr\ plane for three different redshift bins in Figure \ref{fig:XCOmap}: $z$\,=\,0 -- top row; $z$\,=\,1 -- second row; $z$\,=\,2 -- bottom row. We do this explicitly because the mapping of metallicity into the \mstar\ versus \sfr\ plane following \citet{mannucci10} is such that the variation of the metallicity-dependent \aCO\ we adopt for 2-SFM framework is not self-similar (i.e. independent of stellar mass and redshift as was the case for \sfe\ and $f_{\rm mol.}$; see Figure \ref{fig:loon}) and hence cannot be represented with a single, \ssfr-dependent recipe.\\
The individual panels of Figure \ref{fig:XCOmap} show the variation of \aCO\ in \mstar-\sfr\ space for the total SFG population, for MS galaxies and for SBs (columns 1 to 3). We have superimposed contours of constant \aCO\ and in particular indicated the isolines for Milky-Way-like and ULIRG-like conversion factors with a bold red and blue line, respectively. Due to the increasing normalization of the MS with redshift, metallicities at fixed stellar mass decrease with redshift \citep[this reflects the well-established, measured evolution of the mass-metallicity relation, e.g.,][]{kobulnicky04, erb06, liu08, zahid13}. While this evolution to lower enrichment is expected to be quite strong at the smallest stellar masses plotted in these figures (\mstar\,$\sim$\,10$^9$\,\msun) the evolution is less strong for those galaxies of stellar mass \mstar\,$\sim$\,$3{\times}10^{10}$\,\msun\ that contribute most to the cosmic SFR density over the redshift range considered here \citep[e.g.,][]{karim11}. As a consequence, we expect that the conversion factor of such galaxies remains quite similar to the classic Milky Way value of 4.4\,\msun\,(K\,km/s\,pc$^2)^{-1}$ over the range 0\,$<$\,$z$\,$<$\,2. Specifically, for a galaxy of stellar mass \mstar\,$\sim$\,$3{\times}10^{10}$\,\msun\ that is located directly on the average MS locus, the recipes developed in Section \ref{sect:SB-XCO} predict \aCO\,$\simeq$\,3.8 in the local universe and \aCO\,$\simeq$\,4.5 at $z$\,$\sim$\,2. Note that this prediction is not the coincidental outcome of choosing a specific slope and/or normalization of the relation between \aCO\ and metallicity in Equation \ref{eq:alphaofZ_Wolfire}. It would also hold for any of the other relations shown in Figure \ref{fig:FMPadapt} as these all attain quite similar, Milky-Way-like conversion factors around solar metallicity. For our calculations we have assumed that the same relation between \aCO\ and metallicity holds at all redshifts. Under these circumstances, the weak positive evolution of \aCO\ predicted here is purely due to the lower metallicities of high-redshift galaxies. This evolutionary trend could vanish entirely or even tend toward lower values of the conversion factor if large fractions of the star-forming ISM in high-redshift galaxies have densities significantly higher than GMCs in the Milky Way \citep[see, e.g.,][and references therein]{bolatto13} or if conversion factors scale inversely with CO surface intensity, as proposed by \citet{narayanan11} based on simulations. Lower conversion factors in high-redshift galaxies would bring CO-based gas mass measurements into better agreement with those estimates based on the gas-to-dust ratio technique that are systematically lower \citep[e.g.,][but see also \citealp{magdis12b}]{santini14, scoville14}. However, it is presently unclear if these reports on lower gas masses should be interpreted as evidence that systematic overestimates of the conversion factor produce too high gas masses when these are derived from CO-data. For example, \aCO\ measurements for {\it individual} galaxies (rather than stacked populations) at 1\,$<$\,$z$\,$<$\,2 \citep[e.g.,][]{magdis12b, magnelli12} made with the gas-to-dust ratio technique give values that are inconsistent with a strong decrease of conversion factors at high redshift. Likewise, dynamical constraints on the CO-to-H$_2$ conversion factors in BzK-galaxies \citep{daddi10a} favor Milky-Way-like values. Future studies of high-redshift galaxies with ALMA should be able to clarify whether or not the \aCO\ of normal galaxies has a similar metallicity-dependence at all redshifts.\\
For variations of the conversion factor of starbursting sources in the \mstar\ versus SFR plane (second column of Figure \ref{fig:XCOmap}) we expect two different regimes to exist. At high stellar masses, the \aCO\ values of the parent, MS population vary little (both across the MS with a given bin of stellar mass and between stellar mass bins); the boost-dependent \aCO\ decrement alone hence determines the value of SB conversion factors. As a consequence, lines of equal SB \aCO\ are nearly parallel to the MS locus. At low stellar masses, lines of constant SB \aCO\ run nearly perpendicular to the isolines on the MS locus. This is due to the rapid variation of \aCO\ for normal galaxies, which has the effect that ever higher boost amplitudes are required in order for SBs of successively lower stellar mass to reach equal absolute values of \aCO\ (e.g. the standard local ULIRG value 0.8\,\msun\,(K\,km/s\,pc$^2)^{-1}$). Starting at an (s)\sfr\ excess of about +3\,$\sigma_{\rm MS}$, SBs begin to dominate the MS population by number. The transition between the MS locus and the SB-dominated part of \mstar-\sfr\ space is  characterized by both a sudden drop of the average \aCO\ (see Figure \ref{fig:XCOvsboost}(b)) and an abrupt increase of the dispersion of \aCO, which is a result of the heterogeneous mixture of starbursting and high-\ssfr\ MS galaxies in this transition region. This is illustrated in the panels in column 4 of Figure \ref{fig:XCOmap} where we plot the evolution of the interquartile range of \aCO\ values measured in the total SFG population (i.e. including both SBs and normal galaxies). On the MS (\ssfr/\ssfrMS\,$\lesssim$\,3) the scatter in \aCO\ is caused by the metallicity dispersion of the of FMR at fixed \mstar\ and \sfr\ plus the dispersion of \aCO\ at fixed metallicity and is hence relatively small. The \aCO\ scatter for SB galaxies is larger than that on the MS locus because it reflects both the dispersion of \aCO\ at fixed \mstar\ and \sfr\ on the MS, and the fact that the shape of the boost function implies that SBs at a given \ssfr/\ssfrMS\ have been boosted to higher (s)\sfr\ starting from a range of positions on the MS.

\section{Discussion: Toward a Simple Description of Molecular Gas in Star-forming Galaxies}
\label{sect:discussion}

\subsection{The Boost Function: Astrophysical Context and Limitations}
\label{sect:boostcontext}

In Section \ref{sect:boostfct} we provided simple arguments for why the existence of a statistical link (due to a so far unspecified process) between the populations of MS galaxies and SBs is a natural expectation. Here we discuss which processes might be responsible for such a link and how SF activity may reflect cosmological accretion of dark matter (DM) and baryons.

\subsubsection{Star Formation Enhancements in Simulations and Observations}

Based on a suite of simulated interacting galaxies (with comparable masses and a representative range of both orbital configurations and morphologies) \citet{dimatteo07} and \citet{dimatteo08} derived the SFR evolution of major mergers as compared to the evolution of identical, isolated galaxies. In Figure \ref{fig:boostcomp} we plotted the maximal SFR enhancements reported in \citet{dimatteo08}, averaged between the Tree-SPH (smoothed particle hydrodynamics) simulations and grid-based N-body simulations carried out by these authors. Both the 2-SFM boost function and the distribution of simulated SFR enhancements have a clearly defined peak. The exact position of this peak (which corresponds to $\langle x\rangle_{\rm BK}$ in the 2-SFM formalism) in the simulations depends on the gas content of the galaxies. In local disk galaxies (simulated total gas fractions between 10 and 30\%) the most frequently encountered maximal SFR excess is approx. a factor of three, while for gas-rich simulated galaxies ($f_{\rm mol.}$\,$\sim$\,50\%) reminiscent of high-$z$ disks it is twice as large, mainly because these tend to become Jeans-unstable and form dense gas clumps when perturbed. The 2-SFM boost function peaks at a roughly four-fold SFR enhancement but it should not be directly compared to the simulation results because not all SBs that contributed to the shape of the underlying (s)\sfr\ distribution in \citet{rodighiero11} can have been ``caught" at the peak of the SB activity. With respect to the results of \citet{dimatteo07, dimatteo08}, observable distributions of \sfr\ boosts for interacting galaxies will likely be modified if minor mergers and fly-bys, as well as the relative timing of SB events, are accounted for. If SFR enhancements in merger-driven SBs depend on the mass ratio of the galaxies involved \citep[e.g.,][]{cox08}, then the boost distribution including minor mergers should be broader. Likewise, fly-bys and asynchronous burst-activity plausibly shift and skew the boost distribution to lower \sfr\ enhancements.\\
Although the distribution of \sfr\ enhancements caused by interactions between galaxies is poorly constrained, observationally, there have been numerous studies aimed at quantifying the integrated contribution of excess SF associated with mergers/interactions to the cosmic \sfr\ density at redshifts $z$\,$<$\,2 \cite[e.g.,][]{robaina09, kampczyk13, kaviraj13}. One attempt is the recent determination of the distribution of \sfr\ enhancements in SDSS galaxy pairs by \citet{scudder12} which we plot in blue in Figure \ref{fig:boostcomp}. Compared to the boost distribution in the simulations of major mergers in \citet{dimatteo08} it is indeed displaced to systematically lower SFR enhancements. In addition to the expected shifting and skewing, the fact that the SDSS pair sample does not include mergers in which final coalescence has already taken place implies that their distribution of SFR enhancements represents a lower observational limit to the total local SFR excess distribution caused by interactions. The simulations of \citet{dimatteo07} and \citet{dimatteo08} encompass a broad variety of orbital configurations of merging galaxy pairs and have been statistically weighted to reflect the dependence of the collision rate on the relative velocities and impact parameters. However, they do not provide information on the contribution of minor interactions, nor are they carried out in a fully cosmological framework that accounts for, e.g., the preferential alignment of galaxies in different locations within the cosmic web \citep[e.g.,][]{hahn10}. This additional step was taken in simulations by \citet{hopkins10}, such that their spectrum of merger-induced surplus \sfr\ should be similar to the measurement of \citet{scudder12}. \citet{hopkins10} incorporated the results of their own high-resolution merger simulations \citep{hopkins09} in a cosmological (DM) framework and were able to predict \sfr\ distributions of secularly evolving and starbursting galaxies \citep[see Figure 7 in][]{hopkins10} that, in qualitative terms, resemble the split into MS and SB activity we proposed in S12 and which seem broadly consistent with the expected modifications to the boost distributions of \citet{dimatteo08} discussed at the end of the last paragraph. For galaxies with a stellar mass of 10$^{11}$\,\msun, where we can compare with our own double log-normal decomposition according to Equation \ref{eq:DGintro}, the approach of \citet{hopkins10} predicts (1) an \sfr\ boost distribution for SBs that is broader and more skewed to low boosts, and (2) typical SFR enhancements that are smaller (for simulated $z$\,$\sim$\,0 and 2 mergers) in comparison to both the outcome of the major merger simulations of \citet{dimatteo08} and also the 2-SFM boost function.\\
To infer that galaxy-galaxy interactions cannot be the sole trigger of SB activity, based only on the mismatch between the 2-SFM and measured or simulated boost distributions, would, however, be premature. The approach of decomposing an \ssfr\ distribution into two components -- as done in Section \ref{sect:boostmath} -- leads to inherently poor constraints on the shape of the boost function at low boosts (see cross-hatched area in Figure \ref{fig:boostcomp}) since galaxies with small \ssfr\ enhancements blend in entirely with the MS population. This ``maximization" of the MS contribution in Equation \ref{eq:DGintro} will hence cause a truncation of the lower part of the boost function. On the other hand, the distinction between secularly evolving and only weakly starbursting systems itself is not clear-cut because minor merger events occur frequently. If all galaxies that have experienced minimal boosting are regarded as SBs, then a boost distribution that goes to zero at a (s)\sfr\ excess of zero (boost\,=\,1) is unrealistic. The 2-SFM boost function should thus best be viewed as the signature of strong boosting where a significant fraction of the ISM fuels SB activity. Despite this limitation it is interesting that the peak position of the 2-SFM boost function at an excess (s)\sfr\ of a factor of four corresponds exactly to the average \sfr\ boost that was measured by \citet{hwang11} for FIR-selected galaxies at $z$\,=\,0 and $z$\,=\,1 undergoing an interaction with a late-type neighbor. In another study \citet{kampczyk13} demonstrate that SF, as traced by optical line emission, is boosted by roughly a factor four for the most closely bound (physical separation $<$30\,$h^{-1}$\,kpc) kinematic pairs at 0.2\,$<$\,$z$\,$<$\,1 from the zCOSMOS survey. \citet{parkchoi09} investigated the interaction-induced \sfr\ boosting in local late-type galaxy pairs and found an increase of the equivalent width of the H$\alpha$-line by an identical factor four when the two galaxies were separated by less than 1\% of the virial radius of the companion's halo. This constancy of the average \sfr\ excess is reminiscent of the evidence for only mild evolution of the shape of the boost function we presented in S12. Furthermore, it is worth noting that notwithstanding the uncertainties concerning the shape of the lower end of the 2-SFM boost function, \citet{bethermin12} successfully used it as the basis for matching observed IR source counts. Their analysis was a good test of the viability of the 2-SFM description of SBs since it employed different IR SEDs for MS galaxies and SBs.

\subsubsection{Link to Dark Matter}

Merging and SF activity -- even in the ``secular" mode -- reflect the accretion of DM and the primordial gas bound to the DM halos. It is thus interesting to check whether there are clear similarities between the distribution of SFRs of galaxies of a given mass and the accretion of DM onto the corresponding parent halos. \citet{dekel09} determined the DM infall rates at the virial radius of $>$100 simulated DM halos with mass 10$^{12}$\,\msun\ at $z$\,$\sim$\,2.5, which typically host $\sim$10$^{11}$\,\msun\ galaxies. For these systems (which have masses comparable to the galaxies used by \citealp{rodighiero11} to construct distributions of sSFR) the DM accretion spectrum shows an extended tail of high accretion rates which is dominated by ``major" merging activity where the mass ratio between accreted and parent DM halo is fairly high. It is obviously tempting to associate this feature of the DM accretion rate distributions in the simulations presented in \citet{dekel09} to the tail of excess SFRs contributed by SB galaxies while the smooth accretion would then fuel the sustained secular mode of SF that is characteristic of SFGs on the MS. This was already proposed by \citet{dekel09}, who also point out that in this context the abbreviation ``SFG" could legitimately stand for ``stream-fed galaxy". In analogy to our split of the sSFR-distribution in Equation \ref{eq:DGintro}, T. Goerdt et al. (in prep.) have decomposed the DM accretion rate distribution for such $z$\,=\,2.5 DM halos with virial mass 10$^{12}$\,\msun\ into two log-normal contributions and find that the one shifted to high accretion rates contributes approx. 10\% to the infalling mass budget. This value is strikingly similar to the 14.2$^{+1.7}_{-1.3}$\% (68\% confidence limits) we inferred in S12 for the contribution of burst-like SF to the total SFRD at $z$\,=\,2.

\subsection{Universal Star Formation Laws and the Distribution of Galaxies in the Schmidt-Kennicutt Plane}
\label{sect:SFEdiscuss}

The tightness of the SF law (dispersion $\sim$0.2\,dex or less than a factor two) that we found using our newly ``homogenized" literature data in Section \ref{sect:MgasvsSFR} is remarkable and points to a very direct and apparently ubiquitous link between the global molecular content of galaxies and how much of it is being converted into stars. It is akin to stating that, for normal disk galaxies out to at least $z$\,$\simeq$\,2.5, once the \sfr\ has been measured the size of the associated molecular gas reservoir can be inferred with high accuracy, and vice-versa. In our calibration of the integrated S-K law in Section \ref{sect:calib}, we adopted a statistical (metallicity-dependent) estimate of the CO-to-H$_2$ conversion factor, \aCO, for 90\% of the normal galaxies in our reference sample when we were translating the observed correlation between \lir\ and \lco\ to a more physical relation between \sfr\ and \mmol. While using an average \aCO\ and neglecting the associated scatter in principle artificially reduces the dispersion of the S-K law, it is not inconceivable that a dispersion in \aCO\ has produced the observed width of the \lir\ versus \lco\ relation. Simulations by \citet{feldmann12b} for example suggest that on kiloparsec-scales and above variations in the conversion factor can be as large as 0.15\, dex. The underlying SF law thus might be intrinsically even tighter than the 0.2\,dex we measure here. To truly test the universality of the ``normal"-galaxy S-K law there are at least two complementary ways forward. On the one hand, it will be important to compile samples of galaxies for which SF and gas estimates rely on strictly identical tracers (e.g. the ground-state transition of $^{12}$CO). On the other hand, it may prove worthwhile to assess in detail (e.g. by means of a full sampling of the SLEDs of molecular gas tracers) how SF proceeds in different phases of the ISM. Having the capability of doing this in a resolved fashion, we also have the potential to reveal what causes the mild \sfr\ dependence of \sfe, which manifests itself as a non-linear slope of the integrated S-K relation (\sfr\,$\propto$\,$M_{\rm mol.}^{1.2}$). \citet{saintonge12} have proposed that the rise of \sfe\ across the MS is due to morphological ``stabilization" of the ISM in bulged galaxies \citep[see][]{martig09}, which are more abundant on the lower part of the MS locus. Resolved studies of the SF law will also be able to reveal whether a similar mechanism is responsible for the \sfe\ increase with redshift in massive galaxies; while it could be due to the increasing absence of bulged galaxies at high redshift \citep[e.g.,][]{oesch10}, it seems just as plausible that SF in an increasingly turbulent medium including massive star-forming clumps would proceed in a more efficient fashion.

\noindent The variations of \sfe\ with \sfr\ among normal galaxies are small compared to the strong \sfe\ enhancements that are observed in starbursting systems. In Section \ref{sect:2-SFM_SFE} we introduced an empirical, supra-linear scaling between the \sfe\ and (s)\sfr\ increase during SB episodes. This relation can be understood very intuitively by the balance between the three quantities involved: SFR and  \mstar, which grow in the burst-phase, and \mmol\ which decreases as gas is converted into stars. \sfe, as the ratio between SFR and \mmol, thus inevitably increases more strongly than \ssfr. An immediate consequence of this is that we expect a more spread-out distribution of SFEs than is observed. In Figure \ref{fig:sSFRnSFEdistrib} we illustrate how, in the 2-SFM framework, the overlapping \ssfr\ distributions of normal galaxies and SBs move apart into a more clearly double-peaked \sfe\ distribution. Discrete recipes for the assignment of CO-to-H$_2$ conversion factors to normal galaxies and SBs are thus not the only way to obtain a bimodal distribution of galaxies in the S-K plane; this can also be achieved with a more physical, continuous description of SF in SBs. Width and depth of the trough we predict between the ``sequence of disks" and ``sequence of starbursts" in the S-K plane depend on the shape of the lower end of the boost function (see discussion in Section \ref{sect:boostcontext}). Our model of an unbiased profile through the S-K plane at fixed gas mass does highlight, however, that observations of large cosmological volumes are necessary to fully sample the actual distribution of galaxies with respect to \sfe: the relative amplitude of the \sfe\ distributions of SBs and MS galaxies in Figure \ref{fig:sSFRnSFEdistrib} is expected to be a factor of 30 and the contrast between the peak of the SB distribution and the trough merely a factor two. A first attempt to construct a representative sampling of S-K space using the COLD-GASS survey was presented in \citet[see their Figure 6b]{saintonge12} and demonstrated just how insignificant SBs are in determining the shape of the SF law for the bulk of the population.

\begin{figure*}
\epsscale{1.13}
\centering
\plotone{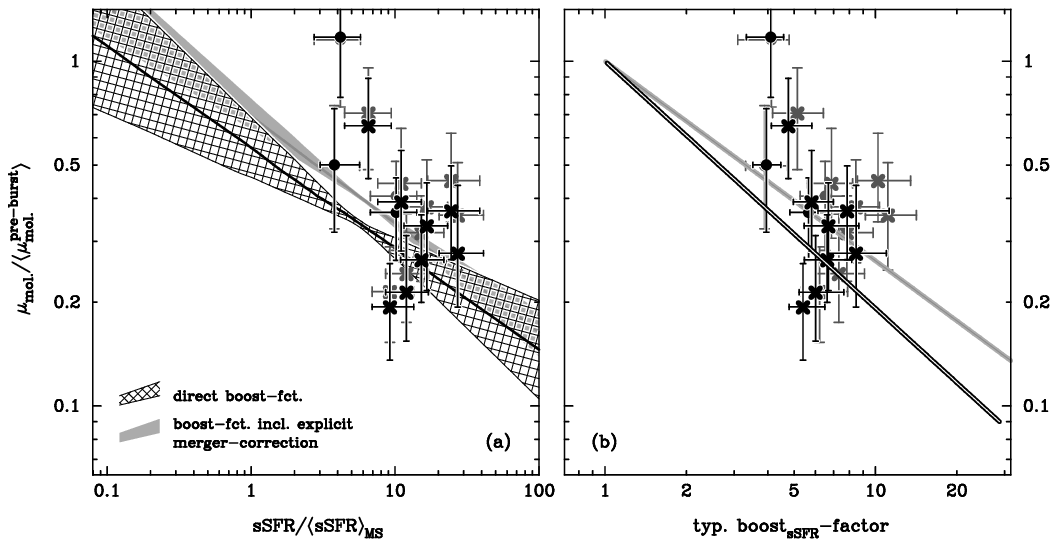}
\caption{\footnotesize ({\it a}) Predicted ratio between the molecular gas-to-stellar mass ratio of starbursts (SBs), $\mu_{\rm mol.}$\,$\equiv$\,$M_{\rm mol.}/M_{\star}$, and the average expected mass ratio $\langle\mu^{\rm pre-burst}_{\rm mol.}\rangle$ prior to the onset of the burst (i.e. during a phase of secular growth on the main sequence (MS)). The variation with \ssfr\ reflects the different \ssfr-dependence, for normal galaxies and SBs, of molecular gas-to-stellar mass ratios predicted by the 2-SFM description (see Figure \ref{fig:loon}(b); dashed and dotted lines, resp.). All symbols are as in Figure \ref{fig:boostconvert}. The pre-burst gas-to-stellar mass ratios of SB galaxies from \citet{downessolomon98} and \citet{magdis12b} were inferred assuming that these sources experienced the median (s)\sfr\ boost expected for sources with the same \ssfr\ excess (see Figure \ref{fig:boostdistrib}). ({\it b}) Dependence of the ratio $\mu_{\rm mol.}/\langle\mu^{\rm pre-burst}_{\rm mol.}\rangle$ on the \ssfr\ boost of SB galaxies (see Equation \ref{eq:fgas_in_SBs}). The black (grey) line shows the 2-SFM prediction for the average trend for the direct (merger-corrected) boost function.
\label{fig:fgasinSB}}
\end{figure*}

\subsection{The Consumption of Gas Reservoirs During Starbursts}
\label{sect:fgas_in_SBs}

As a consequence of the high efficiency with which gas is converted to stars in SB episodes, the gas reservoir in the host galaxy is used up more quickly than it can be replenished by accretion from the intergalactic medium. In Section \ref{sect:obs_fgas} (see Figures \ref{fig:loon} and \ref{fig:normfgas}) we showed that the overall gas fractions (i.e. gas fractions taking into account the molecular and stellar mass content throughout the whole starbursting galaxy) of SBs in our reference sample are indeed in general lower than the average gas fraction of galaxies which reside on the MS. We can use the 2-SFM description of SBs to explicitly calculate how we expect the gas content of starbursting galaxies to change once they have left their MS state. Given that each of the SBs in our calibration sample was observed in a different stage, this should be viewed as a comparison between the gas fraction prior to the onset of the burst and the gas fraction which would be measured approximately half-way through the SB event. If we consider the gas-to-stellar mass ratio $\mu_{\rm mol.}$\,$\equiv$\,$M_{\rm mol.}/M_{\star}$, the relation between pre-burst ($\mu_{\rm mol.}^{\rm pre-burst}$) and mid-burst gas content takes a simple and only boost-dependent form (see Figure \ref{fig:fgasinSB}(b)):
\small
\begin{equation}
\mu_{\rm mol.}/\mu_{\rm mol.}^{\rm pre-burst} = \left({\rm boost}_{\ssfr}\right)^{1-\gamma_{\sfe}}~.
\label{eq:fgas_in_SBs}
\end{equation}
\normalsize
Here we have used that $\mu_{\rm mol.}$\,=\,\nicefrac{sSFR}{SFE} and that the sSFR and SFE of the SB are $\nicefrac{\ssfr}{\ssfr^{\rm pre-burst}}$\,=\,boost$_{\ssfr}$ and $\nicefrac{\sfe}{\sfe^{\rm pre-burst}}$\,=\,(boost$_{\ssfr})^{\gamma_{\sfe}}$, respectively. In Figure \ref{fig:fgasinSB}(a) we plot the variation of the typical ratio between mid-burst and pre-burst gas-to-stellar mass ratio as a function of \ssfr\ excess. This average trend is the result of pairing up each point on the dotted curve for the evolution of $\mu_{\rm mol.}(\ssfr)$ for SBs in Figure \ref{fig:loon}(b) with a position on the corresponding relation for normal galaxies (dashed line in the same figure) by means of the boost-value $b_{\ssfr}^{\rm max}$ at the peak of the \ssfr-dependent boost distribution (see Figure \ref{fig:boostdistrib} and Section \ref{sect:boostspec}). Note that this calculation assumes the stellar mass in the starbursting galaxy and its pre-burst, MS state to be equal. By neglecting the fact that stellar mass has been added to the system during the first phase of the burst, our estimate of the ratio of mid-burst to pre-burst gas fraction effectively represents an upper limit. This simplification also makes the average trends in Figure \ref{fig:fgasinSB} independent of stellar mass and redshift, because all dependence of the absolute value of the pre-burst gas fraction on these two factors (see e.g. Figure \ref{fig:fgasscatter}) is eliminated. The hatched/shaded regions straddling the median trend for $\mu_{\rm mol.}/\mu_{\rm mol.}^{\rm pre-burst}$ reflect the 1\,$\sigma$ uncertainty on the relation between average boost and \ssfr\ (see also Figure \ref{fig:boostconvert}(a)). \\
Based on the theoretical understanding derived from the 2-SFM approach, we are for the first time also able to infer -- in a statistical sense -- the pre-burst gas fractions of the SB galaxies in our reference sample, i.e. of eight local ULIRGs from \citet{downessolomon98} and of three high-redshift SBs studied by \citet{magdis12b}. We do so under the same assumptions as already used to derive the theoretical curve discussed above and superimpose our estimates on the 2-SFM prediction in both panels of Figure \ref{fig:fgasinSB}. Given that the typical \ssfr\ excess of our reference SBs is about a factor of 10, their median $\mu_{\rm mol.}/\langle\mu_{\rm mol.}^{\rm pre-burst}\rangle$ of $\sim$0.35 is in quite good agreement with the 2-SFM prediction for the scenario of the ``direct" boost function (black line in Figure \ref{fig:fgasinSB}(a)). For the most common SBs, which have an (s)\sfr\ boost equal to four, the molecular mass-to-stellar mass ratio half-way through the burst is expected to lie around 35\% (45\%) of its initial value for the direct (merger-corrected) boost function.\\
It is interesting to explicitly compare our constraints on the gas fraction decrease during the SB phase with that expected in the case that the SB is triggered by a major merger. We approximate the \sfr\ evolution during the interaction-induced burst by a top-hat function such that, at a time $t_{1/2}^{\rm merger}$ after the beginning of the burst, stellar mass and gas mass become \mstar\,=\,$M_{\star}^{\rm pre-burst}$\,+\,\sfr${\times} t_{1/2}^{\rm merger}$ and $M_{\rm mol}$\,=\,$M_{\rm mol}^{\rm pre-burst}$\,-\,\sfr${\times} t_{1/2}^{\rm merger}$. We write the mid-burst gas fraction as
\small
\begin{eqnarray*}
f_{\rm mol.} &=& \frac{M_{\rm mol.}}{M_{\rm mol.} + \mstar} = f_{\rm mol.}^{\rm pre-burst}-\frac{{\sfr}\times t_{1/2}^{\rm merger}}{M_{\rm mol.} + \mstar}\\
&=& f_{\rm mol.}^{\rm pre-burst}-f_{\rm mol.}\,\frac{{\sfr}\times t_{1/2}^{\rm merger}}{M_{\rm mol.}}\\
&=& f_{\rm mol.}^{\rm pre-burst}-f_{\rm mol.}\,\frac{\sfe}{\sfe^{\rm pre-burst}}\,\frac{t_{1/2}^{\rm merger}}{\tau^{\rm pre-burst}}~,
\end{eqnarray*}
\normalsize
and rearrange terms to obtain an expression for the ratio of the pre- and mid-burst gas fractions:
\small
\begin{equation}
f_{\rm mol.}/f_{\rm mol.}^{\rm pre-burst} = \left(1+({\rm boost})^{\gamma_{\sfe}}\frac{t_{1/2}^{\rm merger}}{\tau^{\rm pre-burst}}\right)^{-1}~.
\label{eq:roughfgas}
\end{equation}
\normalsize
Here we used that the total mass ($M_{\rm mol}+\mstar$) stays constant and that the \sfe\ before and during the SB are related by the power-law in Equation \ref{eq:SFEcalib_SB}. With a typical boost of approx. a factor of 6 and $\gamma_{\sfe}$\,$\sim$\,1.7 (direct boost function) we find $f_{\rm mol.}/f_{\rm mol.}^{\rm pre-burst}$\,$\sim$\,0.5 for $\tau^{\rm pre-burst}$\,$\approx$\,1\,Gyr (the gas depletion time scale of MS galaxies) and $t_{1/2}^{\rm merger}$\,$\approx$\,50\,Myr \citep[we take this number to be about half the time for which interaction-induced SF is sustained in numerical simulations of galaxy mergers; e.g.,][]{dimatteo08, bournaud11b}. Accounting for the mass and redshift dependence of the conversion\footnote{~The relative gas fraction and gas-to-stellar mass ratio of the pre- and mid-burst state are related by
\begin{equation*}
f_{\rm mol.}/f_{\rm mol.}^{\rm pre-burst} = \mu_{\rm mol.}/\mu_{\rm mol.}^{\rm pre-burst}\,\left(\frac{1+\mu_{\rm mol.}^{\rm pre-burst}}{1+\mu_{\rm mol.}}\right)~.
\end{equation*}
For the massive SFGs discussed here, an initial gas-to-stellar mass ratio $\mu_{\rm mol.}^{\rm pre-burst}$ of $\sim$5-10\% and 50-100\% is expected at low and high-redshift, respectively. The term in brackets should thus vary between roughly 1.1 and $<$2.} between relative gas fractions and gas-to-stellar mass ratios, this corresponds to values of $\mu_{\rm mol.}/\mu_{\rm mol.}^{\rm pre-burst}$ in the range of 0.4--0.5. Obviously, the simple calculation leading up to Equation \ref{eq:roughfgas} will in reality be complicated by, e.g., gas loss and heating in merging systems \citep[e.g.,][]{cox04}, the modified balance between the atomic and molecular hydrogen phase in dense, turbulent media, and IMF variations as have been proposed for SB regions (\citealp[e.g.,][]{baugh05, papadopoulos11}, but see also \citealp{tacconi08, hayward13}). Taken at face value, the reasonable consistency of the estimates of the gas fraction decrease as per eqs. \ref{eq:fgas_in_SBs} and \ref{eq:roughfgas} may indicate that neither of these three factors plays a major role (or that these competing effects compensate each other).\\
Systematic comparisons between the molecular gas fractions of normal and starbursting galaxies will reveal whether the trends we proposed based on our small sample are robust. Further tests of the 2-SFM framework will now be discussed in Section \ref{sect:2SFMoutlook}.

\subsection{Observational Validation of Assumptions and Predictions Made by the 2-SFM Approach}
\label{sect:2SFMoutlook}

The 2-SFM framework as we have developed it so far has produced a remarkably simple description of SFGs over the last 10\,Gyr. One may legitimately wonder whether this simplicity is the true imprint of fundamental laws that govern galaxy formation in a cold DM Universe or the outcome of an incomplete or selective view of the star-forming population due to observational limitations. The answer to this question depends to some extent also on the scope of any investigation. The occurrence of extreme behavior in rare outliers or small scale processes with little impact on global system properties -- while relevant for a complete understanding of all complex aspects regulating SF -- does not imply a general inadequacy of a simpler approach, as we have been advocating here, which aims to provide a panoramic treatment. Further confirmation of the validity of the 2-SFM description will instead involve both (a) revisiting some of its key ingredients and (b) testing its predictions.\\
Concerning point (a), the main focus should lie on verifying our hypothesis that galaxies with stellar mass significantly below \mstar/\msun\,=\,10$^{10}$ follow the same relations that were calibrated on galaxies which are more massive than this threshold. The universality of the S-K law, for example, will soon be routinely tested with ALMA down to low stellar masses and out to high redshift by targeted observations of lensed galaxies. With deeper follow-up of molecular transitions, it will also be possible to identify evolution in the normalization \citep[e.g.,][]{tacconi13} and curvature of the SF laws. A second assumption of the 2-SFM framework is that the double log-normal decomposition of the \ssfr\ distribution is applicable also at $z$\,$\neq$\,2. To ascertain this, tracers of dust-obscured SF are indispensable as extinction-corrected \sfr\ measurements underestimate the true \sfr\ of dusty SBs \citep[e.g.,][]{hughes98, trentham99, buat05, chapman05, daddi07a, casey13} and place these on the locus of the star-forming MS. Obtaining good statistics on the rare starbursting sources (comoving number densities are of the order of 10$^{-5}$\,Mpc$^{-3}$) at the high-end tail of the \ssfr\ distribution hence requires a combination of wide-area IR or radio surveys with complementary deep optical or UV data. A non-universality of the double log-normal decomposition in Equation \ref{eq:DGintro} would introduce more variation in the simple \sfe\ versus \ssfr\ excess relations, etc. than is currently suggested by the fairly limited data. Any evolution in the \ssfr\ decomposition into normal galaxies and SBs would imply a more complex, redshift-dependent behaviour of average scaling relations in the space of normalized molecular gas properties than is shown in Figure \ref{fig:loon}. A more fundamental question is whether the $z$\,$\sim$\,2 sSFR distribution of \citet{rodighiero11}, on which we perform the decomposition to begin with, is accurate. Little is known about the lower tail of the sSFR distribution, but it is unlikely that low-sSFR outliers to the MS should be responsible for a significant amount of SF activity (see S12, and references therein). In the absence of a single SF tracer to map out the distribution of galaxies in the SFR-\mstar\ plane, the two-pronged approach of \citet{rodighiero11} for reconstructing it with two different diagnostics (IR emission for dust-obscured galaxies and UV-emission for the bulk of the MS population) relies on the consistency of the associated SF estimates. Extinction-corrected UV-fluxes and IR measurements at $z$\,$\sim$\,2 are known to agree in an average sense \citep[e.g.,][]{daddi07a}, but the dispersion about the mean extinction correction could potentially contribute to the observed scatter of the MS of SFGs. The analysis of the MS at 0.5\,$<$\,$z$\,$<$\,1.3 by \citet{salmi12} suggests that at least at these redshifts the dispersion of the sequence is mainly intrinsic. Since in \citet{rodighiero11} the sSFR distributions of galaxies with \mstar\,$\geq$\,10$^{11}$\,\msun\ are identical when computed with UV- or IR-emission, this seems to hold for the high-mass end of the $z$\,$\sim$\,2 MS as well.\\
Concerning point (b), the analysis of this paper produced predictions that will be tested in future CO follow-up observations. With a good sampling of the transition region between MS and SB galaxies in the \sfr--\mstar\ plane ($\nicefrac{\ssfr}{\ssfrMS}$\,$\in$\,[3, 5]), these observations will quantify the scatter of, e.g. \sfe\, and determine whether it is indeed larger than elsewhere, as is expected for a heterogeneous mixture of normal galaxies and SBs. While we predict such an increased dispersion to be measurable even using direct observables, e.g., \lco\ and \lir, an estimate of the CO-to-H$_2$ conversion factor \aCO\ is necessary to calculate actual values of \sfe\ and gas fractions. Our predictions for the variation of \aCO\ in the \sfr--\mstar\ plane in Section \ref{sect:XCOmaps} are in principle testable, but obtaining high-confidence measurements of \aCO\ will remain a challenging task that is best tackled using different, complementary strategies in parallel. The gas-to-dust ratio technique employed by \citet{leroy11, magdis11, magdis12b, magnelli12} is powerful, in that it can provide constraints on the conversion factor for large data sets. However, the large scatter in measured gas-to-dust ratios in local, low-metallicity and low-mass galaxies \citep[e.g.,][]{draineli07, galliano08, galametz11,remy-ruyer14} indicates that this method becomes highly inaccurate for $z$\,$>$\,3 galaxies and galaxies with stellar mass \mstar\,$\ll$\,10$^{10}$\,\msun. An alternative approach is to interpret dynamical constraints from CO line profiles in the context of numerical simulations to infer \aCO\ as proposed by \citet{daddi10a}. However, even barring the systematic uncertainties on model DM distributions, the application of this method to large data sets may be impracticable as it requires high signal-to-noise data and a fine spectral sampling of the emission feature.\\
Finally, we note that the (s)\sfr\ boost of SBs, although much less easily determined than their (s)\sfr\ excess with respect to the average of the MS population, is in principle measurable using high-fidelity and ideally also spatially resolved spectroscopy. When compared to the output of stellar evolution models, this kind of data would allow a detailed reconstruction of the SFH of boosted sources prior to the onset of burst activity. It would hence also reveal whether galaxies that show a strong \ssfr\ excess are truly experiencing short-term boosting of their activity at all redshifts or whether they are merely a high-intensity tail of the ``normal" population. The episodic and merger-related nature of ULIRGs at low redshift is well-accepted \citep[e.g.,][and references therein]{sandersmirabel96} but is harder to prove for SBs in the distant universe \citep[e.g.,][]{tacconi06, daddi09, ivison13}. The supporting evidence which has been accumulating in recent years, however (e.g., IR diagnostics, ISM temperatures, host galaxy structure and kinematics; see also our overview in the introduction and the discussion in Section \ref{sect:boostcontext}), is at the basis of our proposed split into SB and normal galaxy populations and the assumption that it provides a valid description of the star-forming population of much of the history of the universe.

\section{Summary}
\label{sect:summary}

The 2-Star Formation Mode (``2-SFM") framework provides a conceptually simple and self-consistent scheme for the prediction of basic properties of the star-forming galaxy (SFG) population. It relies on basic observables -- e.g. the evolution of specific star formation rate (sSFR) in main-sequence (MS) galaxies or their stellar mass (\mstar) distribution -- and their mathematical description -- e.g. the Schechter function parametrization of the stellar mass function or slope and normalization of the Schmidt-Kennicutt (S-K) law -- to produce an analytico-empirical description of the statistical properties of SFGs which can be both predictive and help (re)interpret existing measurements. A central ingredient of the 2-SFM framework is the distinction between ``normal" SFGs that reside on the star-forming MS and starbursts (SBs) that are much rarer and regarded here as a ``perturbation" of the MS state (see Section \ref{sect:2SFMintro}) that is probably dynamically induced or induced by interactions. We recently applied this approach successfully for the prediction of IR luminosity functions at $z$\,$\lesssim$\,2.5 in S12 and of galaxy number counts between 24 and 1100\,$\mu$m and at 1.4\,GHz in \citet{bethermin12}. In this article we have investigated the observational evidence that the molecular gas properties of massive (\mstar\,$\gtrsim$\,10$^{10}$\,\msun) SFGs are amenable to a similarly simplified description as their IR-emission.\\
We use a sample of approx. 130 normal SFGs (see Sections \ref{sect:MSdata_loz} and \ref{sect:MSdata_hiz}) to calibrate scaling relations that allow us to predict -- in anticipation of future complete and unbiased surveys of the ISM content of galaxies -- how molecular gas properties of secularly evolving SFGs at $z$\,$<$\,3 vary depending on their SFR and stellar mass. When all involved quantities are normalized to the value a given observable takes for an average MS galaxy, these trends become strikingly simple (and in general also independent of redshift). In particular, we find that:
\newcounter{saveenum}
\begin{enumerate}
\item All literature measurements of SFR and \mmol\ in massive (\mstar\,$>$\,10$^{10}$\,\msun) MS galaxies at $z$\,$<$\,3 are compatible with the existence of a universal (i.e., redshift-invariant) star formation (SF) law for such systems. This integrated S-K relation is slightly supra-linear (\sfr\,$\propto$\,$M_{\rm mol.}^{1.2}$) and tight (dispersion $\sim$0.2\,dex; see Figure \ref{fig:invSKcalib}). 
\item Star formation efficiency (SFE) varies very little across the MS (see Figure \ref{fig:normSFE}) while the molecular gas mass fractions, \mmol/\mstar, increase almost linearly with (s)SFR for MS galaxies of a fixed stellar mass.
\item Changes in the sSFR of MS galaxies are strongly correlated with changes of the molecular gas fraction, implying that both the dispersion of the MS and the cosmic evolution of sSFR in general reflect variations of the gas content of normal galaxies (see Figure \ref{fig:avfgasevo}).
\setcounter{saveenum}{\value{enumi}}
\end{enumerate}
Based on this characterization of gas in the MS population, we are then able to predict the molecular gas properties of SB galaxies which -- in the 2-SFM approach -- start out as normal galaxies that subsequently experience boosting to higher (s)SFRs. In this paper we go beyond assuming that there are two discrete modes of SF. Instead, we adopt a continuous description of SFE-variations for SBs, in which small SFR enhancements translate to small SFE increases as well. By considering the excess SFR and excess SFE of observed SB galaxies with measured CO-to-H$_2$ conversion factors \aCO, we infer that SFE grows more strongly in the burst-phase than SFR (see Section \ref{sect:2-SFM_SFE}). Taking into account the changing, sSFR-dependent mixture of SB and normal galaxies that constitutes the total star-forming population, this leads to the following expectations:
\begin{enumerate}
\setcounter{enumi}{\value{saveenum}}
\item Normal SFGs and SBs are separated more strongly in the S-K plane than in the space of \mstar\ and SFR (see Figure \ref{fig:sSFRnSFEdistrib}). However, a separation that is as discrete as currently suggested by observations is not expected and is likely the outcome of the incomplete sampling of the S-K plane in surveys explicitly targeting strong SBs and average MS galaxies.
\item Even if SBs are treated as a continuous extension of normal galaxies, with depletion times that decrease in proportion to their burst-related (s)SFR enhancement, a nearly step-like, roughly tenfold increase of the SFE is predicted at the sSFR where starbursting sources begin to outnumber the MS population (see Figure \ref{fig:loon}(a)).
\item A similar, albeit less pronounced step-like behavior is predicted for molecular gas fractions (see Figure \ref{fig:loon}(b)): while these continuously rise across the MS (see point 2 above), the higher SFE of SBs causes their gas fractions to decrease to a value that is smaller than the average observed for a typical MS galaxy. In Section \ref{sect:fgas_in_SBs} we provide recipes for how much gas fractions are expected to drop, depending on the intensity of the SB, if -- as is expected for, e.g., merger-induced SBs -- the timescale for the exhaustion of the molecular fuel reservoir is much shorter than the timescale for accretion of pristine gas from the cosmic web.
\end{enumerate}
Based on the systematic difference between the \lco/\lir\ and \mhtwo/\lir\ ratios of SBs, we derive an empirical recipe for the CO-to-H$_2$ conversion factor, \aCO, of SB galaxies (see Section \ref{sect:SB-XCO}). In combination with an assumed metallicity dependence of \aCO\ for MS galaxies, we are able to predict \aCO\ variations for SFGs throughout the \mstar\ versus SFR plane (see Section \ref{sect:XCOmaps}). Due to the flatness of the mass-metallicity relation at high stellar masses, the conversion factor of Milky-Way-mass galaxies is expected to resemble the canonical Milky Way value even at the cosmic epoch when the SF history of the universe peaked.\\
Our understanding of molecular gas at high redshift will progress rapidly in the near future as the ALMA observatory acquires increasing volumes of data that will quickly outgrow the currently available information. The 2-SFM description of SFGs provides a flexible methodological framework that can adapt to future findings, e.g. by re-calibrating the relation between SFR boosts and SFE enhancements once larger samples of SB galaxies become available, or by recalibrating the Schmidt-Kennicutt relation should new measurements reveal that the relation between SFR and molecular gas mass is more complex than it appears at present. The simple, analytico-empirical description of molecular gas in star-forming galaxies developed in the present work will be used to infer the evolution of molecular gas mass functions and CO luminosity functions, as well as CO source counts in two forthcoming papers.

\acknowledgments
We thank F. Bournaud, A. Cibinel, P. Di Matteo, S. Ellison, R. Feldmann, A. Karim, K. Kraljic, S. Lilly, G. Popping, F. Renaud, D. Riechers and F. Walter for helpful discussions/suggestions, as well as J. Scudder for providing the data shown in Figure \ref{fig:boostcomp} and A. Leroy for kindly sharing new HERACLES measurements with us ahead of publication. We are grateful to both our referees for a careful reading of the paper and valuable suggestions for improvements.\\
M.T.S., M.B. and E.D. acknowledge financial support from the EC through ERC-StG/ UPGAL 240039 and grant ANR-08-JCJC-0008. S.J. was supported by ERC-grant ERC-StG-257720.\\
This article is partly based on observations with AKARI, a JAXA project with the participation of ESA. It has also made use of NED which is operated by the Jet Propulsion Laboratory, California Institute of Technology, under contract with the National Aeronautics and Space Administration. Much of the analysis presented here was carried out in the Perl Data Language (PDL; Glazebrook \& Economou, 1997), which can be obtained from \texttt{http://pdl.perl.org}.

\appendix

\section{Literature measurements of specific star formation rate in MS galaxies}
\label{appsect:sSFR}

\begin{figure}
\epsscale{.6}
\centering
\plotone{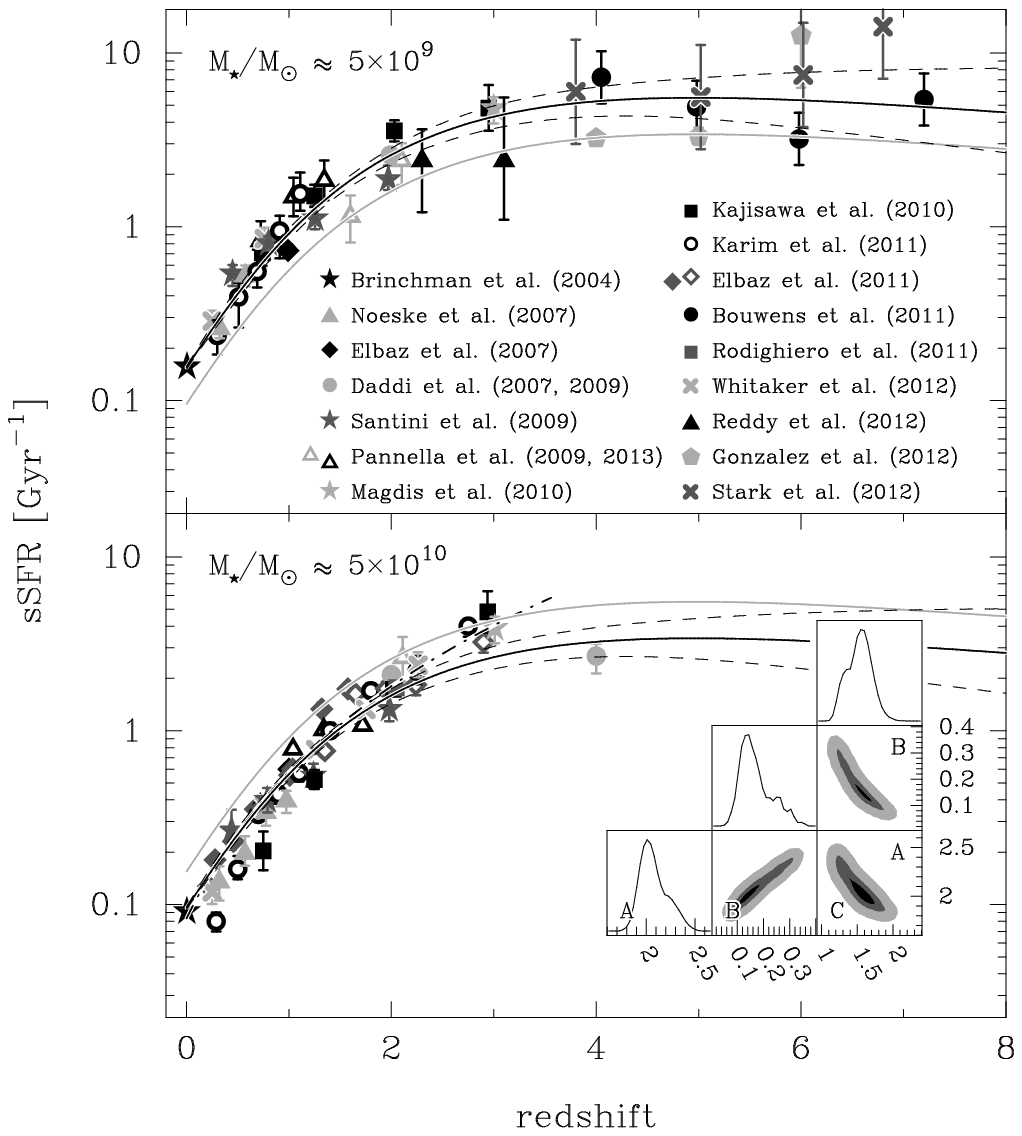}
\caption{\footnotesize Redshift dependence of the sSFR of SFGs with stellar mass \mstar/\msun\,$\approx$\,5$\times$10$^9$ ({\it top}) and 5$\times$10$^{10}$ ({\it bottom}), as published in the recent literature (see legend; where necessary, literature values from adjacent mass bins were used to interpolate to the mass scales displayed here). Measurements derived based on image-stacking are indicated with open symbols and error bars denote the uncertainty on the sSFR average rather than the sSFR scatter in the population. Solid/dashed black lines -- the best-fit evolution of the sSFR -- parameterized as in Equation \ref{eq:sSFRevoeq} (see inset panels on lower right for the covariance between the free parameters of the fit) -- and associated 2\,$\sigma$-errors.; light grey lines -- sSFR evolution in the other of the two stellar mass bins depicted in the figure, for comparison; dot-dashed line -- evolution according to (1+$z$)$^{2.8}$ as used in S12 for the range $z$\,$\lesssim$\,2.
\label{fig:sSFRevo}}
\end{figure}

The locus of the MS is known to depend on sample selection \citep[e.g.,][]{karim11} and, in particular, on how actively star-forming the sample under consideration is. With the aim of deriving a representative, average evolution we gathered measurements from several recent studies of the distribution of SFGs in the (\mstar, \sfr) plane that employed different selection criteria (e.g., different color cuts or selection by morphology, by near-IR flux/mass or by SFR), chose different SF tracers (e.g., UV, IR or radio emission) and/or adopted a variety of measurement techniques (e.g. individual detections versus source stacking). By considering two separate mass scales at $5{\times}10^9$ and $5{\times}10^{10}$\,\msun\ we obtain a constraint on the typical exponent $\nu$ of the \mstar\ dependence of \ssfr, \ssfr\,$\propto$\,\mstar$^{\nu}$. We find that an exponent $\nu$\,$\simeq$\,-0.2 reproduces the systematic shift between the sSFR evolution of galaxies in the two \mstar-bins (see Figure \ref{fig:sSFRevo}). This slope agrees closely with the value $\nu$\,=\,-0.21$\pm$0.04 we adopted \citep[based on the $z$\,$\sim$\,2 MS presented in][]{rodighiero11} in our previous publications investigating the viability of the 2-SFM framework \citep[see S12,][]{bethermin12}. Our literature compilation covers the redshift range $z$\,$<$\,7, with a majority of the measurements tracing the steep rise of \ssfr\ in MS galaxies out to $z$\,$\sim$\,3. At $z$\,$>$\,4, drop-out samples constrain the \ssfr\ evolution at \mstar\,$\sim$\,$5{\times}10^9$\,\msun, but they do not contain enough high-mass galaxies to probe the evolution in our second, more massive bin.\\
We parameterize the evolution of sSFR with a smoothly varying function of redshift with five free parameters which we fit to the data in Figure \ref{fig:sSFRevo}:
\small
\begin{equation}
{\ssfr}(\mstar,\,z) = N(\mstar)\,{\rm exp}\left(\frac{A{\cdot}z}{1+B{\cdot}z^C}\right)~, \text{\normalsize where}\label{eq:sSFRevoeq}
\end{equation}
\begin{eqnarray*}
N(\mstar) &=& N(5{\times}10^{10}\,\msun)\,10^{\nu\,{\rm log}\left(\mstar/[5\times10^{10}\,\msun]\right)}\\
A &=& 2.05_{-0.20}^{+0.33} \\
B &=& 0.16_{-0.07}^{+0.15} \\
C &=& 1.54\pm0.32
\end{eqnarray*}
\normalsize
Here $\nu$ is the slope of the log(\ssfr) versus log(\mstar) relation as above and the normalization at a stellar mass of 5$\times$10$^{10}$\,\msun\ is $N$(5$\times$10$^{10}$\,\msun)\,=\,0.095$_{-0.003}^{+0.002}$\,Gyr$^{-1}$. The quoted uncertainties are 68\% confidence limits as determined by a Monte Carlo Markov Chain (10$^6$ realizations).\\
In Figure \ref{fig:sSFRevo} the 1\,$\sigma$-errors on the average \ssfr\ evolution according to Equation \ref{eq:sSFRevoeq} are marked by dashed lines. Due to the abundant data \citep[e.g.,][]{noeske07, elbaz07, daddi07b, daddi09, pannella09, karim11, rodighiero11, whitaker12} and their generally high fidelity (error bars on the individual literature measurements span the statistical uncertainty on the mean rather than the population dispersion), formal uncertainties at $z$\,$<$\,2 are small but increase steadily thereafter, reflecting the much sparser data at the highest redshifts. The growing formal errors do not include the systematic evolutionary uncertainties at $z$\,$>$\,3, where the {\ssfr}s of drop-out galaxies have been subject to frequent revision on an almost yearly basis. Initial measurements at 4\,$<$\,$z$\,$<$\,8 by \citet{stark09} and \citet{gonzalez10} -- subsequently modified by \citet{bouwens12} to account for dust-extinction -- suggested a much more gradual \ssfr\ evolution than expected by most theoretical models (see, e.g., \citealp{weinmann11}, and references therein). The most recent efforts have focused on quantifying the impact of nebular emission lines \citep[e.g.,][]{schaerer10, stark13, debarros14, gonzalez14} on stellar mass measurements. It is presently unclear, however, whether corrections for nebular emission cause significant deviations from the nearly flat evolution that was found prior to their implementation: line-corrected \ssfr\ values scatter about the \citet{bouwens12} measurements\footnote{~Independent, albeit tentative evidence for a flattening of the \ssfr\ evolution in the range 2\,$<$\,$z$\,$<$\,4 is provided by the similar gas fractions in $z$\,$\sim$\,2 BM/BX-selected galaxies and two $z$\,$\sim$\,3 LBGs by \citet{tacconi10} and \citet{magdis12a}, respectively. See \citealp{magdis12a} and our Section \ref{sect:fgas_evo} for further discussion of the co-evolution of gas fractions and {\ssfr} with cosmic time.} at $z$\,$<$\,6 and only then become consistently larger than non-corrected ones. Our analytical parameterization of the \ssfr\ evolution does not trace the apparent increase at $z$\,$\geq$\,6 but these high redshifts are not the main focus of this article.

%%%%

\end{document}